\documentstyle[]{l-aa}
\input epsf
\def\lsim{\lower.4ex\hbox{$\;\buildrel <\over{\scriptstyle\sim}\;$}}

\def\bib{\bibitem{}}
\newcommand{\xia}{\overline{\xi}}
\newcommand{\rhob}{\overline{\rho}}
\newcommand{\gam}{\gamma}

\newcommand{\beq}{\begin{equation}}
\newcommand{\eeq}{\end{equation}}
\begin{document}
%
%
\topmargin=2.5 cm
\renewcommand{\textfraction}{.01}
\renewcommand{\topfraction}{0.99}
\renewcommand{\bottomfraction}{0.99}
\setlength{\textfloatsep}{2.5ex}
\thesaurus{11.12.2 , 11.05.2 , 11.17.3 , 12.12.1}
\title{The Mass and Luminosity Functions of Galaxies and their Evolution}   
\author{Patrick Valageas \and Richard Schaeffer}  
\institute{Service de Physique Th\'eorique, CEN Saclay, 91191 Gif-sur-Yvette, France} 
\maketitle
\markboth{Valageas \& Schaeffer: The Mass and Luminosity Functions of Galaxies and their Evolution}{Valageas \& Schaeffer: The Mass and Luminosity Functions of Galaxies and their Evolution}

\begin{abstract}

We set up a model for the evolution of the galaxy luminosity function, taking advantage of recent work that brought in some better understanding of the mass function for gravitationally condensed objects. We add to this a simple model of star formation that reproduces the behaviour of the Tully-Fisher relation, in order to attach a luminosity to a massive halo with a given velocity dispersion. The physics of cooling of the gravitationally heated baryonic component allows us to distinguish halos that become groups or clusters  from those that eventually form galaxies (possibly within the former objects). With our new mass function and our new application of the cooling criteria - which motivated this paper - we get a satisfactory and natural cutoff at the bright end of the luminosity function, the needed flat slope for faint magnitudes and the correct trend in colors (brighter galaxies are redder) within the framework of the hierarchical clustering picture. This infirms earlier claims that the latter was inadequate to reproduce the former observations. 
We find the velocity dispersion to be a much better parameter than mass or radius to characterize galaxies. This model of the salient features that may describe galaxies allows one to discuss galaxy evolution as a function of redshift, in number as well as in luminosity. We find that bright galaxies form at $z \sim 2$ from mergers with a rather quiet evolution afterwards, whereas small galaxies are the result of a continuous merging process active up to the present epoch. The transition is found to occur at the observed transition between bright spirals and small dwarf ellipticals or irregulars. The galaxy luminosity was larger in a recent past for bright galaxies, as has been observed in the CFRS survey. This is because the mass of gas in a typical $L_*$ galaxy such as the Milky Way is a small fraction of the total baryonic mass and thus star formation is already slowing down. The evolution in number, which is quite well controlled in our model, agrees reasonably well with the counts as a function of apparent magnitude. The quasar multiplicity as a function of redshift is also discussed.

\end{abstract}


\section{Introduction}

The formation and evolution of galaxies can be studied through numerical simulations or analytic (or semi-analytic) models. In principle, the former could follow with great details the numerous processes (growth of density fluctuations, galaxy interactions, cooling processes, star formation histories, effects of supernovae and metallicity...) which govern galaxy formation. However numerical simulations are much less flexible than analytic means, and 
they also have to use physical parameterisations to model key processes like star formation, not only because the required dynamic range would exceed current computational possibilities but also because these physical processes are not yet very well understood. Since the final results depend strongly on such parameterisations (Navarro \& White 1993), analytic or semi-analytic models provide an attractive alternative way to study galaxy formation, as they clearly show the relative importance of different processes and the influence of the available parameters, which would be more difficult to understand from a complex N-body and hydrodynamic simulation. 
	
The standard calculations follow the ideas of Press \& Schechter (1974) (see also Schaeffer \& Silk 1985), taking advantage of the progress made originally by Bond et al.(1991), currently referred to as the ``excursion set formalism''. The results obtained by such models (White \& Frenk 1991; Kauffmann et al.1993; Cole et al.1994; and subsequent articles) show a reasonably good agreement with various observations which suggests that the main features of the standard scenario (galaxies arise from gas which is able to cool within dark matter halos formed by a hierarchical clustering process) are correct (or provide at least a good approximation). However, several problems are still unsolved: these studies cannot recover simultaneously the normalization of the Tully-Fisher relation and of the B-band luminosity function, the slope of the luminosity function at the faint end is too steep, massive and very bright galaxies usually are not redder than fainter ones contrary to observations and the models often predict a cutoff at the bright end which is too smooth (there are too many very luminous objects). We present in this article an analytic model for galaxy formation and evolution, based on a specific description of density fluctuations in the highly non-linear regime and an original application of the cooling constraints, which provides a solution to these issues and allows detailed predictions for many physical galactic properties together with their redshift evolution.

Thus, the motivations of this study are to:

- take advantage of an improvement (Valageas \& Schaeffer 1997, hereafter VS) over the usual Press-Schechter mass function (Press \& Schechter 1974), that originates in the understanding of the density fluctuations in the deeply non-linear regime (Schaeffer 1984; Balian \& Schaeffer 1989; Bernardeau \& Schaeffer 1991) and amounts to include the important {\it subsequent non-linear evolution} by counting directly the overdensities at the epoch of interest rather than in the primordial universe as implied by the Press-Schechter approximation. Since in this approach objects are defined by taking a snapshot of the actual density field, in order to get the mass function of dark matter halos {\it there is no need to follow the merging history of the various objects}. This is done implicitely since the time evolution of the density field is already built-in within this non-linear hierarchical scaling model (and well verified by numerical simulations, see Bouchet et al.1991; Colombi et al.1997; Munshi et al.1998; Valageas et al.1998a). Note however that the merging history of the dark matter halos enters, but only marginally, when the star formation history is considered. This is discussed in Appendix \ref{Redshift evolution and merging of galaxies}.

- implement in a natural way the {\it cooling constraints}. Indeed, the difference between galaxies and groups (that may have similar masses!) is that the former are able to cool rapidly (Rees \& Ostriker 1977; Silk 1977; White \& Rees 1978) whereas the latter are not. We shall argue that for a given velocity dispersion, galaxies are made from {\it the largest possible baryonic patch that can cool within a few Hubble times at formation}. By contrast, a larger patch which cannot cool rapidly in a {\it uniform} fashion (due to its lower overall density which translates into a larger cooling time) will become a group or a cluster of galaxies while several high density sub-units will cool and form distinct galaxies. Indeed, the non-linear scaling model predicts strong {\it sub-clustering} so that large clusters automatically contain numerous very high density spots that with our method we count as galaxies. For a galactic mass $M$ our constraints translate into a well-defined mass-dependent condition on the galaxy radius $R = R(M)$, or equivalently on its density contrast $\Delta = \Delta(M)$, rather than having a universal (i.e. mass-independent) density contrast for all objects. Our approach is, in this respect, {\it fundamentally different from the earlier ones}. 

We apply this formulation to two descriptions of gravitational clustering, based i) on a Press-Schechter-like approach that uses linear theory and ii) on our non-linear hierarchical scaling model (VS). We show that in both cases it leads to a strong bright-end cutoff as required. As a result, we

- show how the previous puzzles (normalizations of the Tully-Fisher relation and B-band luminosity function, flat slope for the latter, colors) can be solved within a hierarchical scenario. We can note that although relating the luminosity to the mass of the galaxy implies a model of star formation with a few adjustable parameters, observational constraints (Tully-Fischer relation, gas/stars mass ratio,...) allow one to derive a mass-luminosity relation for galaxies independently of any consideration of the galaxy luminosity function. Hence the agreement with the observed luminosity function builds confidence in our model.

- obtain detailed predictions for the {\it redshift evolution} of galaxies. Indeed, since there is nothing special about the present epoch the same considerations predict the evolution of the galaxy mass and luminosity functions in the past (as well as in the future!).

- take advantage of the possibilities of our analytic model to get {\it scaling relations}, analogous to the Tully-Fisher relation, between various physical quantities (circular velocity - luminosity - mass - gas/star mass ratio - metallicity - star formation rate) over well-defined ranges of galaxy luminosity (for instance). Obtaining such laws, which requires the use of an analytic approach, provides a deep level of understanding and a clear presentation of the global trends implied by models based on the standard hierarchical scenario, independently of the details of both the star formation process and gravitational clustering. We think this transparancy and flexibility makes it worthwile to develop analytic models like ours (despite the simplifications they involve) in addition to numerical approaches.

- present a model for galaxy formation within a {\it global description} of gravitational structures and astrophysical objects which provides a unified consistent model for very different phenomena: from clusters (Valageas \& Schaeffer 1998) to Lyman-$\alpha$ clouds (Valageas, Schaeffer \& Silk 1998) and galaxies.

The organization of the paper is as follows: in Sect.\ref{The galaxy mass function} we discuss the determination of the mass function, with two models, one, that we call {\it PS approach}, based on an educated guess of which fluctuations identified in the early universe are going to eventually become galaxies, as prescribed by Press \& Schechter (1974) but with a mass-dependent density contrast, and a second one, based on our previous work, using the same considerations applied directly to the non-linear density field at the epoch under consideration, that we call {\it non-linear hierarchical scaling approach}. In Sect.\ref{Galaxy evolution} we discuss the cooling condition, as well as our star formation model. In Sect.\ref{Numerical results} we construct the galaxy luminosity function at the present epoch using the PS model as well as the non-linear scaling approach for a critical and an open universe. The evolution effects are examined in detail in Sect.\ref{Time-evolution}. Finally, in Sect.\ref{Quasar number density}, these considerations about the mass function and its time evolution are applied to derive the quasar distribution as a function of redshift, with an improvement over the Efstathiou \& Rees (1988) model due to our new mass function that replaces the original one derived using the Press-Schechter model. In the last section we summarize the results of this paper and discuss its similarities and differences with other papers on the same subject (White \& Frenk 1991; Blanchard et al.1992; Kauffmann et al.1993,1998; Cole et al.1994). 

The details of our model are presented in Appendix. In particular, we describe our star formation model in App.\ref{Star formation model for an isolated halo} and App.\ref{Redshift evolution and merging of galaxies}. We discuss the values of our parameters in App.\ref{Scalings} and we present scaling relations in App.\ref{Approximate power-law regimes}. We use $H_0=60$ km/s/Mpc throughout this paper.

\section{The galaxy mass function}
\label{The galaxy mass function}

Mass condensations may be characterized by two parameters, such as for instance velocity dispersion and radius, or equivalently mass and radius. As will be explained in Sect.\ref{Galaxy formation: cooling constraints}, we define galaxies as halos of mass $M$ with a radius $R(M)$ given by a condition provided by virialization and cooling constraints. This can be written as a mean density contrast $\Delta(M)$ within the radius $R(M)$, so that we have a one parameter family of halos. The relation $\Delta(M)$ will also depend on redshift, but since all halos we shall consider will have already virialized we have for all $z$ and $\Omega_0$, and for any mass $M$: $\Delta(M) \geq 177$.

\subsection{Press-Schechter approach (PS)}
\label{Press-Schechter approach}

We shall first devise a ``Press-Schechter approach'' (PS), in the sense that we wish to recognize in the early universe, when overdensities still grow according to linear theory, the density fluctuations which will eventually become halos of mass $M$, density contrast $\Delta(M,z)$, at the redshift of interest $z$. When this density contrast is constant and given by the virialization condition (obtained from the spherical collapse model) this is simply the usual Press-Schechter approximation.
\\

Let us first consider the case $\Omega = 1$. According to the spherical model, an overdensity characterized by the density contrast $\delta(t)$ given by the linear theory, $\delta(t) \propto (1+z)^{-1}$, will slowly separate from the general expansion, its radius will reach a maximum $R_m$ at time $t_m$ and collapse at time $2\;t_m$ when $\delta=\delta_{c0} \simeq 1.68$. We shall assume that in fact the halo virializes at this collapse time at the radius $R_m/2$, as implied by energy conservation and virial equilibrium if kinetic energy is negligible at the turn-around. Hence at time $2\;t_m$ the density contrast of the halo is $\Delta_{c0} \simeq 177$. Then we assume that the density of this object does not evolve significantly so that $(1+\Delta(t)) \propto (1+z)^{-3}$. Thus, a given halo of mass $M$, at redshift $z$, with a density contrast $\Delta(M,z) \geq \Delta_{c0}$, collapsed at the redshift $z_c$ such that $(1+\Delta(M,z)) = (1+\Delta_{c0}) \; [(1+z_c)/(1+z)]^3$ and the density contrast attached to this halo by the linear theory in the present universe is $\delta_c(z_c) = \delta_{c0} \; (1+z_c)$. If we note $\nu=\delta_0/\sigma_0(M)$, where $\sigma_0(M)$ is the amplitude of the density fluctuations extrapolated to $z=0$ by linear theory as usual, we get:
\beq
\nu = \frac{\delta_{c0}}{\sigma_0(M)} \; \left( \frac{1+\Delta(M,z)}{1+\Delta_{c0}} \right)^{1/3} \; (1+z)   
\eeq
as a function of mass and redshift. Then, as long as $\Delta(M,z) \gg 1$, we have
\beq
\Omega=1 \; : \;\; \nu \simeq \frac{3}{10} \; \frac{\Delta(M,z)^{1/3}}{\sigma_0(M)} \; (1+z)
\label{nups}
\eeq 
Then we consider that each halo is characterized by its parameter $\nu$ so that the mass fraction in objects between $\nu$ and $\nu+d\nu$ is:
\beq
\mu(\nu) \frac{d\nu}{\nu} = \sqrt{\frac{2}{\pi}} \; e^{-\nu^2/2} \; d\nu   \label{muPS}
\eeq
where we assumed gaussian initial fluctuations and we corrected by a factor 2 as in the traditional Press-Schechter prescription. Then the comoving multiplicity function of dark matter halos is:
\beq
\eta(M) \frac{dM}{M} = \eta(\nu) \frac{d\nu}{\nu} = \sqrt{\frac{2}{\pi}} \; \frac{\rho_0}{M} \; e^{-\nu^2/2} \; d\nu  \label{etalin}
\eeq
where $\rho_0=\rhob(z=0)$ is the mean density of the present universe. When the density contrast $\Delta(M,z)$ is constant and equal to $\Delta_{c0}$ the formulation (\ref{etalin}) is exactly the usual Press-Schechter multiplicity function. In the case $\Omega_0<1 \; , \; \Lambda=0$, we have $\delta(t) \propto D(z)$ where $D(z)$ is the growing mode of the linear theory and the density contrast of a halo which virializes at redshift $z_c$ is $\Delta_c(z_c)$. Hence we now get:
\beq
\Omega_0 <1 \; , \; \Lambda=0 \; : \;\; \nu =  \frac{\delta_c(z_c)}{\sigma_0(M)} \eeq
where $z_c$ is defined by:
\beq
(1+\Delta(M,z)) =  (1+\Delta_c(z_c)) \; \left( \frac{1+z_c}{1+z} \right)^3
\eeq
The multiplicity function of dark matter halos is still given by (\ref{etalin}). Note that we have the normalization condition:
\beq
\int_0^{\infty} \mu(\nu) \; \frac{d\nu}{\nu} = 1
\label{sumnu}
\eeq

\subsection{Non-linear hierarchical scaling model}
\label{Hierarchical scaling model}

We shall now devise a second method which deals directly with the non-linear regime. We define for each halo the parameter $x$ by:
\beq
x(M,z) =  \frac{1+\Delta(M,z)}{\;\; \xia[R(M,z),z] \;\;}
\label{xnl}
\eeq
where 
\[
\xia(R) =   \int_V \frac{d^3r_1 \; d^3r_2}{V^2} \; \xi_2 (r_1,r_2)  \;\;\;\;\; \mbox{with} \;\;\;\;\; V= \frac{4}{3} \pi R^3
\]
is the average of the two-body correlation function $\xi_2 (r_1,r_2)$ over a spherical cell of radius $R$. Then we write for the multiplicity function of these halos at a given redshift $z$ (see VS):
\beq
\eta(M) \frac{dM}{M}  = \frac{\rho_0}{M} \; x^2 \; H(x) \frac{dx}{x}     
\label{etah}   
\eeq
while the mass fraction in halos of mass between $M$ and $M+dM$ is:
\beq
\mu(M) \frac{dM}{M} = x^2 \; H(x) \; \frac{dx}{x}     \label{muh}
\eeq
The function $H(x)$ is a universal function that depends only on the initial spectrum of fluctuations and that has to be taken from numerical simulations although its qualitative behaviour is well-known: $H(x) \propto x^{\omega-2}$ for small $x$ with $\omega \simeq 0.5$ and $H(x) \propto x^{\omega_s-1} \; \exp(-x/x_*)$ for large $x$ with $\omega_s \sim -3/2$ and $x_* \simeq 10$ to $20$. Moreover, it satisfies:
\beq
\int_0^{\infty} x^2 \; H(x) \; \frac{dx}{x} = 1
\label{sumhx}
\eeq
It has been shown in VS that this function $H(x)$ is quite close to a similar scaling function $h(x)$ that is obtained from the counts in cells. Bounds to estimate the difference between these two functions are given in VS while numerical checks are presented in Valageas et al.(1998a). Hereafter we use the function $h(x)$ in place of $H(x)$. The correlation function $\xia$ that measures the non-linear fluctuations at scale $R$ can be modelled in a way that follows very accurately the numerical simulations (see VS for more details).

\subsection{Comparison of the PS model with the non-linear hierarchical scaling approach}
\label{Comparison of the PS model with the hierarchical scaling approach}

These two formulations look very different, especially for evolution effects. The power at which redshift enters the expressions (\ref{etalin}) and (\ref{etah}) is different even in the exponential. It happens, however, that the parameter $\nu$ of the linear theory can be expressed in terms of the parameter $x$ of the non-linear scaling theory. We will shortly summarize this calculation for $\Omega = 1$. A detailed account can be found in VS. Using an analytic expression for the evolution of the correlation function, one can write the non-linear correlation in the stable clustering regime as a function of the extrapolation of the linear correlation function to the epoch under consideration:
\beq
\left\{ \begin{array}{l} {\displaystyle \xia(R,z) = \left[ \frac{10}{3 \alpha} \; \sigma(R_L,z) \right]^3  }   \\  \\  {\displaystyle  R_L^3 = \left[ 1+\xia(R,z) \right] \; R^3 } \end{array} \right.
\eeq
where $\alpha$ is a parameter close to unity. Using this relation one can obtain for a power-law initial spectrum of index $n$:
\beq
\nu = \frac{1}{\alpha} \; x^{\frac{5+n}{6}}   \label{nux}
\eeq
which shows (VS) that the expression (\ref{etalin}) deduced  from linear theory using the Press-Schechter ideas can be brought into a form quite similar to the non-linear scaling one (\ref{etah}) with: 
\beq
h_{PS}(x) = \sqrt{\frac{2}{\pi}} \frac{5+n}{6\alpha} x^{\frac{5+n}{6} -2} \exp \left[ - x^{(5+n)/3} /(2\alpha^2) \right] \label{hPS}
\eeq 
This holds in the highly non-linear regime, that is for $\xia > \Delta_c$ and $\Delta > \Delta_c$. The definition of galaxies we shall use, which considers that galaxies are objects which have already virialized (and have since undergone a significant cooling) ensures that we always have $\Delta>\Delta_c$ for all galaxies at any time. The condition $\xia > \Delta_c$ implies that (\ref{hPS}) will hold until the redshift $z_f \sim 3$ such that $\xia \simeq \Delta_c$. This similar scaling as a function of the parameter $x$, for $z < z_f$, explains at the same time the similarity of the numerical estimates done with the two mass functions, but also their difference since the scaling functions $H(x)$ and $h_{PS}(x)$  are { \it definitely not the same}.
The coefficient in the exponential is $\simeq 0.05$ to $0.1$ in the non-linear scaling case whereas it is $\simeq 0.5$ in the PS approach, which implies a considerably faster fall-off at large $x$. The powers of $x$, typically larger than unity, entering the exponent in (\ref{hPS}) further increase this difference. Thus, as compared to the non-linear scaling mass function the PS prescription overpredicts the number of intermediate objects ($\nu \sim 1$ or $x \sim 1$) and gives fewer extreme halos (very small or very massive objects), see VS for details. For $z<z_f$ the non-linear scaling approach will give more numerous bright galaxies than the PS prescription, but fewer ones at $z>z_f$. A detailed comparison with numerical simulations of the various mass functions one can define within the framework of the non-linear scaling model, and with the PS prescription, is presented in Valageas et al.(1998a).

\section{Galaxy evolution}
\label{Galaxy evolution}

The formalism introduced in the previous section only deals with the formation of dark halos through the action of gravity. To infer the mass function of galaxies we need to precise when such a halo will form a galaxy, that is we have to express explicitely  the physical processes which lead to the formation of a galaxy.

\subsection{Galaxy formation: cooling constraints}
\label{Galaxy formation: cooling constraints}

\begin{figure}[htb]

\centerline{\epsfxsize=8 cm \epsfysize=5.5 cm \epsfbox{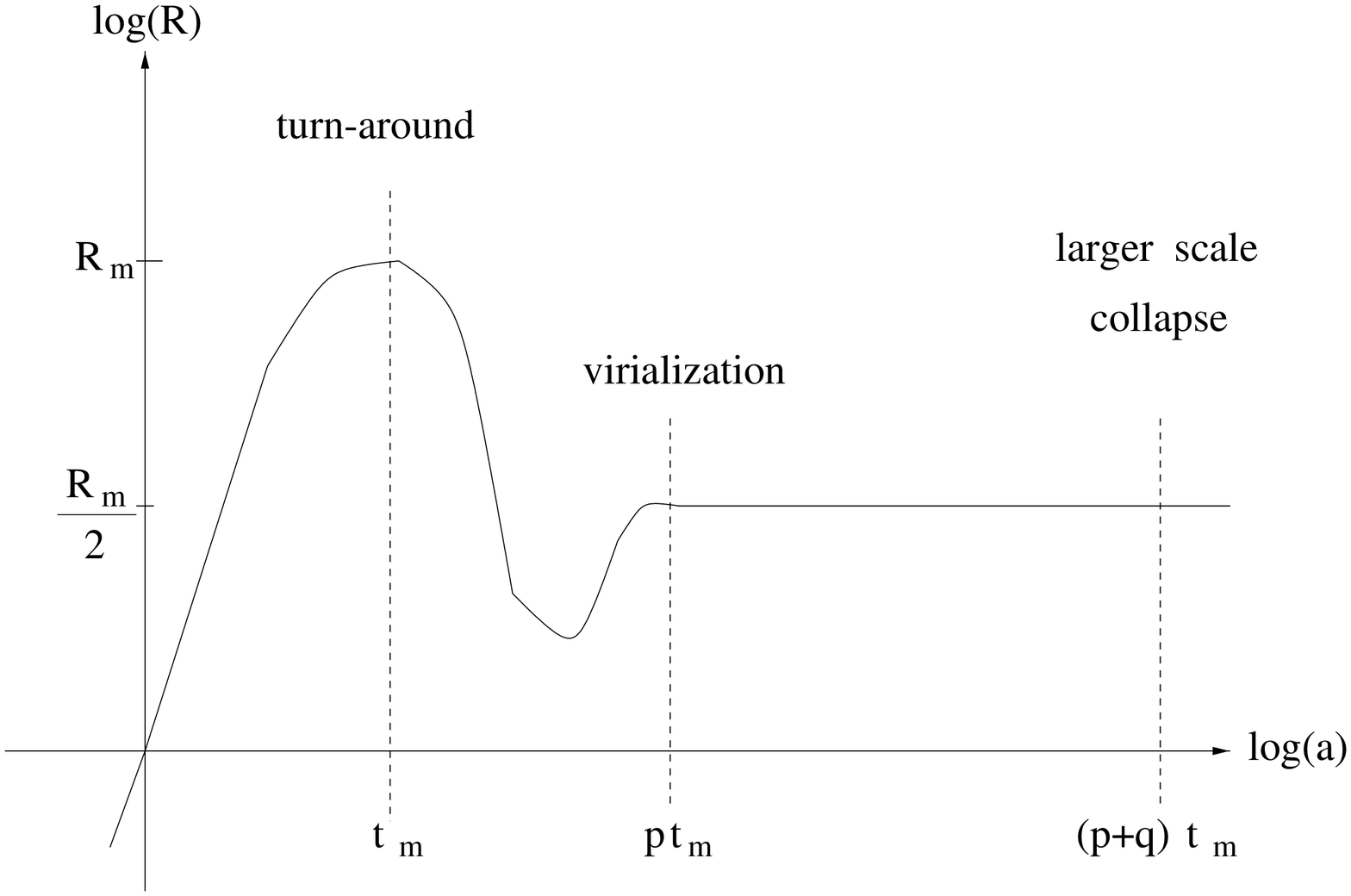}}

\caption{Collapse of an overdensity. Here $R$ is the radius of the external shell, which turns around at $t_m$ and virializes at $p \; t_m$. Matter condensations must {\it virialize} and {\it cool} in order to form a galaxy {\it before} they get embedded, at a time $(p+q)\;t_m$ on average, within a larger condensation.}
\label{figcollap}

\end{figure}

Following the picture of the spherical model, an object with a small overdensity at early times will gradually separate from the general expansion, reach a maximum radius $R_m$ at time $t_m$ and finally collapse at time $2 \; t_m$. One usually assumes that the real halo will virialize at the radius $R_m/2$ at time $p \; t_m$, with $p=2$. We consider that once the halo of dark matter is virialized, the gas can cool and fall into the potential well as it is no longer thermally supported. However, at time $(p+q) \; t_m$ on average we assume that another larger halo containing the previous one will eventually form and will merge the initial overdensity with neighbouring ones. Indeed in the usual hierarchical scenarios, which we consider in this article, larger scales turn non-linear later so that small objects get embedded within increasingly large mass condensations. If the gas contained in the small halo has not cooled sufficiently, it will be distributed over the newly formed overdensity. In other words, {\it the gas has only the time interval $q \; t_m$ to cool and fall into the initial halo in which it was embedded in order to allow the latter to retain its individuality} and possibly form stars to become a galaxy. Fig.\ref{figcollap} displays this sequence of events. Of course cooling does not stop at time $(p + q) \; t_m$. It may go till the present epoch, and the infall of baryons as well as star formation may still be active at much later time.

Thus, we write this cooling constraint as $t_{cool} <  \; q \; t_m$ where $t_{cool}$ is the cooling time. Hence the amount of gas which can cool and fall to the center of the halo, and possibly form a disk and stars, is the mass of gas which is enclosed by the cooling radius $R_{cool}$ where $t_{cool} =  \; q \; t_m$. However, for high redshifts or small temperatures, this radius can be larger than the virialization radius $R_{vir}$, where the averaged density contrast $\Delta$ is equal to $\Delta_c$ (given by $p$). In this case, the actual extension of the halo where the gas comes from is $R_{vir}$, since the surroundings have not collapsed yet. Hence we define the radius of the halos we consider by the constraints: 
\beq
\left\{ \begin{array}{ll}  t_{cool} <  \; q \; t_m  & \hspace{0.5cm} \mbox{curve } {\cal C}_{\Lambda} \\  \;\;\; \mbox{and} \\  \rho >  \; \Delta_c \; {\overline{\rho}} & \hspace{0.5cm} \mbox{curve } {\cal C}_v  \end{array} \right.
\label{Cgal}
\eeq
where $\rhob$ is the density of the universe at the redshift we consider. The parameter $q$ should be of order unity (we shall take $q=2$). Thus, the halos we considered in the definitions (\ref{etalin}) and (\ref{etah}) of the mass function will be characterized by an external radius equal to $R=$Min$(R_{cool},R_{vir})$. This defines their extension and the function $\Delta(M)=$Max$(\Delta_{cool},\Delta_{vir})$ introduced in Sect.\ref{The galaxy mass function}. As explained in Appendix \ref{Characteristics of galactic halos} we do not need to assume a smooth power-law density profile for dark matter halos. Indeed, objects defined by $R_{cool}$ are not defined as being the central core of radius $R_{cool}$ within a larger halo of radius $R_{vir}$ assumed to have a power-law density profile. We directly consider mass condensations satisfying the criteria (\ref{Cgal}) without reference to the properties of a possible larger object. The advantage of this procedure is that {\it clusters automatically contain several galaxies (as they should) instead of only one central huge galaxy}, see also the discussion below in Sect.\ref{Galaxies versus clusters and groups}. The cooling time is given by:
\beq
t_{cool} = \frac{1}{s} \; \frac{3}{2} \; \frac{\Omega_0}{\Omega_b} \; \frac{\mu_e^2 m_p k T}{\mu \rho \Lambda(T)} =  \frac{1}{s} \; t_{cool}(\mbox{mean halo density}) \label{tcool}
\eeq
where $\Lambda(T)$ is the cooling function (in $\mbox{erg.cm}^3.\mbox{s}^{-1}$). We use the cooling function given by Sutherland \& Dopita (1993) for a gas with primordial abundances. The expression (\ref{tcool}) for $s=1$ is the evaluation for the mean halo baryonic density. Cooling may be much more efficient since the local density increases within the halo and the baryon density gets higher during the collapse. This means that the parameter $s$ which enters (\ref{tcool}) should be larger than unity. A crude estimate for $s$ using a mean power-law density profile (with no allowance for the cooling to be increased due to baryon concentration that would lead to even larger a value: such an increase has been shown to exist in numerical simulations, but is naturally limited, see Teyssier et al.1998) shows it may indeed reach several units.

It is convenient to express (\ref{Cgal}) as a relation virial temperature - density, so that the halos we consider can be completely described by their temperature. Hence we can write the curve $\rho_{\Lambda}(T)$ corresponding to the first condition of the system (\ref{Cgal}) as:
\beq
{\cal C}_{\Lambda} \; : \;\;\;   \rho_{\Lambda} = \frac{4 {\cal G}}{3 \pi } \left[ \frac{\Omega_0}{\Omega_b} \; \frac{3 \mu_e^2 m_p k T}{2 s q \mu \Lambda(T)} \right]^2    \label{CLambda}
\eeq
The curve ${\cal C}_v$ corresponding to the second condition of (\ref{Cgal}), which characterizes just-virialized objects, is given by:
\beq
{\cal C}_v \; : \;\;\;    \rho_v = (1+\Delta_c(z) ) \; \rho_0 \; (1+z)^3 
\eeq
where $\Delta_c(z)$ is the density contrast at the time of virialization, at the redshift $z$ we consider, and $\rho_0 \; (1+z)^3$ is the mean density of the universe at that redshift. If $\Omega = 1$ this density contrast is a constant: $\Delta_c \simeq 177$. The general behaviour of these curves is shown in Fig.\ref{figtcool}. The curve ${\cal C}$ which describes the system (\ref{Cgal}) is simply given by:
\beq
{\cal C} \; : \;\;\;\;\;  \rho_{\cal C}(T) = \mbox{Max}[\rho_{\Lambda}(T),\rho_v(T)] 
\eeq
This relation temperature-density defines implicitely the relation $\Delta(M,z)$ we introduced in Sect.2 to determine the mass function of galaxies. Note that there are only two parameters: $p$ and $(s q)$. From the spherical model we choose $p=2$ as a natural value, while the product $(s q)$ will be given by the luminosity of the Milky Way (which implies a well defined mass of baryons), with the constraint that both $s$ and $q$ should be of order of a few units.

\subsection{Galaxies versus clusters and groups}
\label{Galaxies versus clusters and groups}

\begin{figure}[htb]

\centerline{\epsfxsize=8 cm \epsfysize=5.5 cm \epsfbox{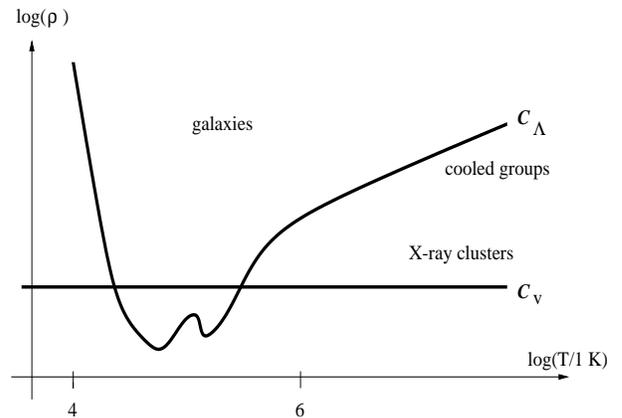}}

\caption{Temperature-density diagram given by the cooling constraint. Galaxies are defined as objects which cool within a Hubble time at birth. Objects which had no time to cool (because of their low density, at a given temperature, which translates into a long cooling time) are either galaxy groups (which have cooled by now) or hot X-ray clusters. They contain galaxies located on ${\cal C}$.}
\label{figtcool}

\end{figure}

The curves ${\cal C}_{\Lambda}$ and ${\cal C}_v$ are shown in Fig.\ref{figtcool}. Objects below either one of these curves violate one at least of the above conditions. Objects above these curves satisfy the criteria, but it is easy to see that slightly larger size objects (and whence lower density objects) also satisfy the criteria. Hence we define galaxies as being, for a given temperature (velocity dispersion), the largest object (hence the one with the lowest density) that satisfies the criteria (\ref{Cgal}). Thus the locus of galaxy formation, curve ${\cal C}$, is the higher of the curves ${\cal C}_{\Lambda}$ and ${\cal C}_v$ in the $\rho$-$T$ diagram. Objects lying below this curve will be groups of galaxies or galaxy clusters. These objects do not cool in a uniform fashion within a Hubble time at formation, but may have cooled by now (groups) or are still cooling (X-ray clusters). The consideration of the latter in some sense justifies our cooling requirement for galaxies: we clearly see clusters cooling at the present epoch but they do not constitute a single galaxy. However, cooling is {\it known} to be necessary for galaxies to form and we would argue that the above condition is the most natural one to avoid the baryonic component of the following merging event to mix completely. Groups and clusters {\it contain some of the galaxies we are considering in the present paper} and they can be described in a consistent way by means of the same methods developed in VS and used here to describe the galaxy distribution. Their multiplicity, counting the same objects that the ones we deal with here but grouped differently, will be calculated elsewhere (Valageas \& Schaeffer 1998). In particular, the normalization conditions (\ref{sumnu}) and (\ref{sumhx}) show that if we integrate from $M=0$ up to $M=\infty$ we recover the total mass of the universe {\it whatever} the curve $\Delta(M)$. The latter only describes how one can divide the matter content of the universe and different choices simply correspond to different classifications: one can count a group as a single large object or as the assembly of several distinct galaxies which are considered as individual entities. In this article we adopt this latter point of view since we are interested in galaxies themselves. In fact, the galaxy mass function does not extend down to $M=0$ because halos with a low virial temperature $T < 10^4$ K do not cool (due to inefficient cooling as well as heating by the UV background). These patches of matter (which fill most of the volume of the universe with voids) form the Lyman-$\alpha$ forest clouds which we describe in details in Valageas et al.(1998b).

The curve ${\cal C}_{\Lambda}$ does not depend on redshift while the curve ${\cal C}_v$ does. At fixed temperature $T$, $\rho_v(T,z)$ as defined by ${\cal C}_v$ increases with $z$ along with the average density $\rhob(z)$ of the universe. Whence the locus of galaxy formation varies with $z$ and at early times it becomes identical to ${\cal C}_v$.

This variation with redshift has important consequences. Galaxies, that at a given epoch settle on the curve  ${\cal C}_v$ because larger objects of lower densities are not virialized, at a later epoch will lie {\it above} the curve ${\cal C}_v$ and whence according to our criteria will be embedded within larger objects. This provides for these galaxies a {\it continuous merging process} with obviously an evolution of the mass as well as the  number of such objects (and indeed an associated star formation). On the other hand, galaxies that settle on the curve ${\cal C}_{\Lambda}$ at a later time will be limited by the same condition on density: larger objects of lower density will not satisfy the above conditions and thus may be groups or galaxy clusters, but cannot become a larger galaxy into which the galaxy we consider gets embedded. So in the framework of our model these galaxies will no longer evolve neither in mass nor in number. We call these galaxies ``isolated galaxies''. Thus we have a population of smaller galaxies that evolve by mergers and a population of larger galaxies that bear only internal evolution. At early times all objects are continuously merging since the only limiting curve at high $z$ is ${\cal C}_v$. As time goes on, the larger objects get limited by cooling, settling on ${\cal C}_{\Lambda}$. So all these now ``quiet galaxies'' have an early phase with strong merging processes. Once this early phase is over, the now quietly evolving galaxy may for instance form a disk. Within this picture, we do not take into account the strong merging processes that take place in the very  dense cores of clusters. Such additional interactions may sufficiently disturb (Balland, Silk, Schaeffer 1998) the galaxies we model here to change them into ellipticals, a plausible explanation of the density morphology relation. This will be accounted for in a subsequent paper about galaxy clusters (Valageas \& Schaeffer 1998).

Note that in our model all galaxies have roughly the same age (of the order of the age of the universe) from the star formation viewpoint, even if they did not exist as distinct objects along this whole period. Indeed, present galaxies are the result of the merging of many smaller and older sub-units where star formation was already active at high $z$. In fact, contrary to the usual statement, although we work within the framework of the standard hierarchical clustering scenario massive galaxies look {\it slightly older} than small ones (see App.\ref{Redshift evolution and merging of galaxies}). Because of their more efficient star formation process (due to their higher density and virial temperature, see App.\ref{Star formation model for an isolated halo}) massive and bright galaxies have a redder and older stellar population.

\subsection{Star formation}
\label{Star formation}

Our scenario describing the history of star formation is rather standard, and kept as simple as possible. Apart from a small component (10\%) of dark baryonic matter (brown dwarfs, planets) that plays a negligible role in the stellar evolution history, we consider four components:

- short lived stars, of total mass $M_{sh}$,  that will be recycled.

- long lived stars, of total mass $M_{lo}$, that will not be recycled. 

- a central gaseous component, of total mass $M_{gc}$, at the sites of star formation, that is deplenished by star formation and ejection by supernova winds, replenished by infall from a diffuse gaseous component located in the dark halo potential well.

- a diffuse gaseous component,  of total mass $M_{gh}$, deplenished by infall and replenished by the supernova winds.

The star formation rate $dM_s/dt$ is proportional to the mass of central gas $M_{gc}$ with a time-scale $\tau_c$ proportional to the dynamical time-scale $\tau_d$:
\beq
\frac{dM_s}{dt} = \frac{M_{gc}}{\tau_c} \hspace{0.5cm} \mbox{and} \hspace{0.5cm} \tau_c \propto \tau_d
\eeq
The mass of gas heated and ejected by supernovae out of the central parts of the galaxy is proportional to the star formation rate and decreases for deep potential wells:
\beq
\left( \frac{dM_{gc}}{dt} \right)_{SN} = - \; \frac{T_0}{T} \; \frac{M_{gc}}{\tau_c}
\label{TSN2}
\eeq
where $T_0$ is a constant and $T$ the halo virial temperature, see (\ref{flowSN}). Finally, the infall of gas from $M_{gh}$ to $M_{gc}$ occurs on a dynamical time-scale and not on the cooling time-scale since $t_{cool} < \tau_d$ because of (\ref{Cgal}):
\beq
\left( \frac{dM_{gc}}{dt} \right)_{IN} = \frac{M_{gh}}{\tau_d}
\eeq
Since $\tau_d \sim 1/\sqrt{{\cal G} \rho} \sim t_m$ we shall use:
\beq
\tau_d = \beta_d \;  t_m  \hspace{0.5cm} , \hspace{0.5cm} \tau_c = \beta_c \;  t_m  \label{tauctaud}
\eeq
where $\beta_d$ and $\beta_c$ are parameters of order unity. We assumed $\tau_c \propto \tau_d$, as it may describe gravitational instabilities within galaxies as well as the influence of neighbours. Moreover, the time-scale $\tau_c$ disappears for faint galaxies, as we shall see below, and our model would still be approximately valid even for bright galaxies if one has in fact $\tau_c \ll \tau_d$, since it would simply correspond to a change of $\beta_d$ (then the system is governed by the longest time-scale among $\tau_c$ and $\tau_d$). This means that other forms for $\tau_c$, for instance $\tau_c \propto 1/\rho$ or $\tau_c \propto 1/\rho_g$ (where $\rho_g$ is the gas density in the core) would give similar results, as we checked numerically.

The details of this model of star formation are presented in App.\ref{Star formation model for an isolated halo} and App.\ref{Redshift evolution and merging of galaxies}. We solve the corresponding equations (\ref{systar}) numerically but we also give there approximate analytic solutions that allow one to follow the behaviour of these various components as a function of time and galactic mass. We can stress here that the fraction of non-luminous stars we adopt (10\%) is consistent with observations (see discussion in Mera et al.1998) while the values used in some other studies (50\% in Kauffmann et al.1993; 63\% in Cole et al.1994 for instance) are much too large. Using a smaller fraction of non-luminous stellar-like objects however would increase the luminosity of the galaxies obtained in the latter models and would lead to stronger discrepancies with the observed luminosity function. 

We also describe the evolution of the metallicity (\ref{metal}) with three different values: $Z_s$ for stars, $Z_c$ for the central gaseous component and $Z_h$ for the diffuse gaseous component.

We summarize here the free parameters of our model:

- $(sq)$: proportionality factor in the cooling constraint (\ref{CLambda}).

- $\beta_c$: proportionality factor for the star formation time-scale (\ref{tauctaud}).

- $\beta_d$: proportionality factor for the definition of the dynamical time-scale (\ref{tauctaud}).

- $T_0$: supernovae efficiency, (\ref{T0SN}) and (\ref{TSN2}).

- $\eta_d= 10\%$: fraction of non-luminous objects (brown dwarfs) (\ref{etad}).

- ``yield'' $y=0.019$: mass of metals produced per mass of stars through SNII, see (\ref{metal}).

The parameters $q$, $\beta_c$ and $\beta_d$ must be of order unity and they are constrained by the luminosity, mass, gas/star mass ratio of the Milky Way, while $T_0$ is constrained by the supernovae rate of similar galaxies. The yield $y$ is given by the metallicity of the solar neighbourhood. Thus all parameters of our model are adjusted on other observations than those related to the luminosity function we seek to reproduce.
Note moreover that for bright and massive galaxies ($T \gg T_0$) $T_0$ is irrelevant: the effect of supernovae is negligible since the potential well is very deep so that gas cannot be ejected efficiently. For faint galaxies ($T \ll T_0$) $\beta_c$ is irrelevant because a quasi-stationary regime sets in very quickly where the galaxy evolution is governed by the equilibrium between the ejection of central gas by supernovae and the infall from the diffuse component (see App.\ref{Analytical approximations}) so that the star formation rate is given by:
\beq
\frac{dM_s}{dt} = \frac{M_g}{\tau_0} \hspace{0.5cm} \mbox{with} \hspace{0.5cm} \tau_0 \simeq \left( 1+\frac{T_0}{T} \right) \; \tau_d  \label{SFRtau0}
\eeq
where $M_g$ is the total gas mass. For massive galaxies the derivation leading to this relation is no longer valid but (\ref{SFRtau0}) still gives a reasonable description of the star formation rate because in our model $\tau_d \sim \tau_c$ so that for $T \gg T_0$ there is only one time-scale which is also given by $\tau_0$ since in this case $\tau_0 \simeq \tau_d$. The value of the main parameters $(sq)$ and $\beta_d$ is further discussed in App.\ref{Scalings}.

\section{Numerical results}
\label{Numerical results}

Now we can compute numerically the galaxy luminosity function, as defined in the previous section, as well as the physical characteristics of the halos we consider.

\subsection{$\Omega = 1$}
\label{Omega=1}

\subsubsection{Halo properties}
\label{Halo properties}

\begin{figure}[htb]

\centerline{\epsfxsize=8 cm \epsfysize=5.5 cm \epsfbox{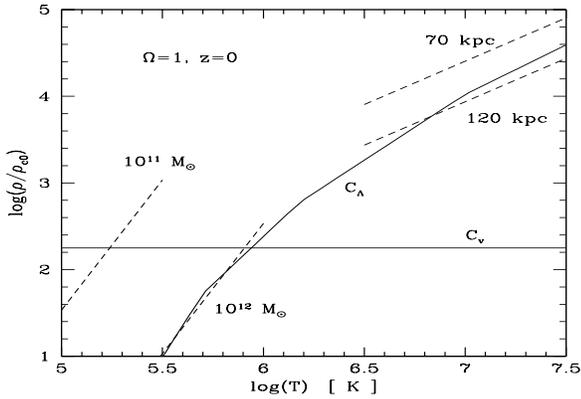}}

\caption{Temperature-density diagram given by the cooling and virialization constraints in the case $\Omega=1, \; z=0$. The short-dashed lines correspond to halos of fixed mass, on the left, and halos of fixed radius, on the right. We note $\rho_{c0}$ the critical density at $z=0$.}
\label{figTrho1}

\end{figure}

In the case $\Omega=1$, Fig.\ref{figTrho1} shows the relation temperature-density which corresponds to the cooling and virialization constraints (Fig.\ref{figtcool}), at redshift $z=0$. We see that the typical mass and radius of the halos we get are close to the observed values for galaxies in the present universe. For $T \sim 10^6 \;$K the cooling constraint translates into a nearly constant mass $M \simeq  10^{12} \; M_{\odot}$, while at large masses, or large temperatures $T > 10^7 \;$K, it translates into a fixed radius $R \simeq 100 \;$kpc. Indeed, the curve ${\cal C}_{\Lambda}$, see (\ref{CLambda}), can also be written:
\beq
R = (s q) \; \frac{\Omega_b}{\Omega_0} \; \sqrt{ \frac{(\gamma-1) \mu \Lambda(T)^2} {2 {\cal G}^2 \mu_e^4 m_p^3 \; k T} }
\eeq
At high temperatures, we can write $\Lambda(T) \simeq \Lambda_0 \sqrt{T/T_0}$ with $T_0=10^7$ K and $\Lambda_0=10^{-23} \; \mbox{erg.cm}^3.\mbox{s}^{-1}$. This gives a constant radius $R$:
\beq
R = (s q) \; \frac{\Omega_b}{0.04} \; \frac{1}{\Omega_0} \; 8.5 \; \mbox{kpc}
\eeq
The fact that we know the radius of the halos we consider to be at least 60 kpc (because a few rotation curves measured in large spiral galaxies remain flat at least up to this radius) requires that $(s q) \geq 7$. However, as we explain in App.\ref{Scalings} the product $(sq)$ is determined independently by the luminosity of the Milky Way. This leads to $(s q) \simeq 12$ in the case $\Omega=1, \; \Omega_b=0.04$. Hence both constraints can be satisfied simultaneously. Since  we use $p=2$ as given by the simple spherical collapse picture and we expect $q \sim p$ and we noticed in Sect.\ref{Galaxy formation: cooling constraints} that $s > 1$, we choose $q=2$ and $s=6$. This gives $R=102$ kpc in the limit of very large virial temperatures. 

Galaxies with small virial temperature $T<10^6$ K and rotation velocity $V_c<100$ km/s are irregular or (dwarf-) elliptic galaxies, dominated by their continuous merging history, while massive galaxies with $V_c>100$ km/s are disk galaxies which have evolved through a much calmer history since they reached the curve ${\cal C}_{\Lambda}$. Indeed, they remain unchanged, or slowly accrete some mass if clustering is not exactly stable, so that a disk can form. This boundary $V_c=100$ km/s (which corresponds to a B-band magnitude $M_B=-18$) is indeed close to the observed transition between spirals and (dwarf-) ellipticals or irregulars (Sandage et al.1985). 

We can see that in the near future ($z \leq 0$) two halos of the same mass or the same radius can have different temperatures and densities. This implies that, within the framework of this model, neither the mass nor the radius are good variables to describe the halos we consider, while the velocity dispersion (or the temperature) is. Note that the latter is very close to our variable $x$ that we think is the true parameter characteristic of mass condensates.

We notice that light and faint galaxies, which correspond to small temperatures or velocity dispersions, are young objects characterized by a density contrast $\Delta \sim 200$. Indeed, these small galaxies are on the curve ${\cal C}_v$ in the cooling diagram (see Fig.\ref{figTrho1}). This may look surprising, since the observed density contrasts of these objects are usually very large. However, this is simply explained by the fact that the observed effective luminous radius $R_{lum}$ is much smaller than the radius we consider which describes the actual size of the dark matter halo from where the gas content of the galaxy originated. This large difference in radius translates into the density contrast we obtain.  The small size of the observed luminous radius of these objects, that is not modelled here, can be explained by several effects. For instance a large fraction of  gas is ejected, since the potential well is too weak to retain baryons as efficiently as in large galaxies. Also star formation is not very efficient so that it cannot take place very far from the center of the galaxy because the density gets quickly too small.

We do not attempt to model here the radius and the mass within the luminous part. This has already been done in a similar context (Bernardeau \& Schaeffer 1991). It was seen there by phenomenologically modelling the baryon
squeezing due to cooling, that the observed behaviour of the luminous radius calls for a dark matter halo radius that behaves nearly as $M^{1/3}$ for the smaller masses (i.e. a constant density contrast) but for an almost constant dark halo radius for the large masses (due to a change in the $M/L$ ratio in these two regimes). These findings, indeed, encouraged us to undertake the present work. For instance, if $R_{lum} \propto L^{0.5}$ and the density profile of galactic halos is a power-law $\rho(r) \propto r^{-\gam}$ ($\gam \simeq 1.8$), then for small galaxies on ${\cal C}_v$ where $L \propto V^3$ the density contrast within the luminous radius scales as $(1+\Delta)_{lum} \propto L^{-\gam/6}$ while for bright galaxies on ${\cal C}_{\Lambda}$ which have a nearly constant dark matter radius we get $(1+\Delta)_{lum} \propto L^{-(3\gam-4)/6}$. Hence the density within the luminous radius decreases for brighter galaxies even though the density within the much larger dark matter halo increases. However, a precise estimate of these relations would require a detailed model of the behaviour of the gas after cooling which we do not consider in this article.

\subsubsection{Gas/star mass ratio}
\label{Gas/star mass ratio}

\begin{figure}[htb]

\centerline{\epsfxsize=8 cm \epsfysize=5.5 cm \epsfbox{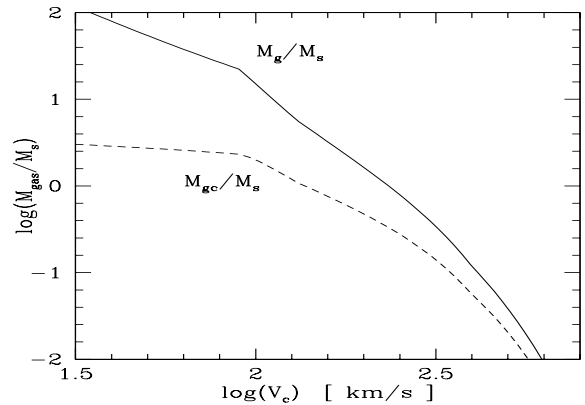}}

\caption{Ratios of the total mass of gas $M_g$ (solid line) and of the mass of star-forming gas $M_{gc}$ (dashed line) over the mass $M_s$ of luminous stars as a function of the circular velocity $V_c$.}
\label{figMgO1}

\end{figure}

The variation of the gas/star mass ratio, which is closely related to the ratio (star formation time-scale)/(age of the galaxy), see (\ref{MgoMs}), is shown in Fig.\ref{figMgO1} as a function of the galaxy circular velocity. Small $V_c$ galaxies which have a small star formation rate $\tau_0^{-1}$, because of their low density and virial temperature, have a high gas/star mass ratio $M_g/M_s \gg 1$. On the other hand, high temperatures at the external radius of the halo, or large $V_c$, correspond to high densities. Hence these galaxies have undergone a very efficient star formation, since their star formation rate $\tau_0^{-1}$ is large. As a consequence, they have already lost most of their gas content, which turned into stars, as we can see in Fig.\ref{figMgO1}, and $M_g/M_s \ll 1$. As expected on general grounds, see (\ref{MgoMs1}), we verify that $M_g/M_s \sim 1$, corresponding to a galaxy similar to the Milky Way, implies $\tau_0 \sim t$ (we have $M_{gc}/M_s = 0.37$ and $M_g/M_s \simeq 1$ since there is some gas in the halo: $M_{gh} \simeq 2 \; M_{gc} \simeq 0.008 \; M$). This value of the gas/star mass ratio leads us to choose for our last parameters $\beta_c=2$, $\beta_d=3$ and $T_0=3\;10^6$ K. We can see in Fig.\ref{figMgO1} that while most of the gas is in the dense component $M_{gc}$ for galaxies brighter than the Milky Way (with a larger rotation velocity), it is mostly in the diffuse phase $M_{gh}$ for faint galaxies, as we can see in (\ref{McMh}), because supernovae are very efficient in these weak gravitational potentials and eject the gas out of the star-forming regions. In particular, from (\ref{McMh}), (\ref{Msh1}) and (\ref{Mlo1}) we see that for faint galaxies $M_{gc} \sim M_s$ while $M_{gc} \simeq M_g$ for bright galaxies.

One may distinguish three different regimes (App.\ref{Approximate power-law regimes}) in which approximate power-law relations can be derived.

1) Very faint galaxies, located on ${\cal C}_v \; : \;\; \Delta \simeq 200$ and $T < 10^{5.5}$ K,  with a { \it constant density contrast}.

2) Faint galaxies, located on ${\cal C}_{\Lambda}$ : $R \simeq 120$ kpc and $10^{5.5} \; \mbox{K} < T < 10^{6.5}$ K,  with a {\it nearly constant radius}.

3) Bright galaxies located on ${\cal C}_{\Lambda} \; : \;\; R \simeq 100$ kpc and $T > 10^{6.5}$ K, which have {\it nearly exhausted all their gas}.

Using these  analytic estimates (App.\ref{Approximate power-law regimes}) it is readily seen that the gas/star mass ratio is a steep function of the circular velocity ($M_g/M_s \propto V^{-2}$ to $M_g/M_s \propto V^{-3}$ and finally it follows an exponential decline). We can see very distinctly in Fig.\ref{figMgO1} these three regimes.
\\

It is important to realize that these relations should be approximatively valid for any model of star formation. The relation (\ref{MgoMs1}) is always correct for $t < \tau_0$, where $\tau_0$ is the star formation time-scale. Indeed, for $t < \tau_0$ we have $dM_s/dt \simeq  M_g/\tau_0$ by definition of $\tau_0$, hence $M_s \simeq t/\tau_0 \; M_g$. Since the Milky Way is characterized by $M_g/M_s \sim 1$, fainter galaxies (which have a higher gas/star mass ratio) verify $t < \tau_0$, $M_g \simeq M_{g0}$ and $M_s \simeq t/\tau_0 \; M_{g0}$, while more luminous galaxies satisfy $t > \tau_0$ and $M_s \simeq M_{g0}$. This means that the relation $V_c \; - \; L$ for these bright galaxies, if the stellar mass/luminosity ratio $M_s/L$ is roughly constant, is directly given by the physical characteristics of the underlying halos - since $L \propto M_s \propto M$ - hence by the cooling curve, and not by the details of star formation processes. 
Hence the slope of the corresponding Tully-Fisher relation (which is a bit shallow) could only be modified by a change of the definition of galactic halos (i.e. using another constraint than the cooling criterion) or by a gas/dark matter mass ratio which would vary with the galactic characteristics (one may argue that deep potential wells could gain some gas from surrounding small halos). On the contrary, the luminosity of faint galaxies $L \propto t/\tau_0 \; M$ depends strongly on the star formation rate. Using $\rho = \Delta_c \rhob$ and $t \simeq t_H$ (where $t_H$ is the age of the universe), we have $L \propto V^3/\tau_0$. Hence the dependence of $\tau_0$ on $V$ is constrained by the observed Tully-Fisher relation. In fact in our model the relation we obtain in this case is $\tau_0 \propto V^{-2}$ but it leads to results which are still consistent with observations. 
Thus, the high uncertainty which lies in any star formation prescription (since this process is rather poorly known) is greatly reduced by these general properties and the observed Tully-Fisher relation. 
It is also apparent through these considerations (and is confirmed by a simple calculation) that the use of a constant star formation time-scale ($\tau = \tau_0$ for all times) would not change these results (at $z=0$), except for small details. As a consequence, we can reasonably expect our results to be fairly general and robust.

\subsubsection{Luminosity}
\label{Luminosity}

\begin{figure}[htb]

\centerline{\epsfxsize=8 cm \epsfysize=5.5 cm \epsfbox{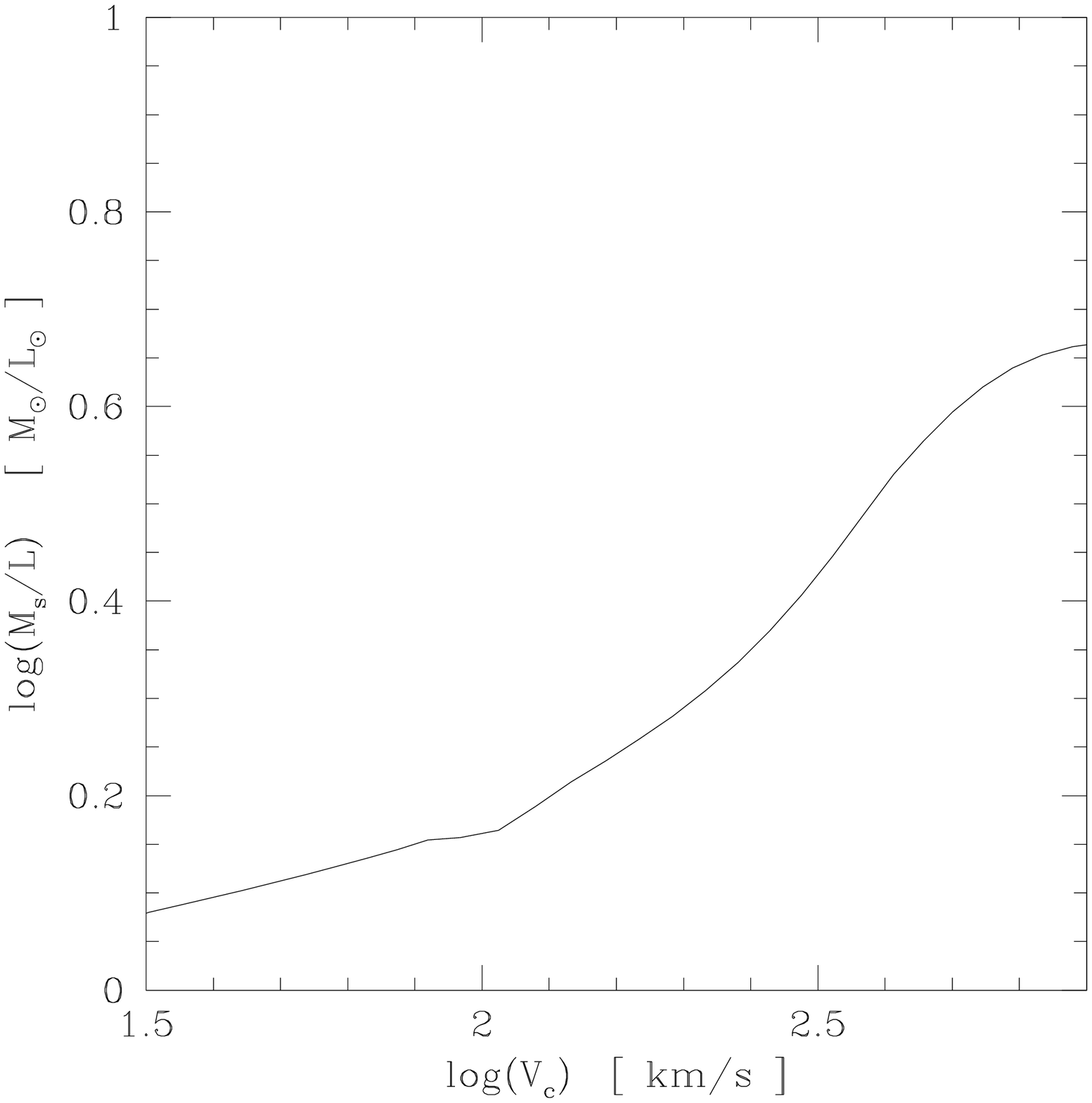}}
\centerline{\epsfxsize=8 cm \epsfysize=5.5 cm \epsfbox{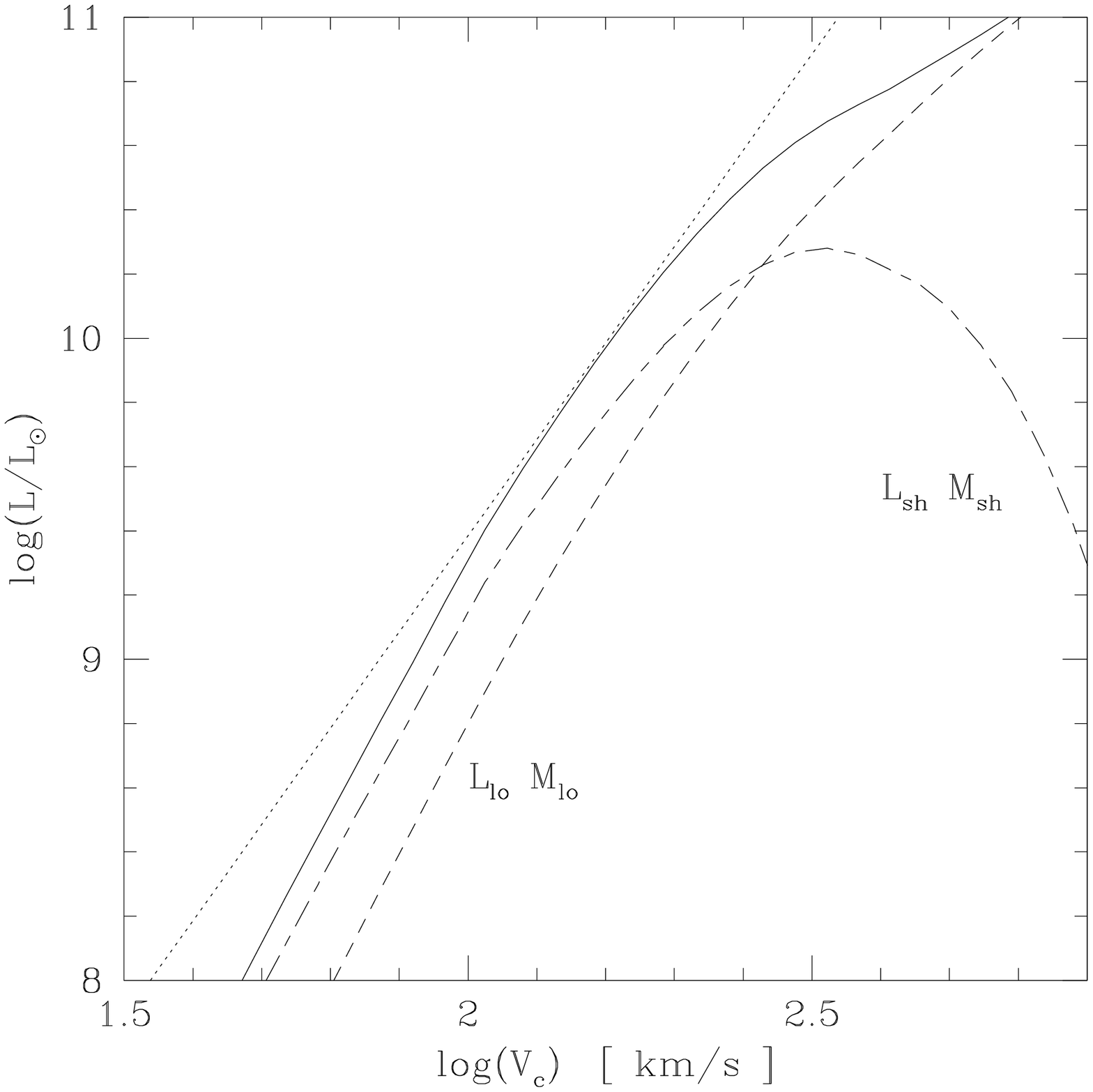}}

\caption{Upper figure: ratio of the total mass of luminous stars $M_s$ over the galaxy luminosity $L$. Lower figure: luminosity $L$ of the galaxy as a function of $V_c$. The thin dotted line corresponds to the power-law $L \propto V_c^3$ normalized to the Milky Way. The short-dashed curve is the luminosity due to long-lived small stars ($L_{lo} M_{lo}$) while the dot-dashed curve is the luminosity due to short-lived massive stars ($L_{sh} M_{sh}$).}
\label{figTF1}

\end{figure}

Galaxies characterized by high temperatures and large $V_c$ also correspond to the most luminous galaxies, and their luminosity is mainly due to small stars which have a long life-time, since there is not much gas left to create new generations of massive short-lived stars, as it appears in Fig.\ref{figTF1}. We can see in the lower panel that the Tully-Fisher relation is approximately satisfied, although the slope of the relation $V_c \; - \; L$ gets shallower at high luminosities. For B-band luminosities, observations give $L \propto V_c^{2.7}$ (Kraan-Korteweg et al. 1988; Pierce \& Tully 1988), while the infrared relation is steeper: $L \propto V_c^{3.5}$ (Pierce \& Tully 1992). 
In our model, these relations are obtained (App.\ref{Approximate power-law regimes}) as a smooth transition from $L \propto V^5$ at the very faint end to $L \propto V^2$ at the bright end.

Hence the relation $V_c \; - \; L$ we obtain has a strong slope close to the usual Tully-Fisher relation, and gets shallower for galaxies more luminous than the Milky Way. It is interesting to note that Persic \& Salucci (1991) found a similar decrease of the slope of the Tully-Fisher relation for luminous galaxies in observations, although in the infrared H-band. We can note in Fig.\ref{figTF1} that for faint galaxies the slope is close to 3 and not 5 as in the approximation obtained in App.\ref{Approximate power-law regimes} because the ratio $M_s/L$ is not exactly constant and increases for large circular velocities.

As we can see from (\ref{MshoMlo}) and (\ref{MgoMs}), this variation of the mass/luminosity ratio $M_s/L$ is due to the fact that large and bright galaxies (also characterized by a large temperature or circular velocity), which have a high star formation rate $\tau_0^{-1}$, have already consumed most of their initial gas content. Hence, their present star formation rate is relatively small (compared to their past history), and their stellar population presents an increasing proportion of old stars, with a long life-time, which were created during the whole life of the galaxy. Thus, as we can see in Fig.\ref{figTF1} the luminosity in the form of short-lived massive stars is much smaller than the contribution of long-lived small stars. On the contrary, small galaxies which still have a relatively high star formation rate (as compared to their past) because of their large gas content show an important contribution from massive stars which are created at the current epoch. 
In fact, both classes of stars give roughly the same luminosity. This is due to the fact that the luminosity of the galaxy is dominated by stars of intermediate mass $M_* \sim 1 \; M_{\odot}$, as described in App.\ref{Stellar properties of galactic halos}. This variation of the mass/luminosity ratio shows that very massive galaxies should be redder than small ones, which is consistent with observations (Lilly et al.1991; Metcalfe et al.1991). Moreover, this would be enhanced by metallicity effects, since bright galaxies are also the most metal-rich, which we recover in our model as we shall see now.

\subsubsection{Metallicity}
\label{Metallicity}

\begin{figure}[htb]

\centerline{\epsfxsize=8 cm \epsfysize=5.5 cm \epsfbox{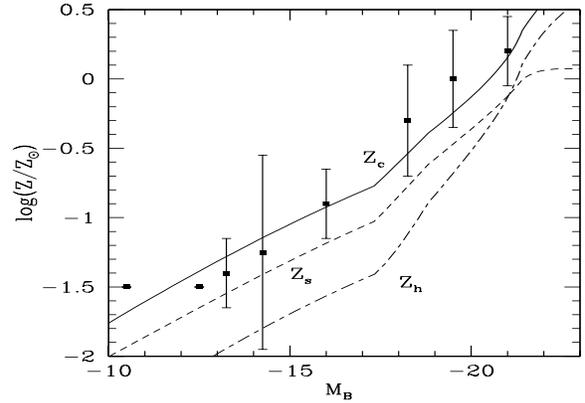}}

\caption{Metallicity $Z$ as a function of the B-band magnitude $M_B$. The solid line corresponds to the gas in the core ($Z_c$), the short-dashed line to stellar metallicities ($Z_s$) and the dot-dashed line to the gas in the halo ($Z_h$). The data points are observations from Zaritsky et al.(1994).}
\label{figZmetO1}

\end{figure}

Next, we can also consider the metallicities $Z_h$ (diffuse gas), $Z_c$ (dense gas) and $Z_s$ (stars), which we introduce in App.\ref{1Metallicity}. They are displayed in Fig.\ref{figZmetO1}, as a function of the B-band luminosity of the galaxy. For faint galaxies we are in the ``stationary'' regime where $Z_h \ll Z_c$ and $Z_s \simeq Z_c$. The variation of the metallicities $Z_c$ and $Z_s$, which are those available to observations, agrees with the data. We chose the value of the yield $y$ so that we get $Z_c = Z_{\odot}$ for the Milky Way. This gives $y = 0.019 \simeq Z_{\odot}$.

As usual, we can consider the 3 regimes we introduced previously to determine approximate relations for the metallicity (App.\ref{Approximate power-law regimes}). For galaxies fainter than the Milky Way we have $Z_c \propto Z_s \propto L^{0.4}$, or $L \propto Z_s^{2.5}$. Galaxies more luminous than the Milky Way have a constant metallicity $Z_s \simeq y$ while $Z_c \propto L^{0.5}$. Note however that the yield may vary with the characteristics of the galaxy, together with the IMF. Nevertheless, we see that our model agrees with observations over 10 magnitudes in $M_B$.

\subsubsection{Stellar history}
\label{Stellar history}

\begin{figure}[htb]

\centerline{\epsfxsize=8 cm \epsfysize=5.5 cm \epsfbox{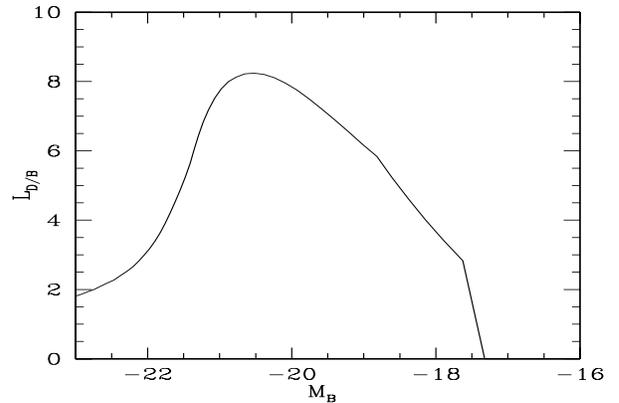}}

\caption{The variation of the disk/bulge luminosity ratio $L_{D/B}$ with the B band magnitude $M_B$.}
\label{figLDoBO1}

\end{figure}

Finally, we can consider (rather crudely) the mean morphological properties of the galaxies we obtain in our model as a function of their B band magnitude $M_B$. We define an approximate disk/bulge luminosity ratio by:
\beq
L_{D/B} = \frac{ L_{sh} M_{sh} + L_{lo} ( M_{lo}(t_H)-M_{lo}(p\;t_m) ) }{L_{lo} M_{lo}(p\;t_m) } 
\eeq
when $t_H > p\;t_m$ ($t_H$ is the Hubble time at the considered redshift and $p \; t_m$ the virialization time of the galaxy). Thus, we evaluate the disk luminosity as the contribution from stars formed after the galaxy reached the cooling curve ${\cal C}_{\Lambda}$, and the bulge luminosity as the contribution from earlier stars. For galaxies which have not reached ${\cal C}_{\Lambda}$ yet we use $L_{D/B} = 0$ since there is no disk. This approximation is only very crude, as the disk may not form exactly at the virialization time $p\;t_m$, and some stars probably form in the bulge after this date. The variation of this ratio $L_{D/B}$ with $M_B$ is shown in Fig.\ref{figLDoBO1}.

Faint galaxies $M_B>-18$ have not reached ${\cal C}_{\Lambda}$ yet, hence they have no disk and are elliptical or irregular galaxies. For $M_B<-18$ the disk/bulge luminous ratio first increases with luminosity, since brighter galaxies (which also have a higher density, virial temperature and a larger mass) have been on ${\cal C}_{\Lambda}$ since a higher redshift. However, this ratio declines for very luminous galaxies $M_B<-21$, which correspond to large densities and deep potential wells, hence to very efficient star formation. As a consequence, these galaxies transformed a large part of their initial gas content into stars at very high redshift, while they were still divided into several sub-units, before they reached the curve ${\cal C}_{\Lambda}$ which marked the end of their merging phase of formation. 
Hence, the mass of stars formed since this latter epoch is increasingly small as compared to the stellar population formed during the merging phase, as the galaxy parameter $x$ increases. Thus, it appears naturally in our model that spiral galaxies only correspond to an intermediate range of luminosities, $-22<M_B<-18$, while brighter and fainter galaxies should be ellipticals or irregulars. Hence we get old bright ellipticals as a straightforward outcome. This is quite similar to the observed dependence of the dominant Hubble type on the luminosity (Sandage et al.1985). However, the interactions between galaxies in clusters would certainly add several effects which we did not take into account explicitely but could have important consequences on the mean luminosity-morphological type relation.

\subsubsection{Milky Way}

The Milky Way corresponds to $V_c \simeq 220$ km/s, and in our model to $R \simeq 140$ kpc, $M \simeq 2.6 \; 10^{12} \; M_{\odot}$, $\Delta \simeq 2200$, $T \simeq 3.6 \; 10^6$ K , $L \simeq 2 \; 10^{10} L_{\odot}$, $M_B \simeq -20.4$, $M_{gas} \simeq 5.5 \; 10^{10} \; M_{\odot}$ and $M_{gc}/M_s \simeq 0.37$. Its present supernovae rate is $R_{SN} \simeq 0.025 \; \mbox{year}^{-1}$ which is consistent with observations: van den Bergh \& Mc Clure (1994) find $R_{SN} = 0.021 - 0.024 \; \mbox{year}^{-1}$. The above value of $M_{gc}/M_s$ is consistent with the values $\sigma_{gas}/\sigma_s$ of the disk surface densities measured in the solar neighborhood: $\sigma_{gas}/\sigma_s = 0.27$ (Bienayme et al.1987), 0.47 (Kuijken \& Gilmore 1989), 0.5 (Gould et al.1996). It implies a gas/(gas+stars) mass ratio of $\sim 0.3$ which is closer to the actual observations than the value $\sim 0.1$ traditionally taken in many calculations of the chemical evolution in the solar neighborhood. The Milky Way also corresponds to a star formation rate $dM_s/dt \simeq 5 M_{\odot}/$year, which agrees with usual estimates. Note that the present ratio (stellar mass)/(age of the universe) $M_s/t_0 \sim 4$ (in $M_{\odot}$/year) implies that any model with the correct luminosity -hence the correct stellar mass- will give a present-day star formation rate of the right magnitude.

\subsubsection{$\Omega = 1$ , CDM}
\label{Omega=1,CDM}

\begin{figure}[htb]

\centerline{\epsfxsize=8 cm \epsfysize=5.5 cm \epsfbox{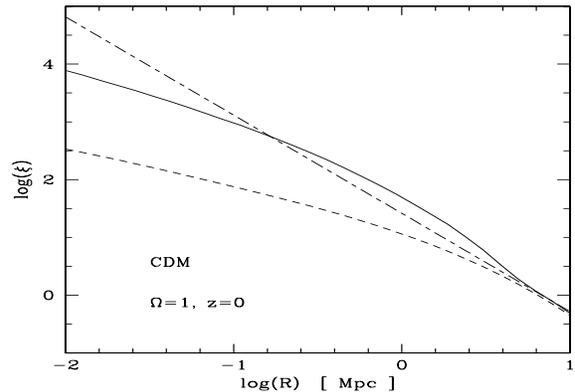}}

\caption{The evolved correlation function $\xia(R,z)$ at $z=0$ (solid line), and the linear extrapolation $\sigma(R)^2$ (dashed line), in the case $\Omega=1$ and $P(k)$ is a CDM power-spectrum. The dot-dashed line is the power-law $(R/R_0)^{-\gamma}$ with $\gamma=1.8$}
\label{figXiCDMO1}

\end{figure}

Now, to get the mass function or luminosity function of galaxies, we need the value of the initial power-spectrum $P(k)$, or the correlation functions $\sigma(R)^2$ and $\xia(R)$. We first consider the case where $P(k)$ is a CDM-like power-spectrum. More precisely, following Davis et al.(1985) we use:
\beq
P(k) = A \; k \; (1 + \alpha k + \beta k^{1.5} + \gamma k^2)^{-2}  \;\; \mbox{Mpc}^3   \label{PkCDM}
\eeq
where
\[
l=(\Omega_0 h^2)^{-1} \; , \; \alpha=1.7 \; l  \; , \; \beta = 9 \; l^{3/2} \; , \; \gamma = l^2
\]
and $A$ is a normalization constant such that $\sigma_8=0.5$. Fig.\ref{figXiCDMO1} shows the linear correlation function $\sigma(R)^2$ and the evolved non-linear correlation function $\xia(R)$ at redshift $z=0$. The calculation of the non-linear correlation function $\xia$ from its linear counterpart $\sigma^2$, using the spherical model normalized by the numerical calculation performed by Jain et al.(1995), is detailed in VS.

\begin{figure}[htb]

\centerline{\epsfxsize=8 cm \epsfysize=5.5 cm \epsfbox{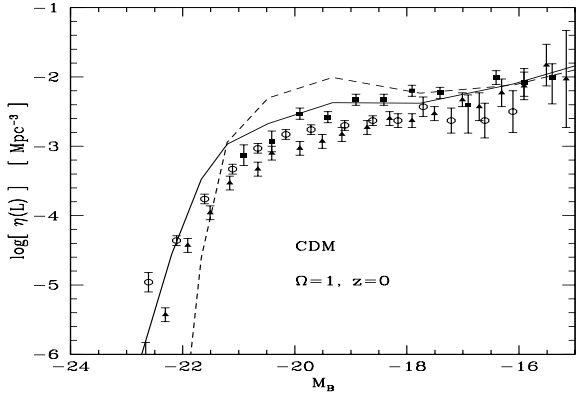}}

\caption{The galaxy luminosity function $\eta(L) \; dL/L$. The graph shows the quantity $\eta(L)$ as a function of the magnitude in the B-band for the non-linear scaling approach (solid line) and the PS prescription (dashed line). The data points are observations from Loveday et al.(1992) (circles), Ellis et al.(1996) (filled squares) and Zucca et al.(1997) (triangles).}
\label{figLCDMO1}

\end{figure}

We can see that the evolved non-linear correlation function is close to a power-law $\xia(R) \propto R^{-\gamma}$, with $\gamma=1.8$, in the range 50 kpc to 1 Mpc, which covers all the values that the radius of the halos we consider can take. As we saw in Sect.\ref{The galaxy mass function}, we can now get the luminosity function of galaxies, provided we know the function $h(x)$ we introduced in that section. We choose the function $h(x)$ given by Bouchet et al.(1991):
\beq
h(x) = a \; \frac{1-\omega}{\Gamma(\omega)} \; \frac{x^{\omega-2}}{(1+b x)^c} \; e^{-x/x_*}    \label{hCDM}
\eeq
with:
\[
a=1.8 \;\; , \;\; b=3.6 \;\; , \;\; c=0.8 \;\; , \;\; \omega = 0.4  \;\; , \;\; x_* = 12.5
\]
The galaxy luminosity function we get in this way is shown in Fig.\ref{figLCDMO1} as a function of the B-band magnitude ($M_B = 5.48 - 2.5 \log(L/L_{\odot})$). The short-dashed curve is the prediction of the PS approach.

We explained in Sect.\ref{The galaxy mass function} that the PS approach leads to the multiplicity function $\eta(L) dL/L$ with:
\beq
\eta(L) = \eta(\nu) \frac{\mbox{dln}\nu}{\mbox{dln}L} = \sqrt{\frac{2}{\pi}} \; \frac{\rho_0}{M} \; \nu \; e^{-\nu^2/2} \; \frac{\mbox{dln}\nu}{\mbox{dln}L}
\eeq
and $\nu$ given as a function of $M$ by (\ref{nups}), the mass $M$ being in turn related to $L$ by the model of star formation. As we could see in Fig.\ref{figTrho1}, the typical masses of our galactic halos are $M \sim 2\;10^{12} \; M_{\odot}$. As a consequence, since the cutoff of the multiplicity function given by the PS approach occurs for $\nu \sim 1$, which should correspond roughly to the cutoff of the observed luminosity function, the number density of galaxies is of the order of:
\beq
\eta(L) \sim  \frac{\rho_0}{M} \; e^{-1/2} \; \sim 0.03 \;  \mbox{Mpc}^{-3}
\eeq
which is much larger than the value $\eta(L_*) \simeq 3 \; 10^{-3} \mbox{Mpc}^{-3}$ given by observations. Hence the luminosity function implied by the PS approach is much larger than the observed values at luminosities smaller than its cutoff, as we can see in Fig.\ref{figLCDMO1}. A galaxy similar to the Milky Way, with a mass $M \simeq 2.6 \; 10^{12} \; M_{\odot}$, a density contrast $\Delta \simeq 2200$ and a B-band magnitude $M_B \simeq -20.4$, corresponds to a linear parameter $\nu \simeq 1.6$. Hence it is already at the cutoff of the PS multiplicity function while the cutoff of the observed luminosity function corresponds rather to $M_B \simeq -21$. Thus, the luminosity function given by the PS approach falls down at luminosities smaller than what is observed, as we can see in Fig.\ref{figLCDMO1}. Moreover, in the case where $P(k) \propto k^n$ (on galactic scales $n \simeq -2$) we have:
\beq
\sigma(M) \propto M^{-(n+3)/6}
\eeq
This allows us to get (App.\ref{Approximate power-law regimes}) the shape of the luminosity function implied by the PS approach in the three regimes we have already considered $\eta(L)  \propto L^{-0.5} e^{- L/L_s }$, but with a rather small value for $L_s$. Thus, as seen in Fig.\ref{figLCDMO1}, the luminosity function we get in this way has a shape somewhat similar to the observations (a power-law with an exponential cutoff), but, as already discussed by VS, its normalization is too high and its cutoff is too strong.

On the other hand, the non-linear scaling approach leads to:
\beq
\eta(L) = \eta(x) \frac{\mbox{dln}x}{\mbox{dln}L} = \frac{\rho_0}{M} \; x^2 \; h(x) \; \frac{\mbox{dln}x}{\mbox{dln}L}
\label{lumhx}
\eeq
with $x$ given by (\ref{xnl}) and the function $h(x)$ by (\ref{hCDM}). Now, the cutoff at $L_*$ must correspond to $x \simeq x_*$, so:
\beq
\eta(L_*) \sim \frac{\rho_0}{M} \; x_*^2 \; h(x_*) \sim   10^{-3} \;  \mbox{Mpc}^{-3}
\eeq
with a typical mass $M \sim 2\;10^{12} \; M_{\odot}$. This is close to the observed values, hence the normalization of the luminosity function implied by the non-linear scaling approach is consistent with observations, as we can see in Fig.\ref{figLCDMO1}. We can also look at the shape of the luminosity function in the 3 regimes we have already considered. We get (App.\ref{Approximate power-law regimes}) $\eta(L)  \propto L^{-0.46}$ at the faint end and $\eta(L)  \propto L^{-1.4} e^{- L/L_* }$ for the bright galaxies, with a much larger value of $L_*$ than with the PS approximation. Hence we see, Fig.\ref{figLCDMO1}, that the luminosity function we get in this way is quite close to observations: it shows an exponential cutoff for bright galaxies $M_B<-21$ and a power-law behaviour for faint galaxies $M_B>-17$. The existence of a flat plateau for $-21<M_B<-17$ is characteristic of our results (for both the PS and non-linear scaling approaches) and holds also for other power-spectra (e.g. $n=-2$). This feature is quite remarkable, as it did not appear in previous models but is in good agreement with observations, which seem to show an upturn at $M_B>-18$ after a flat portion, see for instance Driver \& Phillipps (1996). This sudden change of the slope of the luminosity function around $M_B \simeq -18$ in our model corresponds to the transition from the curve ${\cal C}_{\Lambda}$ to ${\cal C}_v$, which define the global properties of galaxies (mass, radius,...).

As we noticed in Sect.\ref{The galaxy mass function} the PS mass function can be written in the same form as the non-linear scaling one, with a different scaling function $h(x)$ which has a stronger and earlier cutoff for large $x$ and a higher normalization at $x \sim 1$ (see also VS). The difference between both luminosity functions we can see in Fig.\ref{figLCDMO1} is a direct consequence of the difference between these scaling functions $h(x)$. This suggests in turn that the non-linear scaling function describes the actual outcome of gravitational processes more accurately than the scaling function derived in the PS approach, and hence than the usual PS prescription. Indeed, the relation $x-L$ is strongly constrained by the Tully-Fisher relation, which thus enables one to derive strong constraint on the scaling function $h(x)$. Note however that there is an additional factor $\rho_0/M$ in the luminosity functions (\ref{lumhx}), so that the power-law or the normalization of $h(x)$ cannot be constrained without a specific model for galaxies, like ours, which gives the value of $M$ attached to the parameter $x$. Nevertheless, the exponential cutoff of $h(x)$ is quite strongly constrained by the one of the luminosity function. We can note that Kauffmann et al.(1998) also found that a Press-Schechter approach predicts too many intermediate galaxies as compared to the results of N-body simulations. Although their model is significantly different from ours this effect agrees with our analysis of the mass functions (see Sect.\ref{Comparison of the PS model with the hierarchical scaling approach} and VS).

\subsubsection{$\Omega = 1 \; , \; n=-2$ and $n=-1$}
\label{Omega=1,n=-2 and n=-1}

Similarly to the case of a CDM power-spectrum we can also consider the cases of a power-law $P(k) \propto k^n$ with $n=-2$ and $n=-1$. The case $n=-2$ gives results very close to those obtained for a CDM power-spectrum. This is natural since the latter has a local slope $n \simeq -2$ on galactic scales. On the contrary, the case $n=-1$ produces a very strong exponential cutoff which leads to a luminosity function quite far from the observed one. Indeed, we now have $\sigma^2 \propto M^{-2/3}$ instead of $\sigma^2 \propto M^{-1/3}$ (for $n=-2$) which means that bright massive galaxies are much more rare relative to small ones as compared to the previous case $n=-2$, since density fluctuations decrease faster with larger mass. Thus a power-spectrum index $n=-1$ at galactic scales seems to be incompatible with the observed luminosity function, at least within the framework of our model. Moreover, as most galaxy characteristics are strongly constrained by observations (Tully-Fisher relation, gas/star mass ratio, lower limit for galactic masses and radii, ...) it is very likely that no reasonable model would reconcile the observed galaxy luminosity function with $n=-1$.

\subsection{$\Omega_0 = 0.3 \; , \; \Lambda = 0$}
\label{Omega0=0.3,Lambda=0}

In the case $\Omega_0 = 0.3$ and $\Lambda = 0$, we choose $\Omega_b=0.03$ and $\sigma_8=0.77$. The physical properties of the galaxies are close to those in the $\Omega=1$ universe, the analogs of Fig.\ref{figTrho1}, Fig.\ref{figMgO1} and Fig.\ref{figTF1} show the same behaviour as in the previous case. For instance, a galaxy similar to the Milky Way corresponds to $R \simeq 97$ kpc, $M \simeq 1.6 \; 10^{12} \; M_{\odot}$, $\Delta \simeq 14000$, $T \simeq 3.3 \; 10^6$ K and $L \simeq 3 \; 10^{10} \; L_{\odot}$. Note that we obtain a smaller mass and radius, as compared to the case $\Omega=1$, as we explain in App.\ref{Scalings}.
\\

\begin{figure}[htb]

\centerline{\epsfxsize=8 cm \epsfysize=5.5 cm \epsfbox{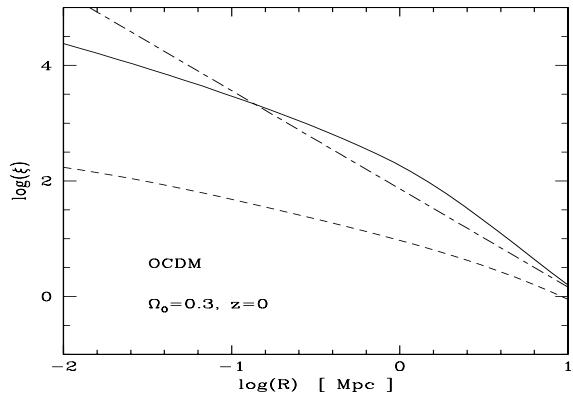}}

\caption{The evolved correlation function $\xia(R,z)$ at $z=0$ (solid line), and the linear extrapolation $\sigma(R)^2$ (dashed line), in the case $\Omega_0=0.3 \; , \; \Lambda=0$ and $P(k)$ is a CDM power-spectrum. The dot-dashed line is the power-law $(R/R_0)^{-\gamma}$ with $\gamma=1.8$}
\label{figXiCDMO03}

\end{figure}

We consider the case of a CDM power-spectrum. Fig.\ref{figXiCDMO03} shows the linear correlation function $\sigma(R)^2$ and the evolved non-linear correlation function $\xia(R)$ at redshift $z=0$.

\begin{figure}[htb]

\centerline{\epsfxsize=8 cm \epsfysize=5.5 cm \epsfbox{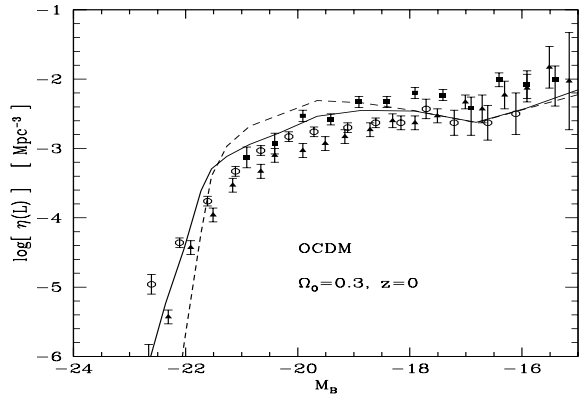}}

\caption{The galaxy luminosity function $\eta(L) dL/L$. The graph shows the quantity $\eta(L)$ as a function of the magnitude in the B band for the non-linear scaling approach (solid line) and the PS prescription (dashed line). The data points are as in Fig.\ref{figLCDMO1}.}
\label{figLCDMO03}

\end{figure}

We see that the evolved non-linear correlation function is still reasonably close to a power-law $\xia(R) \propto R^{-\gamma}$, with $\gamma=1.8$, in the range 30 kpc to 1 Mpc. The scaling function $h(x)$ we need to obtain the galaxy luminosity function in the non-linear scaling approach is not available from the current numerical simulations but as we argued in VS it is expected to be similar to the scaling function obtained in a critical universe with the same power-spectrum. Hence we adopt the function $h(x)$ we used for a critical universe, see (\ref{hCDM}). The galaxy luminosity function we get in this way is shown in Fig.\ref{figLCDMO03} as a function of the B-band magnitude.

As we can see from Tab.1 in App.\ref{Scalings}, when we shift $\Omega_0$ from 1 to 0.3 we should decrease $\Omega_b$ to keep the radius of the Milky Way larger than 60 kpc. In fact, since for a critical universe we had a large radius $R \simeq 140$ kpc, we can keep $\Omega_b$ roughly constant. Thus, we choose $\Omega_b=0.03$ which leads to $R \simeq 97$ kpc. From Tab.1 it implies that the mass of galaxies declines roughly in the same proportion as the radius, and indeed we now have $M \simeq 1.6 \; 10^{12} \; M_{\odot}$ for the Milky Way. The number density of galaxies given by the PS approach is now:
\beq
\eta(L) \sim  \frac{\rho_0}{M} \; e^{-1/2} \; \sim 0.01 \;  \mbox{Mpc}^{-3}
\eeq
which is closer to observations than it was in the case $\Omega=1$, as we can see in Fig.\ref{figLCDMO03}. However, since the relation temperature - luminosity did not change (because it is constrained by the observed Tully-Fisher relation), the cutoff entailed by the PS approach is still too strong. As was the case for a critical universe, the non-linear scaling approach can produce a smoother cutoff at higher luminosities. The analysis made in Sect.\ref{Omega=1,CDM} and App.\ref{Scalings} for the three regimes of galaxies still holds, hence we recover the same slopes. However, in such a low density universe the curve ${\cal C}_v$ is lower relative to ${\cal C}_{\Lambda}$ (as compared to a critical universe) on the analog of Fig.\ref{figTrho1}, hence even for relatively faint galaxies $M_B < -16$ we still are in the regime 2), thus the slope of the luminosity function is very small, for both prescriptions. 
In fact, as we can see in Fig.\ref{figLCDMO03}, the luminosity function is flat between $M_B=-16$ down to $M_B=-20$, which is quite remarkable. Overall, the agreement with observations is still very good.

\section{Time-evolution}
\label{Time-evolution}

The method we presented in the previous sections can obviously be used to derive the galaxy properties at any time, which allows us to get the evolution with redshift of the physical characteristics of galaxies and of their mass function or luminosity function.

\subsection{$\Omega = 1$}
\label{2Omega=1}

\subsubsection{Dark matter properties}
\label{Dark matter properties}

\begin{figure}[htb]

\centerline{\epsfxsize=8 cm \epsfysize=5.5 cm \epsfbox{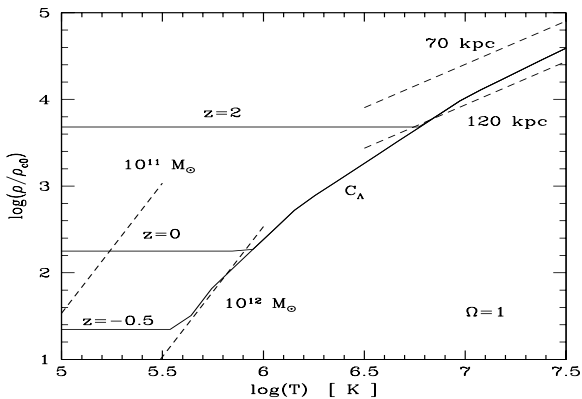}}

\caption{Temperature-density diagram in the case $\Omega=1$ at the redshifts $z=2, \; z=0 \; \mbox{and} \; z=-0.5$ from top down to bottom. The critical density at $z=0$ is noted $\rho_{c0}$. As in Fig.\ref{figTrho1}, the short-dashed lines correspond to halos of fixed mass, on the left, and halos of fixed radius, on the right. Their position does not vary with the redshift.}
\label{figTrho1z}

\end{figure}

The relation temperature-density which corresponds to Fig.\ref{figtcool} is shown in Fig.\ref{figTrho1z} at the redshifts $z=2, \; z=0 $ and $z=-0.5$. Its behaviour is similar to the one we got at $z=0$. As we noticed in Sect.\ref{Galaxies versus clusters and groups} the curves ${\cal C}_v$, hence ${\cal C}$, depend on $z$. This changes the shape of ${\cal C}$, which will modify the shape of the luminosity function $\eta(L)$ as we shall see.

We can notice in Fig.\ref{figTrho1z} that as the redshift increases halos get smaller and less massive. However, since ${\cal C}_{\Lambda}$ does not change with $z$ some halos have fixed physical characteristics below a certain redshift, as long as they stay on ${\cal C}_{\Lambda}$ (in the approximation of stable clustering, which holds for $\xia > 200$). We may consider the evolution of halos in the extreme regimes 1) and 3):
\\

1) Faint galaxies located on ${\cal C}_v \; : \;\; \Delta = $ constant.

At fixed temperature $T$ we get $M \propto (1+z)^{-3/2}$ and $R \propto (1+z)^{-3/2}$. These halos become smaller and less massive, and we have:
\beq
\nu \propto (1+z)^{(1-n)/4} \;\;\; \mbox{and} \;\;\; x \propto (1+z)^{3-3\gamma/2}   \label{nuzxz}
\eeq
where we assumed stable clustering. Thus, as long as the exponential cutoff plays no role ($\nu<1$ or $x<x_*$), the temperature function given by the PS approach verifies $\eta(T) \propto (1+z)^{(7-n)/4}$, that is $\eta(T) \propto (1+z)^{2.25}$ if $n=-2$, while the non-linear scaling approach gives $\eta(T) \propto (1+z)^{1.6}$ if $\gamma=1.8$ and $\omega=0.3$. Hence the comoving number density of halos at these small temperatures increases with $z$ in both prescriptions, but somewhat more strongly within the framework of the PS approach. However, since $\nu$ and $x$ increase the exponential cutoff will eventually lead to a decrease of the comoving number density.
\\

3) Bright galaxies, located on ${\cal C}_{\Lambda} \; : \;\; R = $ constant

The physical properties of these halos remain constant with the redshift while their density contrast decrease as $(1+\Delta) \propto (1+z)^{-3}$. This holds as long as $\Delta > 177$. Their parameters $\nu$ and $x$ (if the clustering is stable) remain constant too so the temperature function does not evolve. This implies that the changes of the comoving luminosity function in the range corresponding to these halos will only be due to the variation of their luminosities (pure luminosity evolution).
\\

As the redshift increases, the regime of galaxies corresponding to a fixed temperature changes: $ 3) \rightarrow 2) \rightarrow 1)$. Finally, for $z \geq 5$ all halos belong to the regime 1).\\

\subsubsection{Star formation history}
\label{Star formation history}

\begin{figure}[htb]

\centerline{\epsfxsize=8 cm \epsfysize=5.5 cm \epsfbox{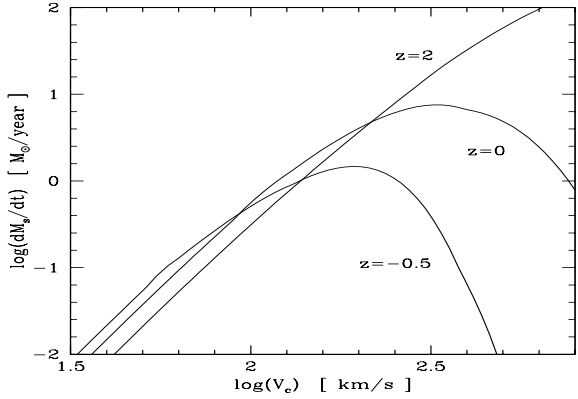}}

\caption{Star formation rate in $M_{\odot}/$year as a function of the circular velocity $V_c$ in km/s, for the redshifts $z=2, \; z=0 \; \mbox{and} \; z=-0.5$ from top down to bottom.}
\label{figSFRO1z}

\end{figure}

Fig.\ref{figSFRO1z} shows the star formation rate $dM_s/dt$ for various redshifts as a function of $V_c$. Small galaxies, with a low circular velocity, have a small star formation rate because their star formation time-scale $\tau_0$ is very long ($\tau_0 \gg t$) since their density and temperature are small. On the other hand, very large and luminous galaxies also have a small star formation rate in the present universe because their star formation time-scale is short as compared to their age ($\tau_0 \ll t$), so they have already consumed most of their gas. In the past they had a higher star formation rate since their gas content was larger.

More precisely, we get for the regimes 1) and 3):
\\

1) Faint galaxies located on ${\cal C}_v$ :

We have $M_g \simeq M_b \propto V^3$ and $\tau_0 \propto V^{-2}$. Since $dM_s/dt \simeq M_g/\tau_0$ we obtain $dM_s/dt \propto V^5$.
\\

3) Bright galaxies, located on ${\cal C}_{\Lambda}$ :

For galaxies which have already consumed most of their gas content we have $M_g \propto V^2 \; \exp(-V/V_0)$ and $\tau_0 \propto V^{-1}$. Hence $dM_s/dt \propto V^3 \; \exp(-V/V_0)$.\\

\begin{figure}[htb]

\centerline{\epsfxsize=8 cm \epsfysize=9 cm \epsfbox{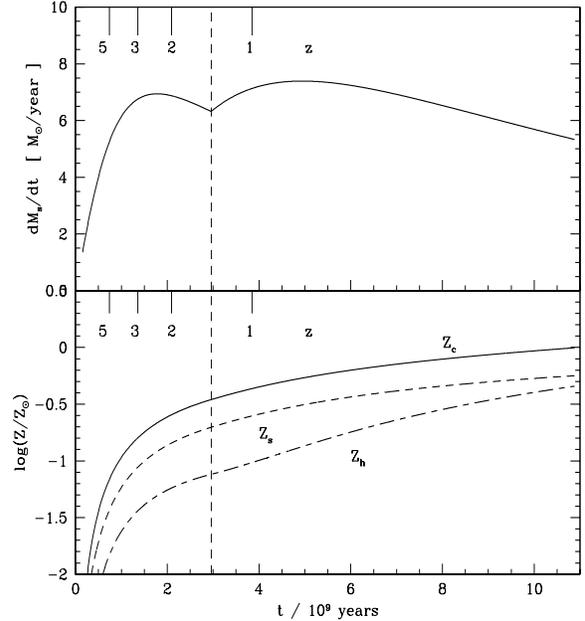}}

\caption{Upper figure: evolution with time of the star formation rate $dM_s/dt$ of the Milky Way. Lower figure: evolution of the metallicities $Z_c, Z_h$ and $Z_s$ of the Milky Way. The vertical dashed-line represents the time $p \; t_m$. The marks on the upper left side of both figures show the redshift.}
\label{figVLt}

\end{figure}

Fig.\ref{figVLt} shows the time-evolution of the star formation rate $dM_s/dt$ (upper panel) and of the metallicities $Z_c, Z_h$ and $Z_s$ (lower panel) of a galaxy similar to the Milky Way. The slight decrease with time of the star formation rate after $5 \; 10^9$ years is due to the smaller gas content of the Galaxy, as its gas is gradually converted into stars. This variation by about a factor 2 is consistent with observations, which imply in fact that the star formation rate of the Milky Way did not vary by much more than this amount. Thus, another prescription with a stronger evolution would increase the global comoving star formation rates or the luminosity functions we shall obtain at $z \sim 1 - 2$, which would improve the agreement with observations, but it would not satisfy the mild evolution observed for the Milky Way. 
One could build for instance such a scenario by using for the star formation time-scale $\tau_c$ a different prescription: $\tau_c \propto 1/\rho_g$ (where $\rho_g$ is the gas density in the core) instead of $\tau_c \propto \rho^{-1/2}$. Such a parameterization, quite plausible, leads naturally to a stronger dependence on time, through the density, and to the effects we described above, as we checked numerically. To keep things simple we shall keep our original prescription, which satisfies the constraints given by the Milky Way. The slight increase of the star formation rate shortly after the time $p\;t_m$ is due to the fact that some gas is still falling onto the galaxy inner parts from the halo.

The temporary increase with redshift of the star formation rate at small times $t < p \; t_m$ (on the left of the vertical dashed-line) is due to the decrease of the relevant time-scales $\tau_c$ and $\tau_d$ at high redshift: $\tau_c \sim \tau_d \propto \rho^{-1/2} \propto (1+z)^{-3/2}$. Note however that during this epoch the mass which will later form the Milky Way is distributed over several smaller halos. The metallicities (meant as [O/H] rather than [Fe/H] since we do not include SNIs) increase steadily with time, in all three components ($M_{gc}, M_{gh}$ and $M_s$). At the time $p \; t_m$, the metallicities $Z_c$ and $Z_s$ were smaller than their present values by a factor $\Delta[Z]=-0.5$, which agrees with the difference in metallicity between the oldest and youngest stars in the disk. Stars created before this date with a high metallicity $-1<[Z]<-0.5$ are located within the bulge, while even older and less metallic stars would be located in the halo. 
The very steep increase in metallicity at the begining implies that there are very few stars of very low metallicity [Z]$<-2$.

\begin{figure}[htb]

\centerline{\epsfxsize=8 cm \epsfysize=5.5 cm \epsfbox{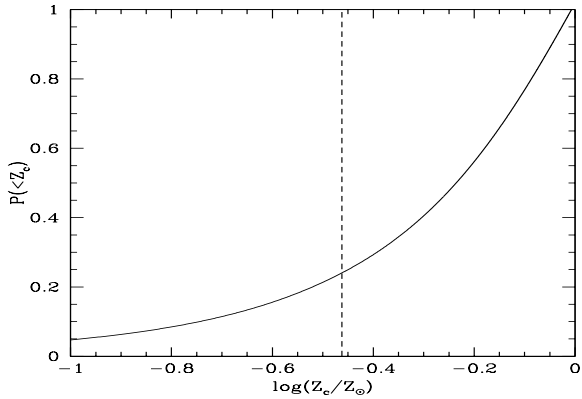}}

\caption{The cumulative stellar metallicity distribution for the Milky Way. The vertical dashed-line represents the metallicity of the newly formed stars at the time $p \; t_m$, when the object switches from the merging regime to the isolated galaxy regime.}
\label{figNZVLO1z}

\end{figure}

Fig.\ref{figNZVLO1z} shows the cumulative stellar metallicity distribution $P(<Z_c)$ for the Milky Way:
\beq
P(<Z) = \frac{M_{lo}(t_Z)}{M_{lo}(t_0)} \;\;\; \mbox{with} \;\;\; Z_c(t_Z) = Z
\eeq
Thus, at time $p \; t_m$, which corresponds to the formation of the disk, the metallicity of the gas, and of the stars being formed, was a factor $0.3$ smaller than the metallicity of stars created today. Moreover, $25 \%$ of present long-lived stars were formed by this date, which is roughly the stellar mass of the bulge. Thus, in our model bulges of disk galaxies contain an important proportion of old stars (with $Z \sim 0.3 Z_{\odot}$ for the Milky Way) which formed before the disk along processes similar to those of elliptical galaxies. This agrees with the interpretations of Ortolani et al.(1995) (who found bulge globular clusters as old as halo globular clusters) and of Jablonka et al.(1996). Note however that some amount of star formation may have kept going on until today. The slow decrease of the metallicity in Fig.\ref{figNZVLO1z} with the fraction of the mass of stars also shows that we have no G-dwarf problem.

\begin{figure}[htb]

\centerline{\epsfxsize=8 cm \epsfysize=5.5 cm \epsfbox{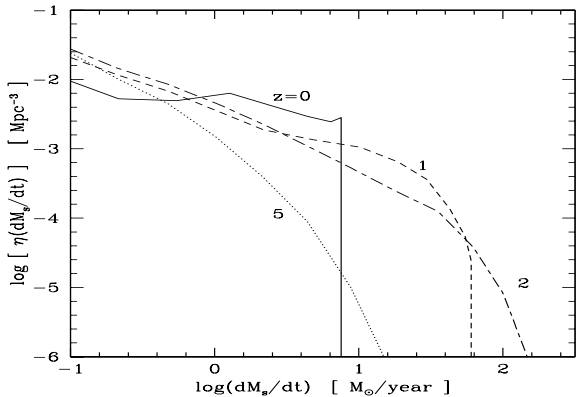}}

\caption{The distribution function of the star formation rates at different redshifts: $z=0$ (solid line), $z=1$ (dashed line) , $z=2$ (dot-dashed line) and $z=5$ (dotted line).}
\label{figdisSFRO1z}

\end{figure}

Fig.\ref{figdisSFRO1z} shows the distribution function of the star formation rates at different redshifts (i.e. the number of galaxies with a given star formation rate per comoving Mpc$^{3}$). It extended to much higher star formation rates in the past, at $z \sim 1-2$, than it does in the present universe, because massive galaxies formed most of their stars at these early epochs (since they have a small star formation time-scale), when they experienced a very active phase, and their star formation rate has steadily declined ever since as their gas content became smaller. This corresponds to the history we developed previously in detail for a galaxy similar to the Milky Way. Moreover, the luminosity (or the galactic mass) of the galaxies characterized by the highest star formation rate was larger in the past (at $z \sim 1$) than it is now. These results agree with the redshift evolution observed by Cowie et al.(1996, 1997).

\subsubsection{Luminosity evolution}
\label{Luminosity evolution}

\begin{figure}[htb]

\centerline{\epsfxsize=8 cm \epsfysize=5.5 cm \epsfbox{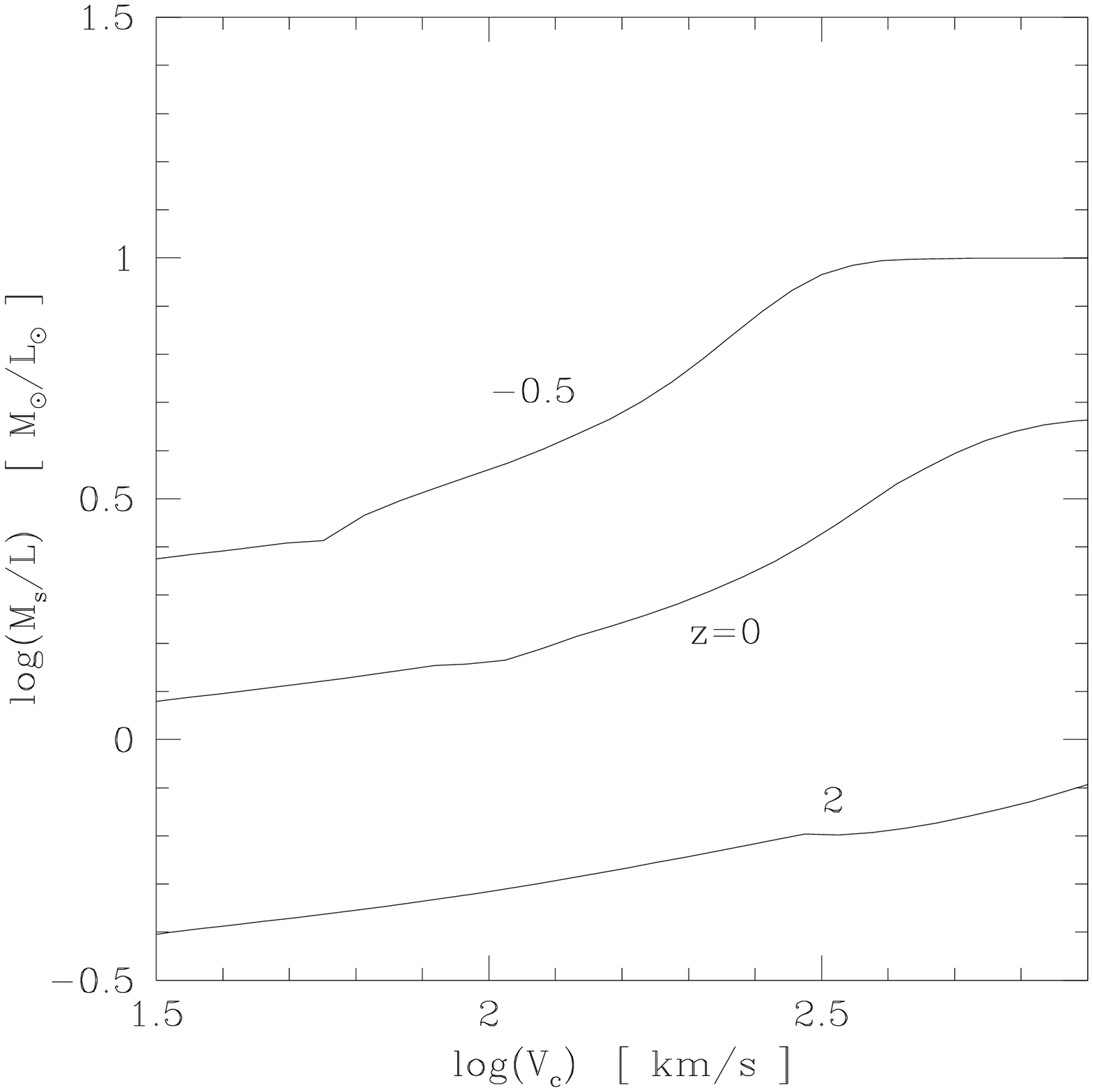}}
\centerline{\epsfxsize=8 cm \epsfysize=5.5 cm \epsfbox{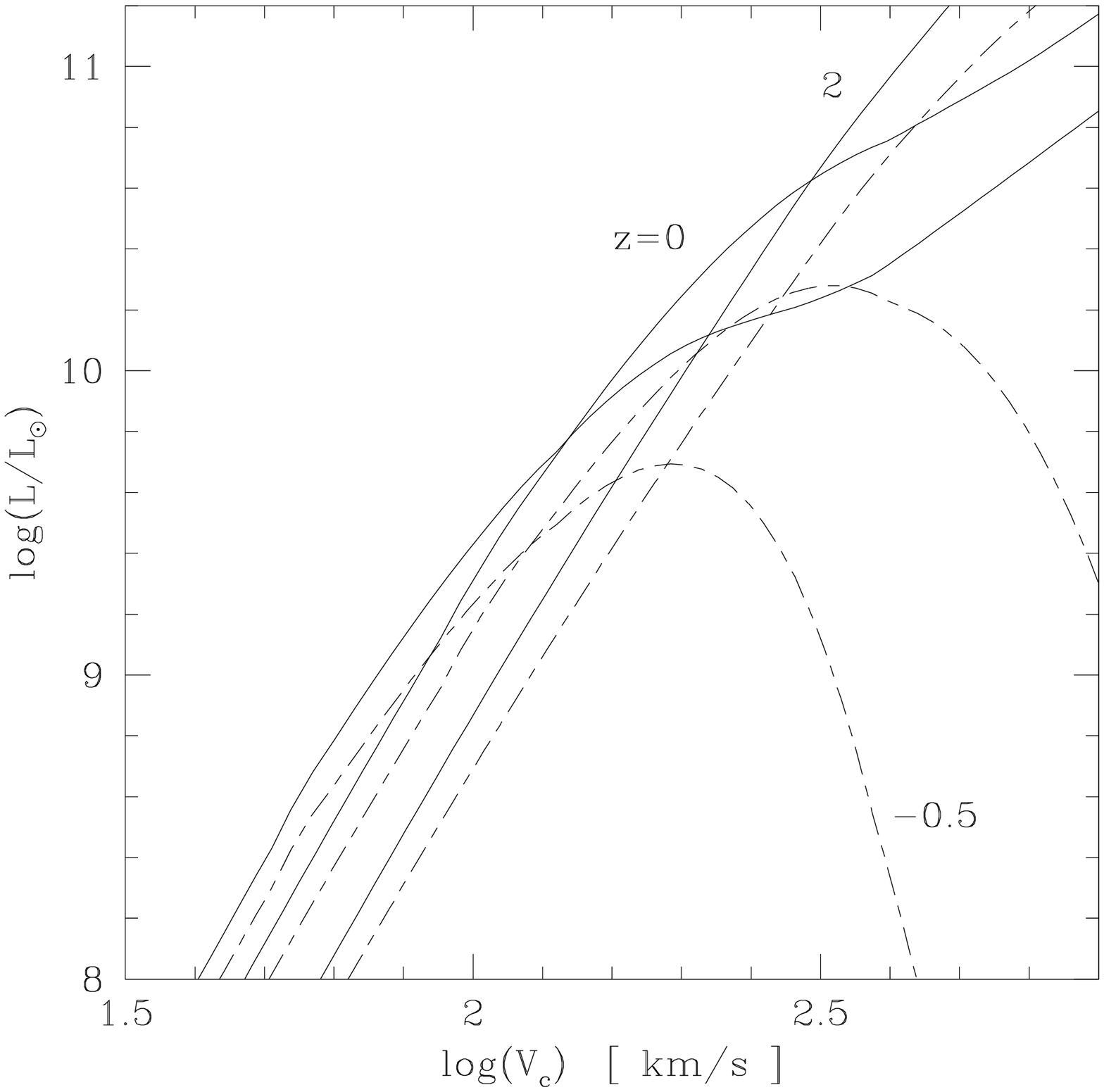}}

\caption{Upper figure: ratio of the total mass of stars $M_s$ over the galaxy luminosity $L$, as a function of the circular velocity $V_c$, for the redshifts $z=2, \; z=0$ and $z=-0.5$. Lower figure: luminosity $L$ of the galaxy as a function of $V_c$. The dot-dashed curves show the luminosity due to massive stars ($L_{sh} \; M_{sh}$). At large $V_c$ a higher $z$ corresponds to larger $L$ and larger $L_{sh} \; M_{sh}$.}
\label{figTF1z}

\end{figure}

We can see in Fig.\ref{figTF1z} that when the redshift increases short-lived stars become more important because these halos are younger, have more gas and a higher star formation rate. Hence the global mass-luminosity ratio gets smaller and, at fixed $V_c$, massive halos which are on the cooling curve ${\cal C}_{\Lambda}$ have a larger luminosity since their mass remains constant (as long as they remain on ${\cal C}_{\Lambda}$). Thus, the slope of the temperature-luminosity relation at high $V_c$ gets stronger. As a consequence, the knee of the luminosity function should move toward larger luminosities in the past, and fainter luminosities in the future, since we noticed above that the temperature function in this region does not evolve with $z$. On the other hand, at small $V_c$ the temperature-luminosity relation keeps the same slope since the analysis developed for $z=0$ is still valid.

However, the normalization decreases at higher redshifts, although the mass/luminosity ratio is smaller, because the mass of these halos decreases. 
Indeed, at higher redshifts the time $\tau_*$, which is the life-time of stars at the boundary between our two classes of stars (short-lived which are recycled and long-lived ones), scales as $\tau_* \propto t_H \propto (1+z)^{-3/2}$, see App.\ref{Stellar properties of galactic halos}. The mass/luminosity ratio of the global stellar population varies more slowly than $\tau_*$ because the IMF contains fewer massive stars:  $M_s/L \propto \tau_*^{1-(x-1)/(\nu-1)} \propto (1+z)^{-3/2 + 3/2 \; (x-1)/(\nu-1)}$ where $x$ and $\nu$ are stellar parameters (IMF, mass-luminosity relation, see App.\ref{Stellar properties of galactic halos}). 
Since for faint galaxies, located on ${\cal C}_v$, we have for a fixed temperature $M_s \propto M \propto (1+z)^{-3/2}$, because the global star formation time-scale $\tau_0$ follows the decrease of the galactic age, the luminosity decreases slowly at high redshifts, as we can check in Fig.\ref{figTF1z}. As galaxies leave the regime 3) to enter the regimes 2) (even when they remain on ${\cal C}_{\Lambda}$, the change is that now their stellar content is small: $M_s \ll M_g$ and $M_g \simeq M_b$) and finally 1), they satisfy the scaling $V_c - L$ described in Sect.\ref{Luminosity} and App.\ref{Approximate power-law regimes} for the regime 1) at $z=0$. This explains why the slope of the Tully-Fisher relation we get remains constant with $z$, and extends up to the bright galaxies at $z \geq 2$ (the high luminosity bend we have for $z \leq 0$ disappears). The regime 1), which only corresponded to the smallest galaxies at $z=0$ is now valid for nearly all galaxies at $z \geq 2$.\\

\subsubsection{Luminosity function}
\label{Luminosity function}

\begin{figure}[tb]

\begin{picture}(230,500)

\epsfxsize=27 cm
\epsfysize=20 cm
\put(-54,-50){\epsfbox{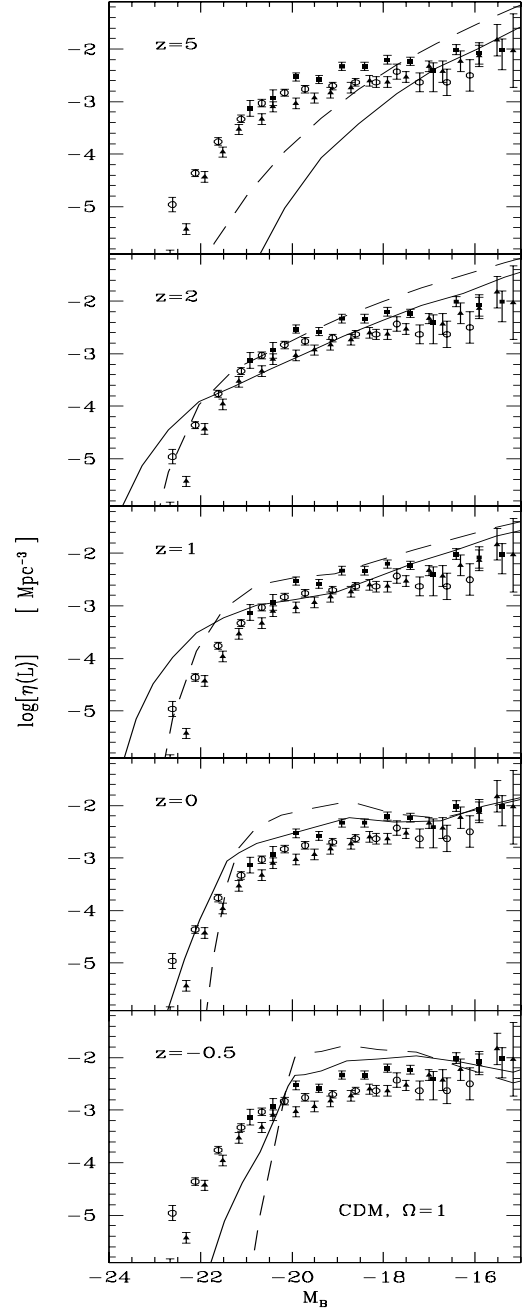}}

\end{picture}

\caption{The comoving galaxy luminosity function $\eta(L) \; dL/L$ at the redshifts 5, 2, 1, 0 and $-0.5$, for the non-linear scaling approach (solid line) and the PS prescription (dashed line). The data points are observational results at $z=0$ as in Fig.\ref{figLCDMO1}.}
\label{figLCDMO1z}

\end{figure}

Fig.\ref{figLCDMO1z} shows the evolution with redshift of the comoving galaxy luminosity function in the case of a CDM power-spectrum. We can see that the knee of the luminosity function moves toward larger luminosities in the past, until $z=1$, and fainter luminosities in the future, while the comoving number density for faint galaxies increases until $z \leq 2$. This is consistent with what we expected from the above analysis. Indeed, for faint galaxies we saw previously that $\eta(T) \propto (1+z)^{2.25}$ for the PS approach and $\eta(T) \propto (1+z)^{1.5}$ for the non-linear scaling prescription. Since the Tully-Fisher relation does not change very much in this range, the luminosity function $\eta(L)$ follows the same behaviour.

On the contrary, for bright galaxies we explained above that while the temperature function $\eta(T)$ does not evolve, the slope of the Tully-Fisher relation gets larger, so that a given temperature corresponds to a larger luminosity in the past, hence the luminosity function cutoff moves toward larger luminosities. We can also notice that the faint-end slope of the luminosity function gets higher in the past. As we said previously, this is related to the change of the curve ${\cal C}$ with $z$. 
Indeed, these faint galaxies leave the regime 2) to enter 1), hence the slope of the luminosity function increases from $\eta(L) \propto L^{-0.2}$ to $\eta(L) \propto L^{-0.5}$ for the PS approach, and from $\eta(L) \propto L^{-0.24}$ to $\eta(L) \propto L^{-0.46}$ for the non-linear scaling prescription (note that although this increase is qualitatively correct, as we can see in Fig.\ref{figLCDMO1z}, the slopes of the luminosity function we obtained in this way are not very accurate because we did not consider the variation of the ratio $M_s/L$). This effect could explain the steepening of the faint-end slope of the luminosity function which is observed in the past at $z=0.5$ or $z=1$. Moreover, the slope of the luminosity function after the cutoff gets smaller, especially for $z \geq 2$. Indeed, we still have (with $n \simeq -2$):
\[
\eta(L) \propto M^{(n-1)/6} \; e^{-(M/M_*)^{(n+5)/3}} \;\; \mbox{or} \;\; \eta(L) \propto M^{\omega_s} \; e^{-M/M_*}
\]
but as galaxies leave the regime 3), where $M_g \ll M_s \simeq M_b$, to enter the regime 2), where $M_s \ll M_g \simeq M_b$, as their gas content increases, the mass-luminosity relation becomes $L \propto M^{5/2}$ instead of $L \propto M$. Hence we get:
\[
\eta(L) \propto L^{(n-1)/15} \; e^{-(L/L_*)^{2(n+5)/15}} \;\;\; \mbox{PS approach} 
\]
\[
\eta(L) \propto L^{2\omega_s/5} \; e^{-(L/L_*)^{2/5}} \;\;\; \mbox{non-linear scaling approach}
\]
With $n=-2$ and $\omega_s=-3/2$ we obtain:
\[
\eta(L) \propto L^{-0.2} \; e^{-(L/L_*)^{0.4}} \;\; \mbox{or} \;\; \eta(L) \propto L^{-0.6} \; e^{-(L/L_*)^{0.4}}
\]
Thus, in both cases the exponential cutoff becomes less sharp in the past (note that the cutoff characteristics $M_*$ or $L_*$ are not the same for the PS and non-linear scaling approaches, as we noticed earlier, in Sect. 4.1.1. for instance).

We can see that the galaxy comoving number density evolves much faster for the non-linear scaling prescription than for the PS approach. Indeed, it leads to a luminosity function which is much higher than the PS one at the bright end for $z \leq 2$ and it suddenly decreases at $z > 2$ to superpose onto the PS prediction at $z \sim 4$ and then gives even fewer galaxies than this latter prescription. The slow evolution for the PS case is due to the fact that $\nu$, hence $\eta(T)$, is constant in the regime 3), and only increases as $(1+z)^{(1-n)/4}$ once the halo enters the domain 1). Moreover, when $\nu$ is not too large, in this regime 1) the prefactor grows as $(1+z)^{(7-n)/4}$ which balances the decrease of the exponential term. On the contrary, in the case of the non-linear scaling prescription $x$ increases slowly until $\xia \sim 177$, since the clustering is not exactly stable. 
Then, when $\xia < 177$ the correlation function $\xia$ decreases suddenly very strongly, and $x$ rises sharply. This produces a sharp decrease of the comoving number density, as the exponential cutoff becomes very important. We must note that for $\xia \ll 177$ the function $h(x)$ should change so that for $\xia \ll 1$ it becomes equal to the result obtained in the quasi-gaussian regime (see Colombi et al.1997). However, the exponential cutoff of the the quasi-gaussian function is stronger than for the non-linear case (Colombi et al.1997), so the predicted decrease of the luminosity function would be somewhat stronger. Thus, it is important to note that up to $z \sim 3$, which is already a rather large redshift, the non-linear scaling approach leads to a much higher luminosity function than the PS prescription at the bright end. In this regime, $\xia$ is still of the order of $\sim 10$ at the onset of the exponential fall-of. 
At such values of $\xia$ it has been seen (Bouchet et al.1991) that $h(x)$ is still given to a good approximation (10\%) by its non-linear form.

\subsubsection{Average comoving stellar properties}
\label{Average comoving stellar properties}

\begin{figure}[htb]

\centerline{\epsfxsize=8 cm \epsfysize=9 cm \epsfbox{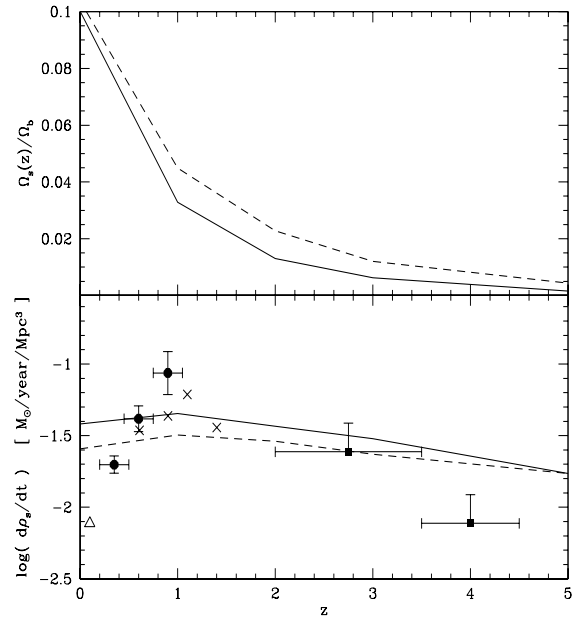}}

\caption{Upper figure: evolution with redshift of the stellar density parameter $\Omega_s(z)/\Omega_b$. Lower figure: comoving star formation rate $d\rho_s/dt$ for $0<z<5$. The solid lines correspond to the case $\Omega=1$ and the dashed lines to $\Omega_0=0.3$. In both cases we only display the non-linear scaling prescription. The data points are taken from Madau et al.(1996) (squares), Lilly et al.(1996) (disks), Gallego et al.(1995) (triangle) and Cowie et al.(1995) (crosses). Note that the points at high redshift ($z>2.5$) are only lower limits.}
\label{figstarz}

\end{figure}

The evolution with redshift of the stellar density parameter $\Omega_s(z)/\Omega_b$, see (\ref{Omegas}), is displayed in Fig.\ref{figstarz} (upper panel). Its decrease in the past is due to two effects. First, we only consider halos with a temperature $T > 10^4$ K (cutoff due to inefficient cooling), which leads to higher parameters $\nu$ and $x$ at high redshifts, see (\ref{nuzxz}), hence to a smaller mass fraction - there is less mass contained in deep potential wells in the past. Second, at larger redshifts galaxies have a higher gas/star mass ratio (smaller $t/\tau_0$), hence most of the mass is in the form of gas. Note that in the present universe, since only galaxies more luminous than the Milky Way have an appreciable stellar content (our Galaxy is just between both regimes as $M_g/M_s \sim 1$) and they form a small part of the total mass (they already are in the exponential cutoff of the luminosity function) the ratio $\Omega_s(z=0)/\Omega_b$ is still small.

We can see in the figure that two thirds of the present mass in stars formed recently at $z<1$. It is also clear that we do not encounter the usual overcooling problem - all the gas cools and is converted into stars within small objects at high redshifts because cooling is very efficient (high densities) - within this framework. As we explain in App.\ref{Analytical solutions}, this is due to the redshift dependence of our star formation time-scale $\tau_0$, which ensures that although the gas may cool at high redshift it cannot be immediately converted into stars ($\tau_0$ does not decrease faster than the age of the universe with redshift). Hence there is still plenty of gas available in the present universe ($\Omega_s \ll \Omega_b$) which is not necessary cold as it has been reheated by supernovae, stellar winds and by the energy released by halo mergings and collapse. 

The comoving star formation rate, see (\ref{drhosdt}), is shown in Fig.\ref{figstarz} (lower figure). It first increases with redshift until $z \sim 1$ because the star formation time-scale $\tau_0$ decreases as $\tau_0 \propto (1+z)^{-3/2}$, as long as $T > T_0$. However, at high redshifts $z>1$ the comoving star formation rate gets smaller because the mass contained in deep potential wells starts to decrease and $\tau_0$ gets larger because of the factor $(1+T_0/T)$. Nevertheless, its variation over the whole range $0<z<5$ is rather small.

\begin{figure}[htb]

\centerline{\epsfxsize=8 cm \epsfysize=5.5 cm \epsfbox{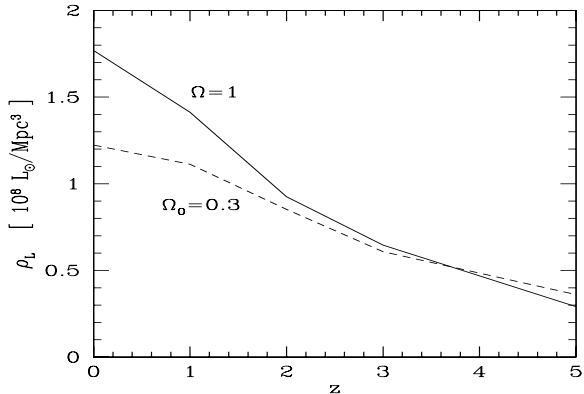}}

\caption{The B-band comoving luminosity density. The solid line corresponds to the case $\Omega=1$ and the dashed line to $\Omega_0=0.3$. In both cases we only display the non-linear scaling prescription.}
\label{figrhoLz}

\end{figure}

The evolution with redshift of the B-band comoving luminosity density is shown in Fig.\ref{figrhoLz}. The luminosity density decreases in the past because the mass fraction within galaxies diminishes and their stellar content is lower. However, the luminosity density remains high until $z \simeq 3$, when it is still equal to one half of its present value. 

On all these figures, we can see that the evolution is slower for a low-density universe, as is well-known.

\subsubsection{Galaxy counts}
\label{Galaxy counts}

Finally, the evolution of the luminosity function allows us to calculate the galaxy number counts as a function of the apparent magnitude $m_B$ and the redshift $z$. The absolute B band magnitude $M_B$ is related to $m_B$ by:
\beq
m_B = M_B + 5 \log\left[ \frac{r_l(z)}{10 \mbox{pc}} \right] + K(z) + E(z)
\eeq
where $r_l(z)$ is the luminosity distance to redshift $z$, $K(z)$ is the usual K-correction and $E(z)$ is the evolution correction. These last two terms vary with the stellar and morphological properties of galaxies, and should be evaluated from our galaxy evolution model, to get a self-consistent result. However, since this would require a detailed description of the stellar properties of galaxies, including their spectra and colors, which we plan to tackle in a future paper, we shall simply take in this article $K(z)+E(z)=0$. Indeed, as can be seen in King \& Ellis (1985) for instance, while $K(z)$ is positive because of the shape of the spectrum, $E(z)$ is negative because galaxies where bluer in the past, and both terms cancel roughly. Of course, the net result varies with the galaxy type, and the evolution model used to get $E(z)$, and for faint magnitudes where the contribution to galaxy counts extends to high redshifts ($z \sim 1$) we may have an error of one magnitude.

However, this approximation should give a fair idea of the implications of our model on the galaxy number counts, and it does not influence the comparison between the PS and non-linear scaling approaches. We note ${\cal N}(m_B,z) dm_B dz$ the number of galaxies per square degree with apparent magnitude $m_B$ to $m_B+dm_B$ and redshift $z$ to $z+dz$:
\beq
{\cal N}(m_B,z) \; dm_B \; dz = \left( \frac{\pi}{180} \right)^2 \; \Phi(M_B,z) \; \frac{dV}{d\omega dz} \; dm_B \; dz
\eeq
where $\Phi(M_B,z) dM_B = \eta(L) dL/L$ is the comoving luminosity function at the redshift $z$. Finally, the number of galaxies $N(m_B) dm_B$ per square degree is the integral over $z$ of ${\cal N}(m_B,z) dm_B dz$:
\beq
N(m_B)  = \left( \frac{\pi}{180} \right)^2 \; \int_0^{\infty} \; \Phi(M_B,z) \; \frac{dV}{d\omega dz}  \; dz
\eeq
Thus, the PS approach leads to:
\beq
{\cal N}(m_B,z) = - \left( \frac{\pi}{180} \right)^2 \; \frac{dV}{d\omega dz} \; \sqrt{ \frac{2}{\pi} } \; \frac{\rho_0}{M} \; e^{-\nu^2/2} \; \frac{\partial \nu}{\partial M_B} 
\eeq
while the non-linear scaling prescription gives:
\beq
{\cal N}(m_B,z) = - \left( \frac{\pi}{180} \right)^2 \; \frac{dV}{d\omega dz} \; \frac{\rho_0}{M} \; x \; h(x) \; \frac{\partial x}{\partial M_B} 
\eeq

\begin{figure}[htb]

\centerline{\epsfxsize=8 cm \epsfysize=5.5 cm \epsfbox{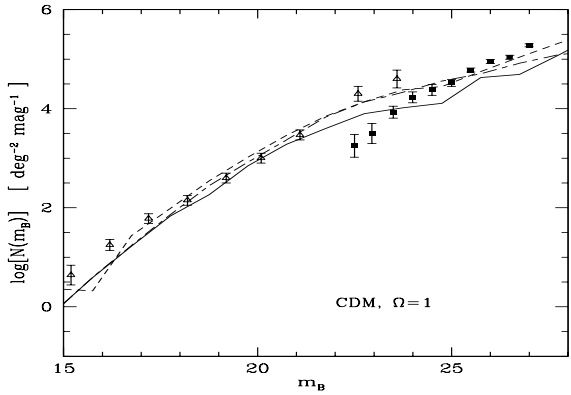}}

\caption{The B band differential number counts. The graph shows $N(m_B)$ (number of galaxies per square degree per apparent magnitude) for the non-linear scaling approach (solid line), the PS prescription (small dashed line), and a no-evolution model (dot-dashed line) such that the comoving luminosity function does not change with $z$. The data points are taken from Lilly et al.(1991) (squares) and Metcalfe et al.(1991) (triangles).}
\label{figNcountCDMO1}

\end{figure}

Fig.\ref{figNcountCDMO1} shows the B counts for the PS and non-linear scaling prescriptions, as well as for a no-evolution model such that the comoving luminosity function does not change with time and is equal to the one given by the non-linear scaling prescription at $z=0$, which allows to distinguish the effects of evolution from those due to geometry (comoving volume element, luminosity distance). Our results agree reasonably well with observations although the number counts given by the non-linear scaling model are somewhat too small at very faint magnitudes $m_B>26$. The counts given by the non-linear scaling approach are lower than those obtained by the non-evolving model because the number of $L_*$ galaxies we get with our model decreases when we look in the past (see Fig.\ref{figLCDMO1z}) and for a fixed apparent magnitude the counts are mainly sensitive to the evolution of the number density of bright galaxies, since we do not see any longer the faintest galaxies. 
This discrepancy with the observational data for $m_B > 26$ is a well-known problem for non-evolving models. In fact, Cole et al.(1992) showed that natural models where the comoving luminosity function evolves in a homogeneous way (evolution of the normalization or of the cutoff) so as to match the B counts will contradict the data in the K band and the observed redshift distribution of galaxies. Hence they concluded that a new population of rapidly evolving blue galaxies fainter than $L_s$ is necessary to fit all the data. Such an effect, even if real, cannot be given by our simple parameterization of star formation. This point deserves a detailed study that will be done elsewhere.

\begin{figure}[htb]

\centerline{\epsfxsize=8 cm \epsfysize=5.5 cm \epsfbox{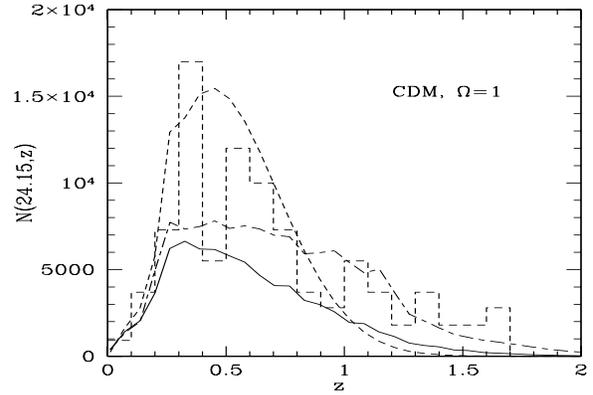}}

\caption{Redshift distribution of galaxies with apparent B band magnitude $m_B=24.15$. The graph shows ${\cal N}(24.15,z)$ for the non-linear scaling approach (solid line), the PS prescription (small dashed line), and the no-evolution model (dot-dashed line). The data points (histogram) are from Cowie et al.(1996) for galaxies in the range $22.9 < m_B < 24.4$.}
\label{figdNodzO1}

\end{figure}

Fig.\ref{figdNodzO1} shows the redshift distribution ${\cal N}(m_B,z)$ of galaxies selected at an apparent magnitude $m_B=24.15$, for the PS and non-linear scaling prescriptions, and for the no-evolution model. Of course we recover the same features as for the integrated number counts: the PS prescription (because of its high normalisation) and the non-evolving model give more galaxies than the non-linear scaling approach. For these last two cases we find a peak at $z \simeq 0.3$ which is consistent with observations (Cowie et al.1996; Colless et al.1993; Colless et al.1990; Broadhurst et al.1988) for this apparent magnitude. Note also the fast evolution of the PS result.

\subsection{$\Omega_0 = 0.3 \; , \; \Lambda=0$}
\label{2Omega0=0.3,Lambda=0}

\begin{figure}[tb]

\begin{picture}(230,500)

\epsfxsize=27 cm
\epsfysize=20 cm
\put(-54,-50){\epsfbox{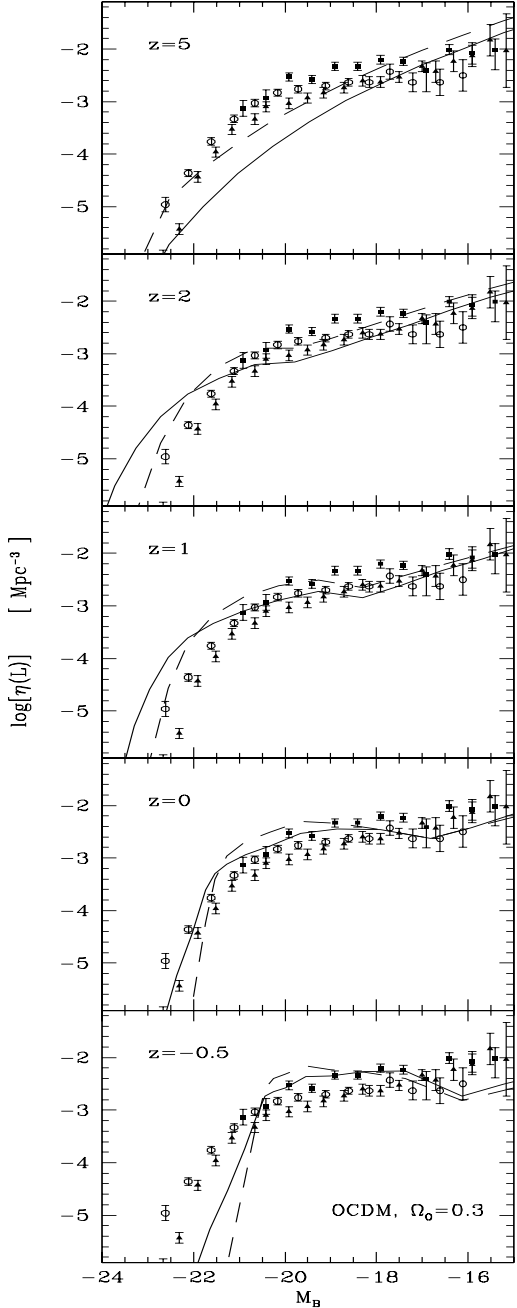}}

\end{picture}

\caption{The comoving galaxy luminosity function $\eta(L) \; dL/L$ at the redshifts 5, 2, 1, 0 and $-0.5$, for the non-linear scaling approach (solid line) and the PS prescription (dashed line). The data points are observational results at $z=0$ as in Fig.\ref{figLCDMO03}.}
\label{figLCDMO03z}

\end{figure}

In the case $\Omega_0=0.3$ and $\Lambda=0$ the analysis developed for a critical universe still holds. Fig.\ref{figLCDMO03z} shows the evolution of the comoving galaxy luminosity function with the redshift in this case for a CDM power-spectrum. As we can see, the evolution is qualitatively similar to what we could see in Fig.\ref{figLCDMO1z} but much slower. This is a well-known property of low-density universes: the evolution of gravitational clustering is slower than for a critical universe.

However, the quantitative difference with the case $\Omega=1$ for the bright end of the luminosity function is quite dramatic, and thus appears to be extremely sensitive on the cosmological parameter $\Omega_0$.

\begin{figure}[htb]

\centerline{\epsfxsize=8 cm \epsfysize=5.5 cm \epsfbox{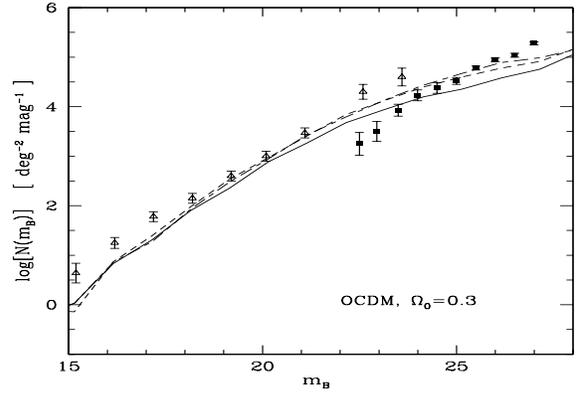}}

\caption{The B band differential number counts. The graph shows $N(m_B)$ (number of galaxies per square degree per apparent magnitude) for the non-linear scaling approach (solid line), the PS prescription (small dashed line), and a no-evolution model (dot-dashed line) such that the comoving luminosity function does not change with $z$. The data points are as in Fig.\ref{figNcountCDMO1}.}
\label{figNcountCDMO03}

\end{figure}

Fig.\ref{figNcountCDMO03} shows the B counts for the PS and non-linear scaling prescriptions, as well as for the no-evolution model. As was the case for $\Omega=1$, the slope of the counts gets smaller at faint magnitudes $m_B>25$.

\begin{figure}[htb]

\centerline{\epsfxsize=8 cm \epsfysize=5.5 cm \epsfbox{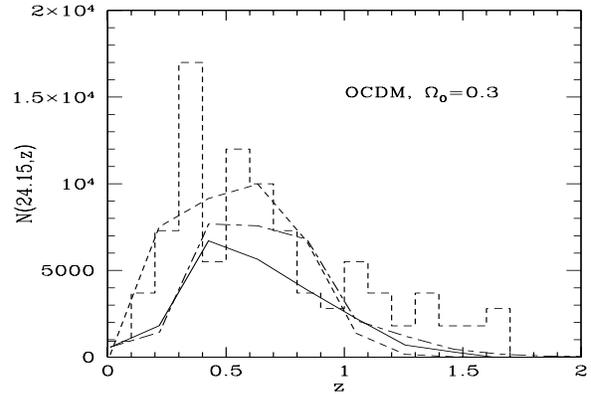}}

\caption{Redshift distribution of galaxies with apparent B band magnitude $m_B=24.15$. The graph shows ${\cal N}(24.15,z)$ for the non-linear scaling approach (solid line), the PS prescription (small dashed line), and the no-evolution model (dot-dashed line). The data points are from Cowie et al.(1996) for galaxies in the range $22.9 < m_B < 24.4$.}
\label{figdNodzO03}

\end{figure}

Fig.\ref{figdNodzO03} shows the redshift distribution of galaxies selected at the apparent magnitude $m_B=24.15$. Once again we can see that the evolution is slower than for $\Omega=1$: the peak moved to higher redshifts $z \simeq 0.4$.

\section{Quasar number density}
\label{Quasar number density}

As quasars are among the first objects to form they provide an interesting probe of the Universe at high redshifts $z \sim 2-5$. Moreover, they constrain the models of gravitational clustering to form objects of large mass at high redshifts. We shall now see whether our formalism can satisfy this requirement. We first assume that the quasar mass $M_Q$ is proportional to the mass of gas $M_{gc}$ available in the inner parts of galaxies:
\beq
M_Q = F \; M_{gc}
\label{MQ}
\eeq
For galaxies which have not already transformed most of their gas content into stars we can use (\ref{MgoMs1}). With (\ref{McMh}) this means that the quasar mass is proportional to the stellar mass $M_Q \sim F \; M_s$. We write the bolometric luminosity $L_Q$ of the quasar as:
\beq
L_Q = \frac{\epsilon \; M_Q \; c^2}{t_Q}
\label{LQ}
\eeq
where $\epsilon \leq 0.1$ is the quasar radiative efficiency (fraction of central rest mass energy converted into radiation) and $t_Q$ is the quasar lifetime. If quasars radiate at most at the Eddington limit at which radiation pressure on free electrons balances gravity (Efstathiou \& Rees 1988; Nusser \& Silk 1993) one has $t_{Q8} \geq 4.4 \; \epsilon$ where $t_{Q8}=t_Q/10^8$ yr. Finally we use a bolometric correction factor $L_B/L_{bol}=0.16$. Since we only consider here very massive and rare quasars we write the quasar comoving number density as:
\begin{eqnarray}
\eta_Q(M_Q) \; \frac{dM_Q}{M_Q} & = & f \; \left[ \eta(M,z) - \eta(M,z[t(z)-t_Q]) \right] \; \frac{dM}{M}   \nonumber \\   & \simeq &  f \; t_Q \;  \frac{\partial}{\partial t} \eta(M,z) \; \frac{dM}{M}  \label{etaQ}
\end{eqnarray}
in a fashion similar to Efstathiou \& Rees (1988) and Nusser \& Silk (1993). Here $\eta(M,z)$ is the galaxy comoving multiplicity function and we assumed a fraction $f \leq 1$ of galaxies actually contains a quasar. Thus we only have two parameters: $(\epsilon \; F / t_{Q8}) \; (\Omega_b / \Omega_0)$ which sets the quasar mass, from (\ref{MQ}) and (\ref{LQ}), and $(f \; t_{Q8})$ which enters the multiplicity function in (\ref{etaQ}). Thus a first estimate of the comoving number density of quasars brighter than $M_B = -26.7$ ($L_B > 8 \; 10^{12} \; L_{\odot}$) is:
\beq
N_{Q1}(>L_B,z) = f \; t_Q \; \frac{dz}{dt} \;  \frac{\partial}{\partial z} N (>M_{min},z) 
\label{etaQ1}
\eeq
where $N(>M_{min},z)$ is the number of galactic halos more massive than $M_{min}$ at the redshift $z$. The minimum galaxy mass $M_{min}$ is given by (\ref{LQ}). However, the mass of gas available in very massive galaxies does not increase linearly with the dark matter mass of the parent halo because in these very dense galaxies star formation was very efficient so that they have already consumed most of their initial gas content, as can be seen from (\ref{Mg}) and (\ref{tau0}). As a consequence, the mass of gas which may power the central black hole reaches a finite maximum for galaxies somewhat more massive than the transition between ${\cal C}_v$ and ${\cal C}_{\Lambda}$. We can note that the maximum quasar mass (or luminosity) obtained in this way {\it decreases} as time goes on, along with the decline in the mass of gas which is progressively turned into stars. This very simple effect will obviously have important consequences, at variance with the predictions of models like (\ref{etaQ1}) where the typical quasar mass keeps increasing with time as larger scales become non-linear. Thus we write a second estimate of the quasar comoving number density as:
\beq
N_{Q2}(>L_B,z) =  N_{Q1}(M_{min},z) - N_{Q1}(M_{max}(z),z)   
\label{etaQ2}
\eeq
where $M_{max}(z)$ is three times larger than the mass which corresponds to the intersection of the curves ${\cal C}_{\Lambda}$ and ${\cal C}_v$ at the redshift $z$. This takes care of the upper bound on the possible quasar mass, as described above. When $M_{max}(z) \gg M_{min}$ (which is always the case at high redshifts) the cutoff at $M_{max}(z)$ has no influence and we recover the number density of quasars given by the previous calculation (\ref{etaQ1}).

\subsection{$\Omega=1$}
\label{3Omega=1}
  
\begin{figure}[htb]

\centerline{\epsfxsize=8 cm \epsfysize=5.5 cm \epsfbox{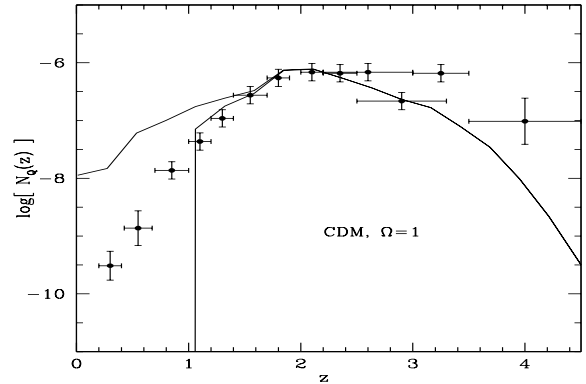}}

\caption{The comoving number density of bright quasars $N_{Q1}(z)$ and $N_{Q2}(z)$. These two estimates superpose at high redshift and $N_{Q2}(z)$ shows a sharper cutoff at low redshift. The observational points are from Pei (1995).}
\label{figquasCDMO1}

\end{figure}

In the case $\Omega=1$ with a CDM power-spectrum, Fig.\ref{figquasCDMO1} shows the comoving number density of bright quasars $N_Q$ as a function of redshift for both prescriptions: $N_{Q1}(z)$ and $N_{Q2}(z)$. We use $(\epsilon \; F / t_{Q8}) = 2.3 \; 10^{-3}$ and $f \; t_{Q8} = 0.018$. This corresponds for instance to $F=0.01$, $\epsilon = 0.1$, $t_{Q8} = 4.4 \; \epsilon$ (Eddington luminosity) and $f=0.04$. Then, the minimum dark matter mass for $M_B<-26.7$ is $3.7 \; 10^{12} \; M_{\odot}$. Note that for nearby galaxies $M_Q \sim 0.006 \; M_{sb}$ (Magorrian et al.1998) where $M_{sb}$ is the stellar mass of the bulge (assumed to have formed at the same time as the central black hole) and that $M_{gc}$ may be smaller than $M_s$ by a factor 2 for the bright galaxies we consider here, see (\ref{MgoMs}).

As exlained above, both prescriptions $N_{Q1}(z)$ and $N_{Q2}(z)$ superpose at high redshift and match the data for $z>2$. We recover the broad maximum between $2<z<3.5$ obtained by observations (see also Hartwick \& Schade 1990 and Warren et al.1994). At low redshifts, the number density of bright quasars given by $N_{Q1}(z)$ declines too slowly as compared with observations. On the other hand, $N_{Q2}(z)$ predicts a very fast decrease so that there are no more bright quasars at $z<1$. Of course, observations show some quasars at $z<1$ but this smoother cutoff could certainly be obtained with a more detailed model which would include some scatter in the mass-luminosity relation (and in the properties of galaxies). Moreover, the physics of quasars is certainly much more complex than the modelisation we used here. Hence we think that these results show that our description is consistent with observations (a better agreement with the data would require a more refined model of quasars themselves to be meaningfull) and provides a very natural simple model for quasars and galaxies. Note also that our prescription for quasars is a straightforward by-product of our model for galaxies: it is its simplest possible extension and we did not have to introduce an ad-hoc redshift dependance in the mass-luminosity relation to obtain a low redshift decline contrary to Haehnelt \& Rees (1993). Note that the counts at high $z$ are very sensitive to the normalization of the power-spectrum and to the mass of the dark matter halos associated to quasars.

\subsection{$\Omega_0=0.3 \;,\; \Lambda=0$}
\label{3Omega0=0.3,Lambda=0}
  
\begin{figure}[htb]

\centerline{\epsfxsize=8 cm \epsfysize=5.5 cm \epsfbox{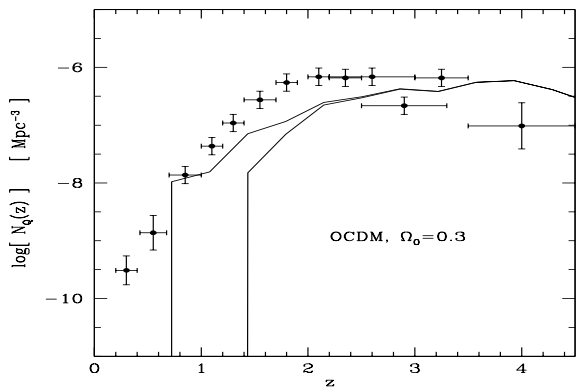}}

\caption{The comoving number density of bright quasars $N_{Q1}(z)$ and $N_{Q2}(z)$. These two estimates superpose at high redshift and $N_{Q2}(z)$ shows a sharper cutoff at low redshift. The observational points are from Pei (1995).}
\label{figquasCDMO03}

\end{figure}

Fig.\ref{figquasCDMO03} shows the comoving quasar number density of bright quasars $N_Q$ as a function of redshift in the case $\Omega_0=0.3 , \; \Lambda=0$ for a CDM power-spectrum. We now use $F=0.007$, $\epsilon = 0.1$, $t_{Q8} = 4.4 \; \epsilon$ and $f=0.07$. This leads to $M_{min}= 2.1 \; 10^{12} M_{\odot}$. Thus the mass of the dark matter halo which corresponds to a given quasar luminosity is lower because the fraction of baryonic matter in this universe is larger. As we can see in the figure we obtain a behaviour similar to the case $\Omega=1$. However, the evolution with redshift of the number density of virialized halos is much slower which implies that the decline of the quasar comoving number density at high $z$ is slower than for a critical universe (the same effect also appeared for the galaxy luminosity function).

\section{Conclusion}

In this article we build a model of galaxy formation and evolution based on the hierarchical clustering picture. 

The counts of non-linear objects are done directly at the epoch under consideration {\it using the actual non-linear density field} (instead of its linear extrapolation) by methods (Schaeffer 1984; Balian \& Schaeffer 1989) that have been shown to represent accurately the density field obtained in numerical simulations (Bouchet et al.1991; Colombi et al.1997). The result, discussed in detail in a previous paper (VS), bears some qualitative resemblance with the Press-Schechter prescription (which explains why the latter appears to be accurate in some cases), but differs quantitatively. Indeed, the Press-Schechter approximation predicts fewer extreme objects (i.e. with a density contrast much larger or smaller than the average fluctuation at fixed mass, or with a mass much smaller or larger than $M_*$ at fixed density threshold $\Delta \sim 200$ where $\sigma(M_*)=1$) and too many intermediate halos (VS). Moreover, our approach allows us to go {\it beyond the traditional Press-Schechter prescription}. Thus we consider mass functions of objects which are defined by a density contrast $\Delta(M)$ which depends on their mass $M$ (see also Valageas et al.1998a).

Indeed we do not assume that galaxies are defined by a constant density threshold resulting from the virialization condition, since with such an assumption one would count induely virialized clusters as galaxies (Schaeffer \& Silk 1985). We add to the virialization condition a cooling constraint (Silk 1977; Rees \& Ostriker 1977) which separates galaxies and clusters. It states that {\it in order to form a galaxy a non-linear object must see its baryonic content cool and settle in its central part before it gets embedded into a larger non-linear halo}. More precisely, galaxies are the largest patch which satisfies these criteria. Virialized objects which do not fulfill this constraint become groups or clusters of galaxies. Note moreover that {\it these groups are divided into several distinct galaxies which we take into account as individual galactic halos} (which satisfy the cooling condition). Then, we can see that in the present universe small and faint galaxies are still defined by the virialization condition, which is more restrictive than the cooling constraint for these small virial temperatures, and are thus the result of many mergers which have kept going on until the present epoch. 
On the contrary, massive $L_*$ galaxies are given by the cooling constraint, which shows that the merging process is no longer effective and their evolution is determined by internal physical processes. 
At earlier epochs (higher redshifts), the proportion of objects which follow the virialization condition, and are thus the result of {\it continuous merging} processes, gets larger. The $L_*$ galaxies reach this regime at $z \sim 2$, a value which is determined solely by the density threshold corresponding to virialization ($\Delta_c \sim 200$) and the cooling processes. This epoch represents the transition from a regime of continuous mergers to a regime where galaxies evolve quietly and can form a disk.\\

To obtain the luminous properties of galaxies we add to this evolution of the number of galactic halos a model of star formation. We use a star formation time-scale proportional to the dynamical time (which is also the period of revolution of a proto-stellar cloud around the galaxy). In addition supernovae winds are assumed to slow down the star formation rate as they can eject or heat some gas (White \& Rees 1978; Dekel \& Silk 1986; White \& Frenk 1991). To this purpose we construct a two-components model where the gas is expelled to the outskirts of the dark halo with an efficiency that is large when the potential well is shallow (which corresponds to faint galaxies) and is not available for star formation until it eventually falls back into the baryonic core. 
This ejected gas may be the source of the metal-rich intergalactic gas in clusters, which we shall study in a forthcoming article (Valageas \& Schaeffer 1998). All unavoidable parameters associated with star formation processes are adjusted to the well-known properties of the Milky Way (mainly its luminosity and the gas fraction in the solar neighbourhood), and not to the luminosity function to be reproduced.\\

Then, {\it our model reproduces the observed Tully-Fisher relation, the characteristic luminosities, masses and radii of galaxies}. Linked with our prescription for the galaxy mass function, it provides a prediction for the galaxy luminosity function which is consistent with observations for a critical universe $\Omega=1$ and a low density universe $\Omega_0=0.3 \; , \; \Lambda=0$ with a CDM (or $n=-2$) initial power-spectrum. However, in this latter case the lack of computational results to which our non-linear scaling model could be compared makes our prediction less certain. On the other hand, this very model with the same initial power-spectrum and star formation processes shows important discrepancies with the observed luminosity function when we use a Press-Schechter-like mass function. This discrepancy {\it cannot be healed without giving away the agreement with the above observations}, and was well-known in this type of approach (inconsistency with the Tully-Fisher relation, too many small galaxies, incorrect color evolution with mass). This, obviously, is no longer a problem in our approach, the improvement arising from our better understanding of the mass function and our more logical assumption that baryon concentrations separate when they are too large for cooling  to occur within a Hubble time at formation.
 
One interesting new feature which arises from this quite classical model of star formation, but less classical with respect to the mass function and the definition of galaxies, is that {\it massive galaxies are more evolved and redder than small ones}, as is observed, although our clustering picture is entirely hierarchical. This was not the case for earlier hierarchical models which often used a constant density threshold to define objects. For this reason, it is often stated that hierarchical models do not lead to the correct colors of galaxies and have difficulties to produce very bright and red galaxies. This statement is shown here not to be related to the hierarchical clustering picture, but to the modelling of galaxies themselves. Note that dust extinction and metallicity effects (not included in this article) would also make bright galaxies redder than would be obtained otherwise (Kauffmann \& Charlot 1998; Somerville \& Primack 1998). The metallicity produced by our star formation process is consistent with the observed correlation with luminosity. It is interesting to note that for the Milky Way the metallicity of stars formed during the early merging regime at $z>2$ is $[Z]<-0.5$ which is indeed the metallicity cut between disk and halo stars.\\

Finally, our analytic model also enables us to derive explicit approximate scaling relations between various properties of galaxies. This clearly shows the consequences of the phenomenological prescriptions adopted for star formation or feed-back processes for instance, as well as the variation of galactic characteristics with the cosmological parameters. In addition, we recall here that our model is part of a global description of structure formation in the universe, which deals with clusters, Lyman-$\alpha$ clouds and reionization processes (detailed in other articles) as well as with the galaxies we studied in this article. Thus, our goal is to provide {\it a simple realistic analytic model} which can describe in a {\it unified} consistent fashion these various phenomena. Despite the simplifications envolved in such an approach, this allows one to obtain reasonable quantitative estimates of these physical processes, to draw an explicit link between these different aspects of the same underlying density field and to get a complementary tool to numerical calculations.\\

Our approach differs from earlier works by many aspects. White \& Frenk (1991) use the Press-Schechter approximation to count overdensities with a fixed density threshold, and within each of these halos consider the cooling condition to determine the fraction of gas that is able to cool, arguing the latter will eventually turn into stars. Cooling, in this approach, does not modify the number of objects, but simply the properties of the latter. Thus they do not solve the old problem (Schaeffer \& Silk 1985; White et al.1987) of predicting too many low-mass galaxies and extremely massive large galaxies (more precisely, they only obtain a ``halo luminosity function'' where the most massive objects are not galaxies but groups or clusters). For similar reasons, in order to avoid the formation of too many huge galaxies Kauffmann et al.(1998) and Somerville \& Primack (1998) impose an ad-hoc cutoff ($V_c < 500$ km/s or $V_c < 400$ km/s) to the halos which are allowed to cool. This problem is cured in our model by the separation of galaxies and galaxy clusters by means of a mass-dependent density threshold as implied by our new implementation of the cooling constraint. 
The higher density threshold insures the proper falloff at a few $10^{12} \; M_{\odot}$ and the slope around $10^6$ K of the cooling curve (see Fig.\ref{figtcool}) provides for the flat slope of the luminosity function. We still get a rather steep slope for very small luminosities,
 similar to the one observed (see comments by Driver \& Phillipps 1996). 
	
Blanchard et al.(1992) discuss the problem of overpredicting the small objects, and mention the Press-Schechter prescription might not be the exact answer (but give no specific alternate prescription), blaming thus the estimate of the mass function. However, the same problems appear in numerical simulations (White et al.1987). In fact, we find that although the estimate of the mass function by the non-linear scaling formulation does reduce the number of faint galaxies, our solution is mainly in implementing the proper effects of cooling. Blanchard et al.(1992) on the other hand suggest another way out, by reheating the universe so as to destroy the objects with the shallower potential. With their solution, the smaller the object the more efficient the destruction mechanism, and the slope of the galaxy luminosity function may be expected to be flat down to the faintest galaxies, whereas for our solution the slope eventually gets steep ($\alpha \sim -1.3$) for the very faint objects.

Kauffmann et al.(1993) follow White \& Frenk (1991) but focus their attention on the individual histories of mass condensations. They note that a fixed density threshold implies that most massive galaxies generally formed recently and consequently have young stellar populations and blue colors, at variance with what is observed. With our views, in total agreement with the assumption of hierarchical clustering, the more massive objects have a higher dark matter density contrast (which does {\it not} imply that their density within their luminous radius is larger than for faint galaxies) and formed at high redshift. Since then, they have formed stars as almost isolated systems and have been able to exhaust their gas, whence exhibiting older stellar populations. In fact, as far as star formation is concerned, massive galaxies have roughly the same age as other galaxies, that is about the age of the universe, since before they reached the curve ${\cal C}_{\Lambda}$ gas was already being transformed into stars, within sub-units which later merged to create the large galaxies we can observe today. 
In this sense, the ``age'' of these bright galaxies (that is the time available for star formation) is even slightly larger than the one of small galaxies, because within our model these massive objects are the final results of the merging of deep potential wells where star formation started earlier since cooling was more efficient there than in weaker halos. However, the main reason in our model (with the fact that all galaxies have roughly the same ``age'') for the old stellar population of luminous galaxies (as compared to faint ones) is that their star formation process is more efficient because of their high density and virial temperature. This trend is also apparent (but to a lesser degree) in the model of Cole et al.(1994).

The study of the galaxy counts as a function of apparent magnitude, where evolution in luminosity and in number are intrinsically tied (Guiderdoni \& Rocca-Volmerange 1990) is not fully satisfactory within our simple description. It deserves a detailed study using our methods but including properly color evolution, and remains to be done with the needed care. Also, the fate of galaxies in the dense parts of clusters, when dark halos start to loose their individuality, would be worth a thorough consideration. Clearly, however, the obvious outcome of such a model where mass and luminosity evolution are treated consistently is the study of the galaxy populations at high redshift. We have here only unveiled some of the applications of our approach.

\begin{acknowledgements}

We thank J. Silk for discussions during various stages of this work.

\end{acknowledgements}

\appendix

\vspace{1cm}

{\bf APPENDIX}

\section{Characteristics of galactic halos}
\label{Characteristics of galactic halos}

Any dark matter halo of radius $R$ can be characterized by its overall density contrast $\Delta$ or equivalently by the parameter $x$ introduced in (\ref{xnl}). We assume that these potential wells have a mean density profile of the form $\rho(r) \propto r^{-\gam}$ where $\gam$ is the slope of the two-points correlation function $\xia$ (we shall come back to this point below). Since $\gamma \simeq 1.8$ is close to $2$, the halo is close to isothermal and the velocity dispersion $\sigma^2 \propto M(R)/R \propto R^{2-\gamma}$ is nearly constant, which agrees with observations. With such a density profile, the velocity dispersion of the dark matter verifies:
\[
\frac{d}{dR} (\rho \; \sigma^2) = - \; \frac{{\cal G} M \rho}{R^2}
\]
which leads to 
\beq
\sigma^2(R) = \frac{1}{2\gamma - 2} \; \frac{{\cal G} M(R)}{R}
\eeq
The energy equipartition $k T = \mu m_p \; \sigma^2$, where $\mu m_p$ is the mean molecular weight of the gas and $m_p$ the proton mass, gives for the gas temperature:
\beq
k \; T(R) = \frac{1}{2\gamma - 2} \; \frac{{\cal G} \mu m_p M(R)}{R}
\label{kT}
\eeq
which also leads to the hydrostatic equilibrium for the gas, using $P=n_b kT=\rho_b/(\mu m_p) \; kT$, where $n_b$ and $\rho_b$ are the baryon number density and mass density, if $\rho_b \propto \rho$. Hence we shall assume that the gas initially follows the spatial distribution of the dark matter in the virialized halo, before it starts cooling, which is consistent with numerical simulations (Evrard 1990). More precisely we assume:
\[
\rho_b = \frac{\Omega_b}{\Omega_0} \; \rho
\]
where $\Omega_b$ is the present ratio of the baryon density to the critical density. We shall take $\Omega_b=0.04$ in the case $\Omega=1$, and $\Omega_b=0.03$ in the case $\Omega_0=0.3$. Both values are consistent with the bounds given by primordial nucleosynthesis (Walker et al. 1991). Finally, we note:
\[
\mu = \frac{\rho_b}{n_b m_p}  = 0.59  \;\;\; \mbox{and} \;\;\; \mu_e = \frac{\rho_b}{n_e m_p} = 1.14 
\]
which correspond to halos of primordial composition (with an helium mass fraction $Y=0.24$), where $n_e$ is the electron number density. We note $V$ the circular velocity of the galaxy at its external radius $R$ :
\beq
V^2 = \frac{{\cal G} M}{R} \;\;\;\; \mbox{and} \;\;\;\; k \; T = \frac{1}{2\gamma - 2} \; \mu m_p \; V^2  \label{V2T}
\eeq
We also define a luminous radius $R_c$, and the circular velocity $V_c$ at $R_c$, by:
\beq
R_c = \sqrt{ \frac{L}{1.9 \; 10^{10} \; L_{\odot}} } \;\; 17 \; \mbox{kpc} \;\;\; \mbox{and} \;\;\; \frac{V_c}{V} = \left(\frac{R_c}{R}\right)^{1-\gamma/2}
\eeq
We shall use $V_c$ to compare our model to the observed Tully-Fisher relation, which relates the luminosity to the circular velocity at a radius $\sim R_c$. Since $\gamma \simeq 2$ we have $V \simeq V_c$ (the rotation velocity is roughly constant throughout the halo).

The time $t_m$ introduced in Sect.\ref{Galaxy formation: cooling constraints} corresponding to the turn-around epoch of a given halo (or to the time when it became non-linear) is given by the spherical collapse model (Peebles 1980):
\beq
\left\{ \begin{array}{ll} R = A (1-\cos \theta) \\ & \;\;\; \mbox{with} \;\;A^3={\cal G} MB^2 \\ t = B(\theta-\sin \theta) \end{array} \right.
\eeq
The time of turn-around is simply $t_m = \pi B$, and the radius of maximum expansion is $R_m = 2 A$. When the overdensity virializes, at $p \; t_m$, its radius is $R_v = R_m/2 = A$, and the averaged density within $R_v$ is $\rho = M/(4 \pi/3 \; R_v^3) = 3 \pi/(4 {\cal G} t_m^2)$. Thus we obtain:
\beq
t_m = \sqrt{ \frac{3 \pi }{4 {\cal G} \rho} }
\eeq
which gives the time of turn-around $t_m$ of the halo as a function of its density.

The slope $\gam$ of the dark matter halos mainly enters our calculations as a numerical factor of order unity in (\ref{kT}). Hence we do not need a precise description of the detailed density profile of the objects we deal with. Moreover, we explicitely consider that very large halos defined by $\Delta \sim 177$ which correspond to clusters of galaxies (at low $z$) contain several higher-density sub-units which are identified as distinct galactic halos. In fact, we recognize directly these smaller objects without dealing (in this article) with their larger host halo.

\section{Star formation model for an isolated halo}
\label{Star formation model for an isolated halo}

With the prescription introduced in Sect.\ref{Galaxy formation: cooling constraints} we are able to obtain the mass function of galaxies, or the temperature, radius functions, but we do not have the luminosity function yet. Indeed, the luminosity of galaxies depends on the processes which govern star formation, which we need to take into account explicitely. We shall adopt a very simple model for the formation of stars within the halos we previously considered.

\subsection{Model}
\label{Model}

For each galaxy we divide stars in two populations: a first class of massive stars, with a life-time $\tau_{sh}$ shorter than the age $t$ of the galaxy, and a second class of small stars with a life-time $\tau_{lo}$ larger than $t$. Thus, the gas used to form these small stars has not been recycled in the ISM, since these stars are still on the main sequence, while the large stars consist of many successive generations which have continuously recycled part of their mass through supernovae explosions, planetary nebulae and winds. We consider explicitely the mass recycled by SNII, and by stars in the AGB phase, but we neglect the mass ejected by SNI (believed to be associated with white dwarf coalescence and explosion) which is very small as compared to the one involved in the former processes. 
To include SNI would be straightforward, but it would not change our results as long as we do not consider the history of Si or Fe. We note $M_g$, $M_{sh}$ and $M_{lo}$ the total mass in the form of gas, short-lived stars and long-lived stars. Moreover, we consider two gaseous phases: some low-density diffuse gas $M_{gh}$ spread over the halo, and some dense gas $M_{gc}$ within the core of the galaxy in the form of clouds which turns into stars with a time-scale $\tau_c$. Initially, we have $M_{gc}=0$ and $M_{gh} = M_b$ with $M_b=\Omega_b/\Omega_0 \; M$ the mass of baryons in the halo (hence we assume that at the time of virialization the baryon fraction within any halo is representative of the universe: there has been no prior segregation between baryonic and dark matter). The diffuse phase $M_{gh}$ is continuously replenished by stellar winds and supernovae, which eject and heat part of $M_{gc}$. Meanwhile, the diffuse gas settles in the central parts of the galaxy and forms dense clouds over a dynamical time-scale $\tau_d$. 
Indeed, the gas ejected by supernovae or ionized by stellar winds, or initially hot after virialization, cools over a time-scale $t_{cool}$ and then falls back into the center of the potential well over a time $\tau_d \sim t_m$. By definition of our halos, the constraints (\ref{Cgal}) ensure that $t_{cool} < \tau_d$ hence $\tau_d$ is the only one relevant time-scale. In fact we even have $t_{cool} \ll \tau_d$ at high redshifts or small temperatures when $\rho_v \gg \rho_{\Lambda}$, in this case most of the gas $M_{gh}$ is cold. This would be different for galaxies in clusters, where the gas may be spread all over the cluster, cooling is less efficient, and where the potential wells of the galactic halos are expected to be modified by the additional cluster gravitational energy. The properties of galaxy clusters, however, will be treated in a forthcoming paper. Finally, we can write: 
\beq
\:\:\:\:\:\:\:\:\:\:\:  \left\{ \begin{array}{rcl}   \frac{dM_{gh}}{dt} & = & - \left( \frac{dM_{gc}}{dt} \right)_{SN} \; - \; \frac{M_{gh}}{\tau_d} \\ \\  \frac{dM_{gc}}{dt} & = & \left( \frac{dM_{gc}}{dt} \right)_{SN} - \; \frac{M_{gc}}{\tau_c} \; + (1-\eta') \; \frac{M_{sh}}{\tau_{sh}} \; + \; \frac{M_{gh}}{\tau_d} \\ \\  \frac{dM_{sh}}{dt} & = & \eta \; \frac{M_{gc}}{\tau_c} - \frac{M_{sh}}{\tau_{sh}}  \\  \\  \frac{dM_{lo}}{dt} & = & (1-\eta) \; \frac{M_{gc}}{\tau_c}   \end{array} \right. \label{systar}  
\eeq
where $\eta \ll 1$ is the mass fraction of the IMF corresponding to short-lived stars, for $1 M_{\odot}$ of gas which is processed into stars, and $\eta' \sim 0.1$ is the fraction of mass which is locked within white dwarfs or neutron stars after the death of these massive stars, and is not recycled in the galaxy to form other generations of stars. We have $\eta \ll 1$ because in the case of usual stellar initial mass functions (IMF) most of the mass is within low mass long-lived stars. In a fashion similar to Kauffmann et al.(1993) we write the mass of gas heated and ejected by supernovae as:
\beq
\left( \frac{dM_{gc}}{dt} \right)_{SN} = - \frac{2 \epsilon E_{SN} \mu m_p \eta_{SN}}{3 \; kT \; m_{SN}} \; \frac{M_{gc}}{\tau_c} = - \frac{T_0}{T} \; \frac{M_{gc}}{\tau_c} \label{flowSN}
\eeq
where $\epsilon \sim 0.1$ is the fraction of the energy $E_{SN}$ delivered by supernovae transmitted to the gas, and we defined:
\beq
T_0 =  \frac{2 \; \epsilon \; E_{SN} \; \mu m_p \; \eta_{SN}}{3 \; k \; m_{SN}} \sim 10^{6} \; \mbox{K}  \label{T0SN}
\eeq
using $\eta_{SN}/m_{SN} \simeq 0.005 \; M_{\odot}^{-1}$ and $E_{SN} = 10^{51}$ erg. This value of $\eta_{SN}/m_{SN}$ is consistent with the IMF we shall use, and it is constrained by the observed supernovae rates. The effect of stellar winds can also be incorporated into this model through $T_0$. 

Using the fact that by definition of short-lived stars $\tau_{sh} \ll t$ we can make the approximation that $M_{sh}$ evolves in a quasi-static way: $\frac{dM_{sh}}{dt} \simeq 0$. Hence we obtain:
\beq
M_{sh} = \eta \; \frac{\tau_{sh}}{\tau_c} \; M_{gc}
\eeq
Then we solve numerically (\ref{systar}) to get the evolution of $M_{gh}$, $M_{gc}$ and all other quantities.

\subsection{Analytic approximations}
\label{Analytical approximations}

For faint galaxies characterized by a small virial temperature $T \ll T_0$ the system (\ref{systar}) leads to a simple approximation, which also provides usefull insight into the behaviour of more massive galaxies, as can be checked numerically. More precisely, this approximation is valid provided $\tau_c \ll (1+T_0/T)^2 \; \tau_d$. In this case, a quasi-stationary regime sets in very quickly, where the galactic evolution is regulated by supernovae (and stellar winds) and the dynamical time-scale: the sink term for the star-forming gas $M_{gc}$, corresponding to matter ejected by supernovae (which involves the star formation rate and the supernovae efficiency parameterized through $T_0$), balances the source term due to the infall of gas from the extended halo $M_{gh}$ (which involves the dynamical time-scale). Hence the global star formation rate is governed by the interplay of the supernovae efficiency and the dynamical time-scale. Then, $M_{gc}$ evolves in a quasi-static way and follows closely the mass of the reservoir $M_{gh}$:
\beq
M_{gc} = [1+T_0/T-\eta (1-\eta')]^{-1} \; \tau_c/\tau_d \; M_{gh}  
\label{McMh}
\eeq
Note that the condition of validity of this approximation implies that $M_{gc} \ll M_{gh}$. Then, since $M_g \simeq M_{gh}$ we obtain
\beq
M_g = M_{g0} \; e^{ - t_g / \tau_0}    
\label{Mg}
\eeq
where $t_g$ is the age of the galaxy and we defined:
\beq
\tau_0 =  \frac{1+T_0/T-\eta(1-\eta')}{1-\eta(1-\eta')} \; \tau_d  \simeq \left( 1 + \frac{T_0}{T} \right) \; \tau_d   
\label{tau0}
\eeq
Thus, as we explained above, the galactic evolution is governed by $T_0/T$ and $\tau_d$, the time-scale $\tau_c$ does not appear in the global gas mass, or stellar content. The instantaneous star formation rate is
\beq
\frac{dM_s}{dt} = \frac{M_g}{\tau_0}
\eeq
since $M_s+M_g=M_{g0}$, hence $dM_s/dt = - dM_g/dt$ (we neglect the mass in the form of short-lived stars). The mass within massive short-lived stars is:
\beq
M_{sh} = \frac{\eta}{1-\eta(1-\eta')} \; \frac{\tau_{sh}}{\tau_0} \; M_g
\label{Msh1} 
\eeq
while the mass in the form of small long-lived stars is: 
\beq
M_{lo} = \frac{1-\eta}{1-\eta(1-\eta')} \; ( M_{g0} - M_g ) 
\label{Mlo1}
\eeq
The mass locked in star remnants, white dwarfs or neutron stars, is
\beq
M_r = \frac{\eta' \; \eta}{1-\eta} \; M_{lo} \label{Mr}
\eeq

\section{Redshift evolution and merging of galaxies}
\label{Redshift evolution and merging of galaxies}

\subsection{Model}
\label{1Model}

In the previous section we considered non-evolving halos, in the sense that the total mass of the galaxy, and its time-scales $\tau_c$ and $\tau_d$, remained constant with time. However, as we explained in Sect.\ref{Galaxies versus clusters and groups}, the characteristics of the galactic halos we consider evolve with time, following the curve ${\cal C}$ (see Fig.\ref{figtcool}). Thus, we now have to model the evolution of the matter located on ${\cal C}_v$ through its merging history. Similarly to White \& Frenk (1991), we can write for the comoving stellar content of the universe at time $t$:
\[
\eta(x,t) M_s(x,t) = \int_0^t dt' \int_0^{\infty} dx' \; \frac{dM_s}{dt}(x',t') 
\]
\beq
\hspace{4.1cm} \times \; \eta(x',t') \; P(x,t|x',t')
\eeq
where $P(x,t|x',t') \; dx$ is the probability (which we do not know) that matter which was in a halo of parameter $x'$ at time $t'$ will be part of a halo of parameter $x$ - $x+dx$ at $t$, and $dM_s/dt(x',t')$ is the star formation rate of the corresponding halo. We can define halos by their parameter $x$ for both PS and non-linear scaling approaches because we noticed in Sect.2 that the usual linear parameter $\nu$ is a function of $x$ only, see (\ref{nux}). If individual halos are stable once in the non-linear regime (when $\Delta > \Delta_c$ and $\xia \gg 1$), their parameter $x$ remains constant with time, but they accrete some mass as their virialization radius gets larger with time and they may join together to form a larger halo. Hence, we shall make the approximation that the mass which is within halos $x - x+dx$ at $t$ was within halos of the same range in $x$ at all earlier times. 
This also satisfies the constraint that mass be conserved since the mass fraction within halos $x - x+dx$ does not depend on time, and is a function of the sole variable $x$, see (\ref{muh}). In fact, the conservation of mass implies to use a distribution in $x$, or to take $x$ constant which should be a good approximation if this distribution is peaked around a particular value. This also ensures that we do not mix the mass of non-evolving galaxies located on ${\cal C}_{\Lambda}$ with the halos on ${\cal C}_v$, and that the mass at a given time $t$ comes from older less massive halos characterized by a slightly weaker potential well since for a fixed $x$ the virial temperature $T$ is smaller at higher redshifts. This is quite natural since as time goes on potential wells merge to form deeper ones.
In other words, we neglect the scatter in all possible ``merger trees'' and simply use a ``mean history'' defined by compatibility requirements.
This analysis also applies to the PS approach, which would only give a different function $h(x)$, hence a different mass fraction. 

In fact, when $t' \ll t$ one expects that the correlation $x' - x$ disappears, or at least gets weaker, but this is not a problem if the stellar content of a given object at $t$ is dominated by its recent star formation history, which is the case.
Moreover, within the framework of our model for star formation small galaxies
are regulated by the interplay between supernovae (which eject gas) and the infall of gas from the outer parts of the halo and their luminosity is 
dominated by recently formed stars which are more numerous. As a consequence 
the details of their previous stellar history are not very important. On the 
other hand, very massive and bright galaxies are not affected by the supernovae feedback mechanism (since their potential well is sufficiently deep to retain the gas very efficiently) so that their past history matters. Besides, at low redshifts their star formation rate begins to decline significantly as they exhaust their gas content. This means that their luminosity and stellar properties are governed by old stars which formed during ealier and more active periods. However, in our astrophysical model we assume that these galaxies (located on the cooling curve ${\cal C}_{\Lambda}$) do not evolve significantly any more so that our approximation $x'=x$ becomes correct for these objects.
Thus, our approximation is consistent with our model and it should provide a
reasonably good description. Moreover, it allows one to get simple analytic 
insights into the global galaxy formation process, which clearly show the 
general trends implied by any such model based on hierarchical structure formation supplemented by cooling constraints. Note that using a different galaxy formation model Kauffmann et al.(1998) found that their results were not very sensitive on the details of the merging trees of their halos (they obtained similar results with N-body simulations and an extended Press-Schechter theory for the properties of individual galaxies although the merger trees differ in detail). Using (\ref{etah}) we obtain
\beq
\frac{M_s}{M}(x,t) = \int_0^{t} dt' \; \left( \frac{1}{M} \; \frac{dM_s}{dt} \right) (x,t')   \label{Mstime}
\eeq
In fact, this approach simply means that we can still use the sytem (\ref{systar}) to get the proportion of the mass of gas which is converted into stars in a galaxy, but the time-scales $\tau_c$ and $\tau_d$, and the virial temperature $T$, are now functions of time.

\subsection{Analytical solutions}
\label{Analytical solutions}

Although in practice we compute numerically the solution of the system (\ref{systar}), we present now the case of the simplified quasi-stationary regime, corresponding to small galaxies, to give a clear illustration of the effects of this additional time-dependence. Moreover, since the temperature $T$ decreases at higher redshifts, for a fixed $x$, all galaxies follow this regime when they are young.\\
 
Since $M_s+M_g = M_b = \Omega_b/\Omega_0 \; M$, and $dM_s/dt=M_g/\tau_0$ in this case, we obtain:
\beq
1 - \left( \frac{M_g}{M_{g0}} \right) (t) = \int_0^t \frac{dt'}{\tau_0(t')} \left( \frac{M_g}{M_{g0}} \right)  (t')  \label{Mgtime}
\eeq
Thus we have an equation similar to the one describing halos located on ${\cal C}_{\Lambda}$, but now the global star formation time-scale $\tau_0$ depends on $t'$. The density $\rho_v$ of the halos located on ${\cal C}_v$, which virialize at the time $t_{vir}=p \; t_m$, is according to the spherical model:
\beq
\rho_v = \frac{p^2 3 \pi}{4 {\cal G} t_{vir}^2}  
\eeq
Hence $\rho_v \propto t^{-2}$, whatever the value of $\Omega_0$. If clustering is stable, for $\xia \gg 200$, the quantity ${\overline \rho} \xia(R,t)$ is constant, hence $\xia \propto R^{-\gamma} \; {\overline \rho}^{\;-1}$. Thus, for a constant parameter $x=\rho_v/({\overline \rho} \xia)$ we obtain for halos on ${\cal C}_v$:
\beq
R \propto \rho_v^{-1/\gamma} \hspace{0.5cm} \mbox{and} \hspace{0.5cm} T \propto \rho_v^{1-2/\gamma}
\eeq
As a consequence, since $\tau_0 \propto \rho^{-1/2} T^{-1}$ for halos such that $T \ll T_0$, see (\ref{tau0}), we get:
\beq
\tau_0(t) \propto \rho_v^{-3/2+2/\gamma} \propto t^{3-4/\gamma}  \;\;\; \mbox{along} \; {\cal C}_v \; \mbox{at fixed} \; x
\eeq
Hence, if we note $t_H$ the age of the universe at the redshift we consider, and $\tau_0$ the star formation time-scale at this date, we can write:
\beq
1 - \left( \frac{M_g}{M_{g0}} \right) = \int_0^{t_H} \frac{dt}{\tau_0} \left( \frac{M_g}{M_{g0}} \right) \; \left( \frac{t}{t_H} \right)^{-1+\frac{2 (1-n)}{3(3+n)}}
\eeq
using $\gamma = 3(3+n)/(5+n)$. We can note that this integral converges for $t \rightarrow 0$, and $T$ (which measures the depth of the potential well) decreases at higher redshifts, provided $-3 < n < 1$, which corresponds to the range of interest where hierarchical clustering is valid. Hence our analysis applies to all relevant power-spectra $P(k)$. Thus we obtain for the gas mass fraction at any time $t$:
\beq
\frac{M_g}{M_{g0}} =  \exp \left[ - \; \frac{3(3+n)}{2(1-n)} \; \frac{t_H}{\tau_0} \; \left( \frac{t}{t_H} \right)^{2(1-n)/(9+3n)} \;  \right]
\eeq
In the case $n=-2$ we have for the gas mass fraction at time $t_H$, when we calculate $\tau_0$,
\beq
\frac{M_g}{M_{g0}} =  e^{ - t_H /(2 \tau_0) }  \;\;\; \mbox{on ${\cal C}_v$ with $T<T_0$}
\eeq
Thus, we obtain a relation similar to (\ref{Mg}), with an ``effective age'' $t_g$ for the galaxy given by $t_g = t_H/2$. Hence the time-evolution of $\tau_0(t)$ reduced this age by a factor $1/2$, as compared to (\ref{Mg}), since for $n=-2$ the time-scale $\tau_0(t)$ increases at high redshifts, which leads to a less efficient star formation, because the temperature $T$ decreases. This would not be the case for large $n$, where the dominant effect would be the increase of the density. Note however that the variation with $n$ of the factor $1/2$ can simply be incorporated into the parameters $\beta_c$ and $\beta_d$. Galaxies located on ${\cal C}_v$ with a high virial temperature $T > T_0$ have a slightly more complicated history. There is a first stage, at small times, where $T<T_0$ and they follow the behaviour we have just described, and a second stage where $T>T_0$ and $\tau_0(t) \propto t$. 
Hence we divide the integral of (\ref{Mgtime}) in two parts, and we get a similar result, with an effective age: $t_g = [ 1+\ln(T/T_0) ] \; t_H/2$. In fact, the relation (\ref{Mg}) does not apply to these massive galaxies, but it still gives a good estimate of their evolution, and this simple result shows that the effective age is increased because the temperature remains constant for some time, as could be expected, but only by a logarithmic term. For massive galaxies located on ${\cal C}_{\Lambda}$ we have $\tau_0=$ constant down to the time $t_{vir}$ when $\Delta=\Delta_c$ and we switch onto ${\cal C}_v$. At small redshifts $z<z_{vir}$ the fact that $\tau_0$ is constant means that the gas mass fraction follows a simple exponential decline $M_g \propto \exp(-t/\tau_0)$, usually different from the evolution $M_g \propto \exp[-(t/\tau_0)^{2(1-n)/(9+3n)}]$ along ${\cal C}_v$. 
The time of virialization is simply $t_{vir} = p \; t_m$ and the star formation time-scale at this date is $\tau_0$. Hence these galaxies have an effective age $t_g$ which is the sum of their effective age at $t_{vir}$, which we obtained above, and of the time which has elapsed since this date: $t_H - p \; t_m$. Thus, to sum up, for all galaxies the quasi-stationary approximation gives again the relations of the previous section, (\ref{Mg}) to (\ref{Mr}), with an effective age given by:
\[
\left\{ \begin{array}{ll}    T < T_0 : &  t_g = 1/2 \; p \; t_m \; + \; (t_H - p \; t_m) \\  \\   T > T_0 : & t_g = 1/2 \; [1+\ln(T/T_0)] \; p \; t_m \; + \; (t_H - p \; t_m) \end{array} \right.
\]
Note that halos on ${\cal C}_v$ satisfy $t_H = p \; t_m$ by definition. Thus, the gas mass fraction within halos at time $t_H$ is given by a relation of the form $M_g/M_{g0} = \exp(-t_H/\tau_0)$, whatever the precise dependence on $z$ of the star formation time-scale, provided the integral of the right-hand side of (\ref{Mgtime}) converges (star formation does not occur as a sudden burst at high redshifts). Moreover, the variation with $n$ of the factor $1/2$ in the age of the galaxy, when it is on ${\cal C}_v$, is simply incorporated into the parameter $\beta_d$ which enters the definition of $\tau_0$, see (\ref{tau0}). Thus, as far as star formation processes are concerned, all galaxies have roughly the same age (of the order of the age of the universe), even if they did not exist as distinct entities over this whole period of time: they continuously accreted some mass or merged with neighbours.\\

We can see that the time evolution of these virialized halos does not depend on $\Omega_0$. This is due to the fact that the internal dynamics of overdensities given by the spherical model does not depend on the background universe. However, the number of such halos will naturally vary with the cosmological parameters. Besides, the redshift evolution depends on $\Omega_0$, in parallel to the relation time-redshift. For instance, if $\Omega=1$ we have $(1+z) \propto t^{-2/3}$ while $(1+z) \propto t^{-1}$ if $\Omega \simeq 0$ and $\Lambda=0$. Thus, in the case $n=-2$ we have for halos on ${\cal C}_v$ defined by a fixed parameter $x$:
\beq
\left\{ \begin{array}{ll}  \Omega=1 \; : \; & \displaystyle{ \frac{M_g}{M_{g0}} = e^{ - t_0 /(2 \tau_0) \; (1+z)^{-3} } } \\  \\  \Omega \ll 1 \; , \; \Lambda=0 \; : \; & \displaystyle{ \frac{M_g}{M_{g0}} =  e^{ - t_0 /(2 \tau_0) \; (1+z)^{-2} } } \end{array} \right.   
\eeq
where $t_0=t_H(z=0)$ is the present age of the universe. In the case of a low-density universe the redshift evolution is slower, because of the faster expansion which implies that the same multiplying factor in redshift corresponds to a smaller factor in time.

We can note that for halos on ${\cal C}_v$, the time-scale which governs their merging history is $\tau_{d}$. Hence, if star formation is also enhanced by gravitational interactions with surroundings and merging with other halos, the natural time-scale is again $\tau_{d}$ which gives another justification for our star formation time-scale $\tau_c \propto \tau_d$. The gas of halos which undergo this succession of mergings can be reheated to the virial temperature by the energy released during these violent encounters. However, for halos located on ${\cal C}_v$ we have by definition $t_{cool} \ll \tau_d$, hence we can neglect these successive phases of reheating.
\\

The evolution equation (\ref{Mgtime}) we obtained is interesting as it readily shows the influence of the parameterization adopted for the star formation time-scale $\tau_0$ on the history of galaxies. Indeed, since $\rho_v \propto t^{-2}$, we can see that the integral on the right-hand side diverges for $t' \rightarrow 0$ if $\tau_0$ is a strong power-law of the density: $\tau_0 \propto \rho^{-\alpha}$ with $\alpha > 1/2$. In fact, there is a cutoff at high redshifts because for a fixed parameter $x$ the temperature gets smaller in the past, and when $T < 10^4$ K cooling is very inefficient so that star formation is suppressed. However, this large decrease of $T$ by a factor 100 (from $\sim 10^6$ K to $10^4$ K) means that the redshift of cutoff $z_{cut}$ corresponding to present galaxies is rather high: $z_{cut} \sim 4$. 
Thus, such a model for $\tau_0$ would imply that most stars formed at $z \sim 4$ and that the global star formation rate has been very low ever since. This is certainly inconsistent with observations, which show that the star formation rate has not experienced such a dramatic decline since $z \sim 4$ and that star formation is still active in the present universe. If $\tau_0$ varies weakly with $\rho$: $\alpha < 1/2$, the integral converges and star formation is dominated by the latest epochs, and increasingly so for smaller $\alpha$, until all the gas is converted into stars (at some time in the future). The case $\alpha=1/2$, which corresponds to our prescription, is intermediary as the divergence would only be logarithmic, which is not a problem because of the cutoff. In fact, the additional temperature dependence $(1+T_0/T)$ introduces another redshift factor which makes the integral converge. 
This strongly suggests that the star formation time-scale $\tau_0$ should vary as $\tau_0 \propto \rho^{-1/2}$, at least at high redshift, (with a possible additional dependence on $T$), independently of the arguments presented in the previous section, so that star formation is neither dominated by a sudden burst at high redshift when most of the mass reaches $T > 10^4$ K, nor by the very recent epochs, since these both cases are inconsistent with observations. Note that it is correct to consider the quasi-stationary regime in this analysis since we are interested in the earliest stages of galactic evolution when $T \ll T_0$.

\subsection{Total mass in stars and global star formation rate}
\label{Total mass in stars and global star formation rate}

We can also consider the evolution with redshift of the total mass in stars per comoving $\mbox{Mpc}^3$. The mass of baryons within galaxies is $\int \; \Omega_b/\Omega_0 \; M \; \eta(M) \; dM/M$. Using $M_b = M_{g0} = M_g+M_s$ we obtain:
\beq
\Omega_s(z) = \frac{\Omega_b}{\Omega_0} \; \Omega(z) \; \int \;  \left( \frac{M_s}{M_{g0}} \right) \; \mu(M) \; \frac{dM}{M}   \label{Omegas}
\eeq
where $\Omega_s(z) = \rho_s(z)/\rho_c(z)$ is the stellar density parameter and $\mu(M) dM/M$ is the mass fraction in galaxies of mass between $M$ and $M+dM$, given by (\ref{muPS}) and (\ref{muh}) for the PS and non-linear scaling prescriptions. Naturally, we always have $\Omega_s(z) < \Omega_b/\Omega_0 \; \Omega(z)$, as is clearly seen in (\ref{Omegas}), since $\int \mu(M) dM/M = 1$. In fact, there is also a low temperature cutoff at $T=10^4$ K, but up to $z \sim 4$ this only involves a very small fraction of the total mass. 

We obtain the comoving star formation rate in the same way. We have $dM_s/dt=[1-\eta(1-\eta')] \; M_{gc}/\tau_c$, for long-lived stars, hence:
\beq
\frac{d\rho_s}{dt} = \int \;  [1-\eta(1-\eta')] \; \frac{M_{gc}}{\tau_c} \; \eta(M) \; \frac{dM}{M}
\eeq
For the quasi-stationary regime we get for the derivative of the stellar comoving mass density:
\beq
\frac{d\rho_s}{dt} = \frac{\Omega_b}{\Omega_0} \; \rho_0 \; \int \; \frac{1}{\tau_0} \; e^{-t/\tau_0} \; \mu(M) \; \frac{dM}{M}   \label{drhosdt}
\eeq	
It first increases with redshift because the gas content of bright galaxies gets higher (depletion term $\exp(-t/\tau_0)$) and the star formation time-scale $\tau_0$ decreases. Indeed, since $\tau_0 \propto \rho^{-1/2}$, see (\ref{tau0}), and $\rho \propto (1+z)^3$ for most galaxies which are close to ${\cal C}_v$, we obtain $d\rho_s/dt \propto (1+z)^{3/2}$ as long as $T \geq T_0$. However, at high redshifts the comoving star formation rate gets smaller because the mass contained in deep potential wells starts to decrease (influence of the cutoff at $T=10^4$ K), and more importantly because as the virial temperature declines the star formation time-scale $\tau_0$ starts to increase (in the case $n=-2$), see (\ref{tau0}). Naturally, if we neglect the very small apparent mass loss at $T < 10^4$ K, we have: 
\beq
\frac{d\rho_s}{dt} = \frac{d}{dt} [ \Omega_s(z) \; \rho_c(z) ]
\eeq
by construction, as implied by (\ref{Mstime}) (and checked numerically).

We can also look at the evolution with redshift of the star formation rate $dM_s/dt$ of individual halos defined by a fixed temperature $T$. Note that two such halos defined by the same temperature at different redshifts are not necessarily formed by the same matter. Along ${\cal C}_v$, we have $\rho \propto (1+z)^3$, hence $\tau_0 \propto (1+z)^{-3/2}$, and $M \propto (1+z)^{-3/2}$. For $\Omega \simeq 1$ the age of the universe scales as $t_H \propto (1+z)^{-3/2}$, and $\exp(-t_H/2\tau_0) \simeq 1$, hence we obtain $dM_s/dt = 1/\tau_0 \; M_{g0} \; \exp(-t/2\tau_0) \simeq$ constant with time. In a similar fashion, for halos located on ${\cal C}_{\Lambda}$ we also get $dM_s/dt \simeq$ constant as long as $t < \tau_0$. Thus, the star formation rate is roughly constant with time for these objects, which is consistent with observations.

\section{Stellar properties of galactic halos}
\label{Stellar properties of galactic halos}

We can note that the mass in the form of short-lived stars is always much smaller than the mass contained in long-lived stars, and increasingly so for well evolved galaxies (when $t \gg \tau_0$) as could be expected. Indeed these galaxies have already consumed most of their gas content, hence their present star formation rate is relatively small (as compared to their past) and their stellar population is dominated by all stars which were formed all along the galaxy history. Thus, for the quasi-stationary approximation we obtain:
\beq
\frac{M_{sh}}{M_{lo}} \simeq \eta \; \frac{\tau_{sh}}{\tau_0} \; \left[ \exp \left( \frac{t}{\tau_0} \right) -1 \right]^{-1} \; \ll 1   \label{MshoMlo}
\eeq
where we used $\eta \ll 1$, since in the case of usual stellar initial mass functions (IMF) most of the mass is within small long-lived stars, and we have $\tau_{sh} \ll t$. Thus, luminous galaxies, which correspond to large circular velocities (as is observed through the Tully-Fisher relation) and high densities (because of the cooling constraint, see the curve ${\cal C}_{\Lambda}$ in Fig.\ref{figtcool}), hence to a small global star formation time-scale $\tau_0$, have consumed most of their gas and will be redder than faint galaxies. This is an important success of our model as this trend is actually observed (Lilly et al.1991, Metcalfe et al.1991), but usually difficult to get by common models. We can also notice that the mass in the form of white dwarfs or neutron stars is always very small as compared to the total stellar mass from (\ref{Mr}). Indeed we get (with $\eta \sim 0.1$ and $\eta' \sim 0.2$):
\beq
\frac{M_r}{M_s} = \frac{\eta' \; \eta}{1-\eta} \sim 0.02
\eeq
as is the case in any reasonable stellar evolution model. We can check that mass is conserved with time in the sense that:
\beq
M_g + M_{lo} + M_r = M_{g0}
\eeq
In fact we have a slight excess of mass $M_{sh}$, which is negligible since we noticed above that $M_{sh} \ll M_{lo}$ for instance. This is due to our quasi-static approximation $\frac{dM_{sh}}{dt} \simeq 0$. Finally, we also consider that for $1 M_{\odot}$ of gas converted into stars a fraction $\eta_d$ goes into ``invisible'' compact objects, such as brown dwarfs, which we included among the long-lived stars. Since observations seem to show that this mass fraction is rather small we choose $\eta_d=0.1$ in the numerical calculations (in fact we could as well use $\eta_d=0$ or $\eta_d=0.2$, since this parameter has almost no influence as long as it remains small). Then, the mass $M_s$ of luminous stars is in our model $M_s=M_{sh} + (1-\eta-\eta_d)/(1-\eta) \; M_{lo} \simeq M_{lo}$, since we have to remove the part of $M_{lo}$ formed by dark objects, but this fraction is negligible. 
Besides, $\eta \ll 1$, $\eta_d \ll 1$ and we have already noticed that $M_{sh} \ll M_{lo}$. Thus, we obtain: 
\beq
\frac{M_g}{M_s} \simeq  \left[ \exp \left( \frac{t}{\tau_0} \right) -1 \right]^{-1} \label{MgoMs}
\eeq
At small times when the gas content of the galaxy is still important we have:
\beq
\frac{M_g}{M_s} \simeq \frac{\tau_0}{t}  \hspace{1cm} \mbox{for} \; t \leq \tau_0  \label{MgoMs1}
\eeq
It is clear on this expression, which is valid for any model of star formation and does not rely on the quasi-stationary approximation, that the observed ratio $M_g/M_s$ of the Milky Way and its age give directly its star formation time-scale $\tau_0$ (whence the parameter $\beta_d$). 
\\

Thus we now have attached a peculiar stellar content to each halo, or galaxy. To get the luminosity of such a galaxy, we only need to precise the luminosity of its stars. We note $L_{sh}$ and $L_{lo}$ the global luminosity of short-lived and long-lived stars per unit mass, so that the luminosity $L$ of the galaxy is:
\beq
L = L_{sh} \; M_{sh} + L_{lo} \; M_{lo}
\eeq
Now we have to precise the values of the quantities $\eta, \; \eta', \; \tau_{sh}, \; L_{sh}$ and $L_{lo}$. We shall derive them from the initial mass function (IMF) of stars and mass luminosity and mass life-time relations. For $1 M_{\odot}$ of matter converted into stars, the number of stars $d{\cal N}_*$ formed in the mass range $m\;-\;m+dm$ is
\beq
d{\cal N}_* = \phi(m) \; dm   \;\;\;\; \mbox{with} \;\;\;\; \phi(m) = a \; m^{-(1+x)}   \label{IMF}
\eeq
where $m$ is the star mass in units of $1 M_{\odot}$ and $a$ the normalization constant. We use $x=1.35$ for $m<1$ and $x=1.6$ for $m>1$, which is similar to the IMFs given by Salpeter (1955) and Scalo (1986). This applies to stellar masses between $m_- = 0.1 M_{\odot}$ and $m_+ = 125 M_{\odot}$. We could change somewhat this IMF (for instance choose $x=1.8$ for $m>1$) without significant variation in our results. Moreover, a fraction $\eta_d$ may go into ``invisible'' compact objects, such as brown dwarfs, so we have
\beq
1 = \eta_d \; + \; \int_{m_-}^{m_+} m \; \phi(m) \; dm   \label{etad}
\eeq
which defines the normalisation of $\phi(m)$. We note $m_*$ the mass which separates the two classes of stars we introduced above. Then, we obtain:
\beq
\eta = \int_{m_*}^{m_+} m \; \phi(m) \; dm 
\eeq
and
\beq
(1-\eta) = \eta_d + \int_{m_-}^{m_*} m \; \phi(m) \; dm
\eeq
The mass $M_{lo}$ is formed of luminous stars, with $m_- < m < m_*$, and of dark objects. Next we use the mass luminosity and the mass life-time relations:
\[
L = L_{\odot} \; m^{\nu} \;\;\; \mbox{and} \;\;\; \tau_l = 10^{10} \; \frac{M}{M_{\odot}} \; \frac{L_{\odot}}{L} = 10^{10} \; m^{1-\nu} \; \mbox{years}
\]
with $\nu=3.3$, which is consistent with the mean observed mass $-$ B band luminosity relation for stars on the main sequence. Hence we have:
\beq
\eta \; \tau_{sh} = \int_{m_*}^{m_+} m \; \tau_l(m) \; \phi(m) \; dm
\eeq
and
\beq
\eta \; \tau_{sh} \; L_{sh} = \int_{m_*}^{m_+} \tau_l(m) \; L(m) \; \phi(m) \; dm
\eeq
Similarly, we get:
\beq
(1-\eta) \; L_{lo} = \int_{m_-}^{m_*} L(m) \; \phi(m) \; dm
\eeq
Using (\ref{IMF}) and the previous mass-luminosity relation we can see that the luminosity of the galaxy is dominated by the contribution of stars of mass close to $m_*$. Finally we use $\eta'=0.2$, and the mass $m_*$ which separates the two classes of stars we introduced above is chosen to be given by:
\beq
\tau_* \equiv \tau_l(m_*) = 0.5 \; t_H
\eeq
so that $\tau_{sh}$ is smaller than the age of the galaxies, which is close to the age of the universe at the time of interest. For instance, in the case $\Omega=1$ a galaxy like the Milky Way, that is with a circular velocity $V_c=220$ km/s, corresponds in the present universe to $t_0=t_H(z=0)=11 \; 10^9$ years, $m_*=1.3 \; , \; \tau_{sh}/t_0=0.11$ and $\eta=0.25$.

Thus we can attach a luminosity to each galaxy, which enables us to get the luminosity function of galaxies from the mass function:
\beq
\eta(L) \; \frac{dL}{L} = \eta(M) \; \frac{dM}{M}       
\eeq

We can also use this model to obtain the supernovae rate in galaxies. Thus, we assume that stars more massive than $m_i = 6 \; M_{\odot}$ will explode as supernovae after they leave the main sequence. Naturally, they must also be part of our class of short-lived stars to explode during the life-time of the galaxy. The mass in the form of these massive stars is:
\beq
M_{SN} = \eta_{SN} \; \frac{\tau_{SN}}{\tau_c} \; M_{gc}
\eeq
Thus, in the case of the quasi-stationary regime we obtain for the number of such stars at any time:
\beq
N_{SN} = \frac{M_{SN}}{m_{SN}} = \frac{\eta_{SN}}{1-\eta(1-\eta')} \; \frac{\tau_{SN}}{\tau_0} \; \frac{M_g}{m_{SN}} 
\eeq
and the supernova rate is:
\beq
R_{SN} = \frac{N_{SN}}{\tau_{SN}} = \frac{\eta_{SN}}{1-\eta(1-\eta')} \; \frac{M_g}{m_{SN}} \; \tau_0^{-1}
\eeq
with
\beq
\frac{\eta_{SN}}{m_{SN}} = \int_{\mbox{Max}[m_*,m_i]}^{m_+} \; \phi(m) \; dm
\eeq
In the case $\Omega=1$, we obtain for a galaxy similar to the Milky Way a supernovae rate of 2.5 explosions per century, which is close to the observed value for type II supernovae.

\section{Metallicity}
\label{1Metallicity}

Finally, we can derive the metallicity from our model (\ref{systar}). The metallicity here is understood as being the abundance of Oxygen, or any other element that is not significantly produced in SNI which are not included in our model. We define $Z_h$ and $Z_c$ as the fractions of metals within the diffuse gas $M_{gh}$ and within the gas in the core of the galaxy $M_{gc}$. We note $Z_s$ the mean stellar metallicity. Thus, we obtain: 
\beq
\left\{ \begin{array}{rcl}   \frac{dZ_h}{dt} & = & \frac{T_0}{T} \; \frac{M_{gc}}{M_{gh}} \; \frac{Z_c-Z_h}{\tau_c} \\ \\  \frac{dZ_c}{dt} & = & \frac{y}{\tau_c} \; + \; \frac{M_{gh}}{M_{gc}} \; \frac{Z_h-Z_c}{\tau_d} \\ \\  \frac{dZ_s}{dt} & = & [1-\eta(1-\eta')] \; \frac{M_{gc}}{M_s} \; \frac{Z_c-Z_s}{\tau_c}   \end{array} \right.  \label{metal}
\eeq
where $y$ is the yield. The fractions of metals in both gas phases $M_{gh}$ and $M_{gc}$ vary because of the exchange of matter between these two components. In addition, the central gas $M_{gc}$ is enriched by stellar ejecta. The mass of metals in stars increases as metals are incorporated from the star-forming gas $M_{gc}$. In the case of the quasi-stationary regime, we obtain for the gas in the diffuse phase:
\beq
Z_h =  \frac{y}{1+T/T_0} \; \frac{t_g}{\tau_0}  \label{Zh}
\eeq
while we get for the dense central phase:
\beq
Z_c =  \frac{y}{1+T/T_0} \; \frac{t_g}{\tau_0} \; +  \; \frac{y}{1+T_0/T}  \label{Zc}
\eeq
where $t_g$ is the age of the galaxy. Since the halo is not enriched directly, but through the mass loss of the galactic core, its metallicity remains for a long time much smaller than the one attached to this central component which receives the stellar ejecta. The second term of $Z_c$, which is constant in time, corresponds to the fact that very quickly a ``quasi equilibrium'' regime sets in where the gain of metals within this central phase $M_{gc}$ from stellar ejecta balances the loss due to the exchange of matter with the diffuse phase, which replaces some gas with the metallicity $Z_c$ by some gas falling from the halo with the metallicity $Z_h \ll Z_c$. 
Since this exchange of gas between both components $M_{gc}$ and $M_{gh}$ is driven by supernovae, or stellar winds, whose importance is parameterized by $T_0/T$, see (\ref{flowSN}), the equilibrium metallicity is smaller for small temperatures (weak potential wells) where these flows of matter are more important. Then $Z_h$ increases linearly with time as $M_{gh}$ receives some metals from $M_{gc}$ at a constant rate (the term in $Z_h$ corresponds to the second term in $Z_c$, simply multiplied by a temperature-dependent factor and time). Naturally, when $Z_h$ reaches the ``stationary'' value described above for $Z_c$, this ``equilibrium'' regime stops, both phases have the same metallicity which increases linearly with time as the enrichment process goes on. The stellar metallicity is:
\beq
Z_s = \frac{y}{1+T/T_0} \; \left( 1 - \frac{t_g/\tau_0}{\exp(t_g/\tau_0) -1} \right) \; + \;  \frac{y}{1+T_0/T}  \label{Zs}
\eeq
The second term in $Z_s$, which is constant in time, corresponds to the ``stationary'' regime we described for $Z_c$. Since stars form from the dense phase $M_{gc}$, their metallicity quickly reaches the surrounding gas metallicity $Z_c$ which is constant with time. Later, when most baryons are incorporated into stars ($t_g > \tau_0$), the stellar metallicity increases to reach $y$ if the previous equilibrium value was smaller than $y$. Indeed, at very long times when all the gas is converted into stars the mean stellar metallicity $Z_s$ must be equal to the yield by definition, since our system is closed (there is no global mass loss nor gain although the mass of individual galaxies evolves with time).

\section{Role of parameters and scalings}
\label{Scalings}

The model we constructed in the previous sections has only two main specific parameters: $(sq)$ and $\beta_d$, which determine respectively the size (and mass) of galaxies and their star formation rate. For massive galaxies, with $T>T_0$, there is also a dependence on $\beta_c$, while for faint galaxies there is a dependence on $T_0$, but this is not critical. Naturally, it also depends on the usual cosmological parameters $\Omega_0, \; \Omega_b, \; h$, and on the power-spectrum $P(k)$ for the multiplicity functions. Now we shall see how we can get the value of $(sq)$ and $\beta_d$, for a given set of cosmological parameters and a fixed $T_0$.

Let us consider a galaxy like the Milky Way located on the curve ${\cal C}_{\Lambda}$ (see Fig.\ref{figtcool}), defined by a fixed temperature $T$ or circular velocity $V$. From the relation $t_{cool}=q \; t_{m}$ we obtain $\rho \propto (sq)^{-2} \Omega_0^2 \; \Omega_b^{-2}$, and then all the characteristics of this galaxy. Thus, we get:
\beq
\frac{M_g}{M_s} \propto \beta_d (sq) \Omega_0^{-1/2} \Omega_b \Delta_c^{1/2} h 
\eeq
and
\beq
L_{lo} \; M_{lo} \propto  (sq) \left( \frac{\Omega_b}{\Omega_0} \right)^2 
\eeq
Besides, since $M_{sh} \ll M_{lo}$ we have $L \sim L_{lo} \; M_{lo}$, and increasingly so for massive galaxies like the Milky Way as we noticed earlier (in fact, for faint galaxies which have not exhausted their gas content $L_{sh} \; M_{sh} \sim L_{lo} \; M_{lo}$ while for bright galaxies which have already consumed most of their gas content $L_{sh} \; M_{sh} \ll L_{lo} \; M_{lo}$). Hence, a larger $(sq)$ leads to a higher gas/star mass ratio and a larger luminosity, which is quite natural since it means a less constraining cooling constraint, whence a broader, more massive and lower density halo. Similarly, a larger $\beta_d$ leads to a higher gas/star mass ratio, since it means a longer star formation time-scale, whence fewer stars. It does not influence the luminosity because the mass in the form of long-lived stars is of the order of the initial mass of gas for these dense galaxies. 
As a consequence, the observed luminosity and gas/star mass ratio of the Milky Way, which corresponds to a circular velocity of $220$ km/s and a temperature $T=3.6 \; 10^6$ K, give the value of $(sq)$ and $\beta_d$, as a function of the cosmological parameters. Thus we obtain:
\beq
(sq) \propto \left( \frac{\Omega_0}{\Omega_b} \right)^2 \hspace{0.5cm} \mbox{and} \hspace{0.5cm} \beta_d \propto \Omega_0^{-3/2} \Omega_b \Delta_c^{-1/2} h^{-1}
\eeq
Now, we can use these values of $(sq)$ and $\beta_d$ to get the variations of all physical characteristics of the galaxies we considered in our model with the cosmological parameters. Some of these scalings are shown in Tab.1. However, these relations are quite general. For instance, the fact that $M_{gc}/M_s \simeq 0.4$ for the Milky Way gives $M_s \simeq 0.7 \; \Omega_b/\Omega_0 \; M$ if the initial ratio of baryons in the halo is representative of the Universe, there has been no loss of baryons since the formation of the galaxy and most of the gas is in the disk. Then, the luminosity $L_G$ of the Galaxy, and its ratio $(M_s/L)_G$, lead to $M \simeq \Omega_0/\Omega_b \; 1.3 \; (M_s/L)_G \; L_G \propto \Omega_0/\Omega_b$. In this way we obtain the mass and the radius of the halo from which the baryons which constitute the Milky Way came, whence the position of the curve ${\cal C}_{\Lambda}$ and the value of the product $(sq)$. 
For $\Omega_0=1, \; \Omega_b=0.04$, we get $M = 2.6 \; 10^{12} \; M_{\odot}$. In Tab.1 we consider two types of galaxies: i) massive galaxies like the Milky Way located on the cooling curve ${\cal C}_{\Lambda}$, which have a high luminosity and have already consumed most of their initial gas, and ii) small galaxies located on ${\cal C}_v$ (hence characterized by the density contrast $\Delta_c$), which are faint and have a small stellar content.

\begin{table}
\begin{center}
\caption{Scalings}
\begin{tabular}{ll}\hline

\hspace{0.2cm} Massive galaxies \hspace{0.5cm} & \hspace{0.5cm} Small galaxies  \\ 
\hline\hline
\\ 

$\hspace{0.2cm} \rho \propto ( \Omega_b/\Omega_0)^2$   &  $\rho \propto ( \Delta_c \; \Omega_0 \; h^2 )$ \\

$\hspace{0.2cm} R \propto  \Omega_0/\Omega_b$   &  $R \propto ( \Delta_c \; \Omega_0 \; h^2 )^{-1/2}$ \\

$\hspace{0.2cm} M \propto  \Omega_0/\Omega_b$   &  $M \propto ( \Delta_c \; \Omega_0 \; h^2 )^{-1/2}$ \\

$\hspace{0.2cm} \rho_0/M \propto \Omega_b \; h^2$   &   $\rho_0/M \propto \Delta_c^{1/2} \; \Omega_0^{3/2} \; h^3$ 

\end{tabular}
\end{center}
\label{table1}
\end{table}

Thus, the values of the cosmological parameters imply well-defined galaxy characteristics. For instance, for a given $\Omega_0$, a larger $\Omega_b$ (that is a higher baryon fraction) leads to a smaller radius $R$ (since the total baryon mass, linked to the luminosity, must not vary), a smaller mass $M$, whence a larger number of objects (measured by $\rho_0/M$). Since the radius of a galaxy similar to the Milky Way should be larger than 60 kpc we get a higher limit on $\Omega_b$, while the observed luminosity function, which gives the number density of galaxies, sets a lower limit. These two limits could have been incompatible, since for $\Omega_0=1$ there is no freedom in the choice of the function $h(x)$ which determines the multiplicity function in the non-linear scaling approach. 
Thus it appears to be a remarkable success that for $\Omega_0=1$ it is possible to find a baryon fraction (in our case $\Omega_b=0.04$) which is compatible with the three constraints provided by 1) the nucleosynthesis predictions, 2) the galaxy masses and 3) their luminosity function. Note however that for this latter constraint we still have the choice of the power-spectrum $P(k)$, that is its local index $n$ and its amplitude parameterized by $\sigma_8$. Nevertheless, these two parameters are also constrained by observations and we find that the choice $\sigma_8=0.5$ and $n \simeq -2$ or a CDM power-spectrum (which is in good agreement with observations) provides satisfactory results. In fact, we could also choose a lower value for $\Omega_b$, but this would create some problems for Lyman-$\alpha$ clouds (Valageas, Schaeffer \& Silk 1998). In the case of a low-density universe $\Omega_0=0.3, \; \Lambda=0$, the conditions 1) and 2) imply $\Omega_b \simeq 0.03$.

\section{Approximate power-law regimes}  
\label{Approximate power-law regimes}

From the locus of galaxy formation implied by virialization and cooling
we may distinguish using the quasi-stationary approximation three regimes characterized by a specific power- law behaviour:
\\

1) Very faint galaxies, located on ${\cal C}_v \; : \;\; \Delta \simeq 200 $ 
and $T < 10^{5.5}$ K, with a {\it constant density contrast}.

2) Faint galaxies, located on ${\cal C}_{\Lambda}$ : $R \simeq 120$ kpc and $10^{5.5} \; \mbox{K} < T < 10^{6.5}$ K, with a nearly {\it constant radius}. 

3) Bright galaxies located on ${\cal C}_{\Lambda} \; : \;\; R \simeq 100$ kpc and $T > 10^{6.5}$ K, which {\it nearly have exhausted all their gas}.

\subsection{Star formation}

1) Very faint galaxies

Using (\ref{V2T}), (\ref{tau0}) and (\ref{MgoMs1}) (since these galaxies have only consumed a very small fraction of their gas content) we get:
\beq
M \propto V^3 \hspace{0.5cm} \mbox{and} \hspace{0.5cm} \frac{M_g}{M_s} \propto V^{-2}
\eeq
Moreover, since these galaxies have a small stellar content $M_s \ll M_g$ we have $M_g \simeq M_b \propto M$, hence:
\beq
M_s \propto V^5
\eeq

2) Faint galaxies

We obtain:
\beq
M \propto V^2 \;\;\; , \;\;\; \frac{M_g}{M_s} \propto V^{-3} \;\;\; \mbox{and} \;\;\; M_s \propto V^5
\eeq
 
3) Bright galaxies

Since these galaxies have lost most of their gas, which has already formed stars, we now have $M_s \simeq M_b \propto M$, and using $\tau_0 \propto \rho^{-1/2} \propto V^{-1}$ we get:
\beq
M \propto V^2 \;\;\; , \;\;\; \frac{M_g}{M_s} \propto \exp(- V/V_0)
\eeq

\subsection{Mass/Light ratio and Tully-Fisher relation}

For a constant star-mass/luminosity ratio $M_s/L$ we obtain for the 3 regimes we described above the scaling relations:

1) Very faint galaxies:

\beq
L \propto V^5  \;\;\; \mbox{and} \;\;\; L \propto M^{5/3}
\eeq

2) Faint galaxies:

\beq
L \propto V^5 \;\;\; \mbox{and} \;\;\; L \propto M^{5/2}
\eeq

3) Bright galaxies:

\beq
L \propto V^2 \;\;\; \mbox{and} \;\;\; L \propto M
\eeq

\subsection{Metallicity}

1) Very faint galaxies:

Since we noticed earlier that $L \propto V^5$, we obtain:
\beq
Z_h \propto V^2 \propto L^{0.4} \;\;\; \mbox{and} \;\;\; Z_s \simeq Z_c \propto V^2 \propto L^{0.4} 
\eeq

2) Faint galaxies:

We still have $L \propto V^5$, so we get:
\beq
Z_h \propto V^3 \propto L^{0.6} \;\;\; \mbox{and} \;\;\; Z_s \simeq Z_c \propto V^2 \propto L^{0.4} 
\eeq
The increase of the slope of $Z_h$, as compared to the previous regime, is due to the fact that the evolution time-scale $\tau_0$ gets smaller for more luminous galaxies (their density increases, which implies that their dynamical time decreases), and for a constant galactic age (of the order of the age of the universe) it means that the halo is more evolved, hence more enriched. 

3) Bright galaxies:

These galaxies have already converted most of their initial gas content into stars, which implies that $Z_s \simeq y$. Most of the gas is in the dense component $M_{gc}$, since supernovae are not very efficient and $t_0 > \tau_d$, hence $Z_h \simeq Z_c$ and we recover the usual one component closed-box model, with $Z=y \; \ln (M_{g0}/M_g)$. Thus, in our case we obtain $Z_c \simeq y \; t/\tau_0$. Since $L \propto V^2$ we have:
\beq
Z_h \simeq Z_c \propto V \propto L^{0.5} \;\;\; \mbox{and} \;\;\; Z_s \simeq y 
\eeq

\subsection{Slope of the luminosity function}

Using the PS approximation, we get for an $n = -2$ initial spectrum (that gives results quite close to those obtained with CDM initial conditions) the following behaviour.

1) Very  faint galaxies:

\[
\eta(L) \propto L^{(n-3)/10}  \;\;\; \mbox{hence} \;\;\; \eta(L) \propto L^{-0.5}
\]

2) Faint galaxies:

\[
\eta(L) \propto L^{(n-1)/15}  \;\;\; \mbox{hence} \;\;\; \eta(L) \propto L^{-0.2} 
\]

3) Bright galaxies:

\[
\eta(L) \propto L^{(n-1)/6} e^{- (L/L_s)^{(n+5)/3} } \;\;\; \mbox{and} \;\;\; \eta(L)  \propto L^{-0.5} e^{- L/L_s }
\] 

Using the non-linear scaling approximation, we write $h(x) \propto x^{\omega-2}$ for $x \ll x_*$, where $\omega=0.4$, and $h(x) \propto x^{\omega_s-1} \; e^{-x/x_*}$ for $x \gg x_*$ with $\omega_s=-1.4$.
\\

1) Very  faint galaxies:

\[
\eta(L) \propto L^{(\gamma \omega-3)/5}  \;\;\; \mbox{hence} \;\;\; \eta(L) \propto L^{-0.46}
\]

2) Faint galaxies:

\[
\eta(L) \propto L^{2(\omega-1)/5}  \;\;\; \mbox{hence} \;\;\; \eta(L) \propto L^{-0.24}
\]

3) Bright galaxies:

\[
\eta(L) \propto L^{\omega_s} e^{- L/L_* } \;\;\; \mbox{hence} \;\;\; \eta(L)  \propto L^{-1.4} e^{- L/L_* }
\]

\end{document}